\documentclass[prd,twocolumn,showpacs,amsmath,amssymb,preprintnumbers,nofootinbib]{revtex4}
\usepackage{graphicx}
\usepackage{multirow}
\usepackage{amsmath}
\usepackage{amssymb}
\usepackage{xspace}
\usepackage{gensymb}
\usepackage{hhline}
\usepackage{multirow}
\usepackage{booktabs}
\usepackage{hyperref}
\usepackage{cleveref}
\usepackage{aas_macros}
\usepackage{soul}
\usepackage{xcolor}
\usepackage{array}
\usepackage{adjustbox}
\usepackage{cancel}

\newlength{\wdo}
\newcommand{\stroke}[1]{{$#1$}%
\settowidth{\wdo}{${#1}$} {\kern-\wdo}%
\partialvartstrokedint}

\makeatletter
\newcommand{\fancysep}{%
  \@afterindentfalse
  {\begin{center}
    \resizebox{0.8\linewidth}{0.4ex}{{%
        \fontsize{20}{24}\usefont{U}{webo}{xl}{n}{4}}}%
  \end{center}}\@afterheading}
\makeatother

{\vskip 1.0em
\framebox[\columnwidth][r]{%
\begin{minipage}[c]{\columnwidth}%
\vspace{-1.0em}%
#1%
\end{minipage}}}
{\vskip 1.0em}

\def\XXint#1#2#3{{\setbox0=\hbox{$#1{#2#3}{\int}$}
     \vcenter{\hbox{$#2#3$}}\kern-.5\wd0}}

\newcommand{\beq}{\begin{equation}}
\newcommand{\eeq}{\end{equation}}
\newcommand{\beqa}{\begin{eqnarray}}
\newcommand{\eeqa}{\end{eqnarray}}

\newcommand{\pppd}[1]{\frac{\partial{\hphantom{#1}}}{\partial{#1}}}

\newcommand{\eps}{\ensuremath{\epsilon}}

\newcommand{\Heiii}{\ensuremath{{^3}\textnormal{He}}\xspace}

\newcommand{\Heiv}{\ensuremath{{^4}\textnormal{He}}\xspace}

\newcommand{\Livii}{\ensuremath{{^7}\textnormal{Li}}\xspace}

\newcommand{\nut}{\ensuremath{\nu_\tau}}

\newcommand{\neff}{\ensuremath{N_\textnormal{eff}}\xspace}

\newcommand{\phie}{\ensuremath{\phi_e}\xspace}
\newcommand{\bardens}{\ensuremath{\omega_b}\xspace}
\newcommand{\burst}{{\sc burst}\xspace}

\newcommand{\num}{\ensuremath{\nu_\mu}\xspace}

\newcommand{\bnum}{\ensuremath{\bar\nu_\mu}\xspace}

\newcommand{\tin}{\ensuremath{T_\textnormal{in}}\xspace}
\newcommand{\tstop}{\ensuremath{T_\textnormal{stop}}\xspace}
\newcommand{\tcm}{\ensuremath{T_\textnormal{cm}}\xspace}

\newcommand{\nbins}{\ensuremath{N_\textnormal{bins}}\xspace}
\newcommand{\epsmax}{\ensuremath{\epsilon_\textnormal{max}}\xspace}
\newcommand{\ssamp}{\ensuremath{\langle|\mathcal{M}|^2\rangle}\xspace}
\newcommand{\feq}{\ensuremath{f^\textnormal{(eq)}}\xspace}
\newcommand{\nuratiotol}{\ensuremath{\varepsilon({\rm net}/{\rm FRS})}\xspace}
\newcommand{\epsvals}{$\epsilon$-values\xspace}
\newcommand{\epsval}{$\epsilon$-value\xspace}
\newcommand{\tcmpl}{\ensuremath{T_\textnormal{cm}/T}\xspace}
\newcommand{\np}{\ensuremath{n/p}\xspace}
\newcommand{\nprates}{\ensuremath{n\leftrightarrow p}\xspace rates}

\newcommand{\nue}{{\ensuremath{\nu_{e}}}\xspace}

\newcommand{\bnue}{\ensuremath{\bar\nu_e}\xspace}

\newcommand{\ecuti}{\ensuremath{E_\textnormal{cut}^\textnormal{(1)}}\xspace}
\newcommand{\ecutiii}{\ensuremath{E_\textnormal{cut}^\textnormal{(3)}}\xspace}
\newcommand{\etransii}{\ensuremath{E_\textnormal{trans}^\textnormal{(2)}}\xspace}
\newcommand{\elimi}{\ensuremath{E_\textnormal{lim}^\textnormal{(1)}}\xspace}
\newcommand{\elimii}{\ensuremath{E_\textnormal{lim}^\textnormal{(2)}}\xspace}

\newcommand{\ecutii}{\ensuremath{E_\textnormal{cut}^\textnormal{(2)}}\xspace}
\newcommand{\etransi}{\ensuremath{E_\textnormal{trans}^\textnormal{(1)}}\xspace}

\newcommand{\eout}{\ensuremath{E_\textnormal{out}}\xspace}
\newcommand{\qout}{\ensuremath{q_\textnormal{out}}\xspace}
\newcommand{\ein}{\ensuremath{E_\textnormal{in}}\xspace}
\newcommand{\qin}{\ensuremath{q_\textnormal{in}}\xspace}

\newcommand{\pmax}{\ensuremath{p_\textnormal{max}}\xspace}
\newcommand{\pmin}{\ensuremath{p_\textnormal{min}}\xspace}
\newcommand{\pmed}{\ensuremath{p_\textnormal{med}}\xspace}
\newcommand{\spl}{\ensuremath{s_\textnormal{pl}}\xspace}
\newcommand{\stot}{\ensuremath{s_\textnormal{tot}}\xspace}

\newcommand{\ben}{\begin{enumerate}}
\newcommand{\een}{\end{enumerate}}

\font\FermiSmallfont=cmssq8 scaled 1200

\def\LANLppthead#1{
\null 
\begin{center}\vskip -1.0truein{\hbox to 7.0truein {
\hfill
\vbox to 1in {\vfill \FermiSmallfont
              \hbox{#1}
              \vfill}
}}\vskip-0.0truein\end{center}}

\begin{document}

\preprint{LA-UR-15-29163}

\title{Neutrino energy transport in weak decoupling and big bang
nucleosynthesis}

\author{E. Grohs$^{1,2}$}
\author{G. M. Fuller$^1$}
\author{C. T. Kishimoto$^{1,3}$}
\author{M. W. Paris$^4$}
\author{A. Vlasenko$^{1,5}$}

\affiliation{$^{1}$Department of Physics, University of California,
San Diego, La Jolla, California 92093, USA}
\affiliation{$^{2}$Department of Physics, University of Michigan, Ann Arbor, Michigan
48109, USA}
\affiliation{$^{3}$Department of Physics and Biophysics, University of
San Diego, San Diego, California 92110, USA}
\affiliation{$^{4}$Theoretical Division, Los Alamos National
Laboratory, Los Alamos, New Mexico 87545, USA}
\affiliation{$^{5}$Department of Physics, North Carolina State University,
Raleigh, North Carolina 27695, USA}

\date{\today}

\begin{abstract} 
   We calculate the evolution of the early universe through the epochs of weak
   decoupling, weak freeze-out and big bang nucleosynthesis (BBN) by
   simultaneously coupling a full strong, electromagnetic, and weak nuclear
   reaction network with a multi-energy group Boltzmann neutrino energy
   transport scheme.
   The modular structure of our code provides the ability to dissect the
   relative contributions of each process responsible for evolving the dynamics
   of the early universe in the absence of neutrino flavor oscillations.  Such
   an approach allows a detailed accounting of the evolution of the $\nu_e$,
   $\bar\nu_e$, $\nu_\mu$, $\bar\nu_\mu$, $\nu_\tau$, $\bar\nu_\tau$ energy
   distribution functions alongside and self-consistently with the nuclear
   reactions and entropy/heat generation and flow between the neutrino and
   photon/electron/positron/baryon plasma components.
   This calculation reveals nonlinear feedback in the time evolution of
   neutrino distribution functions and plasma thermodynamic conditions (e.g.,
   electron-positron pair densities), with implications for: the phasing
   between scale factor and plasma temperature; the neutron-to-proton ratio;
   light-element abundance histories; and the cosmological parameter \neff.
   We find that our approach of following the time development of neutrino
   spectral distortions and concomitant entropy production and extraction from
   the plasma results in changes in the computed value of the BBN deuterium
   yield.  For example, for particular implementations of quantum corrections
   in plasma thermodynamics, our calculations show a $0.4\%$ increase in
   deuterium.  These changes are potentially significant in the context of
   anticipated improvements in observational and nuclear physics uncertainties.
\end{abstract}


\maketitle

\section{Introduction} 
\label{sec:intro} 


In this paper we concurrently solve for the evolution of the neutrino
and matter/radiation components in the early universe.  A key result
of this work is that there is, in fact, nonlinear feedback
between these components during the time when the neutrinos go from
thermally and chemically coupled with the plasma of
photons/electrons/positrons/baryons, to completely decoupled and free
streaming. This feedback can be important for high precision
calculations of the primordial light element abundances emerging from
big bang nucleosynthesis (BBN). The work we describe here builds on
the many previous studies of the evolution of the neutrino energy
distribution functions in the early universe (see
Refs.\ \cite{1982PhRvD..26.2694D,1982NuPhB.209..372C,
1989ApJ...336..539H,1991PhRvD..44..393R,1992PhRvD..46.3372D,
1995PhRvD..52.1764H,Dolgov:1997ne,Gnedin:1998ne,
1999NuPhB.543..269D,1999PhRvD..59j3502L,2000NuPhB.590..539E,
Mangano:3.040,neff:3.046,2015NuPhB.890..481B} and Appendix
\ref{trans-app:overview}). Higher precision in theoretical
calculations of neutrino transport and nucleosynthesis in the early
universe is warranted by recent and anticipated improvement in the
precision of cosmological observations.

The advent of high precision cosmological observations will demand a
deeper understanding and higher precision in modeling the known
microphysics of the standard model relevant during early universe
through the neutrino weak decoupling (or simply ``weak decoupling'')
and BBN epochs. For example, future cosmic
microwave background (CMB) polarization experiments promise increased
sensitivity to issues closely associated with relic neutrino energy
distribution functions such as the ``sum of the light neutrino
masses'' and measures of the radiation energy density
\cite{Abazajian201566}.  Additionally, the advent of extremely large
optical telescopes, with adaptive optics, can improve the precision in
primordial abundance determinations \cite{TMT,GMT,EELT}.

Leveraging the increased observational precision to achieve
better probes of and constraints on beyond standard model (BSM)
physics will demand higher precision in simulation of {\it standard
model} physics. Many possible BSM scenarios ({\it e.g.} sterile
neutrinos, light scalars, out-of-equilibrium particle decay, etc.)
could affect weak decoupling and, hence, nucleosynthesis in
subtle but potentially measurable ways.  Accurate and
self-consistent treatments of the standard nuclear and particle
physics furthers the objective of a clear interpretation of
potential BSM issues (see Refs.\
\cite{FKK:2011di,2015PhRvD..91h3505N,2015arXiv151000428V}).

Neutrino kinetics affect the neutrino distributions and primordial 
nuclide abundances in the early universe in three principal respects.
First, the transfer of entropy from the photon/electron/positron
plasma to the neutrino seas cools the plasma temperature
relative to the case of no transport.
The cooler temperature alters the ratio of comoving to plasma energy scales
from the canonical value $(4/11)^{1/3} \approx 0.7138$
\cite{1990eaun.book.....K,Dodelson:2003mc,2008cosm.book.....W}.

The second out-of-equilibrium effect is the distortion of the thermal
Fermi-Dirac (FD) spectrum of high energy neutrinos. Upscattering of
low energy neutrinos and the production of neutrino-antineutrino pairs
contribute to this distortion through a variety of mechanisms.  An
important consequence of this mechanism is the effect that the
high-energy distortion has on the neutron-to-proton ratio ($n/p$). A
running theme throughout the present study is that such changes
induced by the distortion of the neutrino distributions away from
equilibrium have effects that must be calculated concurrently with the
evolution of the nuclide abundances.  In this way, we reveal
nonlinearities in feedback mechanisms between the neutrino transport
and the thermodynamics of the plasma. These changes to the temperature
evolution have an effect on relative changes in the nuclide abundances
through the reaction rates and the sensitive dependence of, for
example, Coulomb barriers on them.

The third out-of-equilibrium effect is entropy production. The Boltzmann H
theorem implies that the entropy of a closed system is a non-decreasing
function of time. In this paper, we investigate the conventional
assumption \cite{1990eaun.book.....K,2008cosm.book.....W} of comoving entropy
conservation. We find that there is a small change in the total entropy of the
universe due to the non-equilibrium kinetics of the neutrinos, which generates
entropy.
In essence, out-of-equilibrium neutrino energy transport and associated entropy
flow changes the phasing between scale factor and plasma temperature evolution.

A common feature of past works is that the effect of these transport/entropy
issues on the primordial abundances is small, typically on the order of 0.05\%
for helium-4 and lithium (in particular, see Ref.\ \cite{Dolgov:1997ne},
hereafter DHS).  Our work shows that the magnitude of these effects can be
significantly larger, depending on assumed microphysics.

The present work employs a non-perturbative method to calculate the evolution
of active neutrino occupation probabilities $f_{\nu_i}(p,t)$ for flavor
$i=e,\mu,\tau$. Homogeneity and isotropy has been assumed to restrict the
dependence of the $f_{\nu_i}$ to only the magnitude of the three-momentum $p$
and the comoving time $t$.
The evolution is computed in the presence of two-body to two-body ($2\to 2$)
collisions, the rates of which are given by the collision integrals
$C_{\nu_i}[f_j]$, where $f_j$ refers to the occupation probabilities of
neutrinos, antineutrinos, and charged leptons.  These are functionals of the
set of neutrino and antineutrino occupation probabilities $f_{\nu_j}$ and
evolve, within the Boltzmann equation approach, as
\begin{align}
\label{trans-eqn:boltz}
  \left[\pppd{t} - H(a) p \pppd{p}\right]
  f_{\nu_i}(p,t) &= {C}_{\nu_i}[f_j],
\end{align}
where $H(a)$ is the Hubble expansion rate at scale factor $a$.
We define the independent variable $\epsilon\equiv E_{\nu}/\tcm$ using
the neutrino energy $E_\nu$ and the comoving temperature parameter
\tcm.  The comoving temperature parameter is not a physical
temperature.  It is simply an energy scale that redshifts like the
energy of a massless particle in free fall with the expansion of the
universe and is, in essence, a proxy for inverse scale factor.
Therefore, we can write $\tcm(a) = \tin a_{\rm in}/a(t)$ as a function of
scale factor, where \tin and $a_{\rm in}$ are the plasma temperature and
scale factor at an initial epoch of our choosing.  For neutrinos in
the range of plasma temperatures 3 MeV $\gtrsim T \gtrsim$ 10 keV,
$\epsilon$ is equivalent to the commonly used quantity
$\tilde\epsilon=p/\tcm$.  Equation \eqref{trans-eqn:boltz} can be
cast in terms of $\epsilon$ as
\begin{align}
\label{eqn:boltz-feps}
\frac{d}{dt} f_{\nu_i}(\eps,t)
&= {C}_{\nu_i}[f_j].
\end{align}
The independent variable $\eps$ is chosen so that energy conservation
takes the simple form $\epsilon_1+\epsilon_2=\epsilon_3+\epsilon_4$
for the scattering process $1+2\leftrightarrow3+4$.

The evaluation of the collision integral in Eq.\ \eqref{trans-eqn:boltz} or
\eqref{eqn:boltz-feps} for the weak-interaction processes of interest is
numerically intensive.  However, the required integrations (described in detail
in Sec.\ref{sec:nquad} and Appendices \ref{trans-app:nunu} and
\ref{trans-app:other}) are performed in parallel with the code \burst
(BBN/Unitary/Recombination/Self-consistent/Transport) in Fortran 90/95 under
{\sc openmpi}.  We have developed a routine to evaluate the collision term for
the Boltzmann equation in \burst (using methods detailed in the Appendices)
which reduce the number of required integrations to two. Numerical integration,
effected under a combination of quadrature techniques (detailed in
Sec.\ref{trans-sec:wd}), has been tested by ensuring conservation of lepton
number; it is satisfied at the level of $10^{-14}$ (see Sec.\ref{sssec:nlnc}).

The code has been developed to address the problem of weak-decoupling collision
terms and for self-consistent coupling to nuclear reactions assuming that a
Boltzmann equation treatment is sensible.  The ``embarrassingly parallel''
structure of the problem allows for the simultaneous evaluation of the
occupation probabilities $f_{\nu_i}$ for each energy, implying a nearly linear
scaling of code performance with the number of cores. The present calculational
approach is readily generalizable to treat the full neutrino quantum kinetic
equations (QKEs) developed in Ref.\ \cite{VFC:QKE} and therefore neutrino
flavor oscillations (see Refs.\
\cite{1991NuPhB.349..743B,AkhmedovBerezhiani,1993APh.....1..165R,2005PhRvD..71i3004S,2007JPhG...34...47B,2013PhRvD..87k3010V,2013PrPNP..71..162B,Gouvea,2014PhRvD..90l5040S,2015PhLB..747...27C}
for discussion on the QKEs).  As mentioned, the present work neglects neutrino
flavor oscillations. A detailed calculation that concurrently solves the
neutrino QKE equations, incorporating both effects of flavor oscillations and
energy transport, and the primoridal nucleosynthesis is required and currently
underway. An example of the need for such a calculation is indicated by the
high sensitivity of the $n/p$ ratio at weak freeze-out to the electron neutrino
energy distrubtion (see Sec.\ref{trans-ssec:wdcalcs}).  One of the primary
effects of flavor oscillation, whose subsequent effect on primordial
nucleosynthesis is difficult to estimate in a self-consistent approach in the
dynamic environment of the BBN-epoch of the early universe, is the suppression
of the $\nu_e + n\rightarrow p + e^-$ rate. This suppression occurs when an electron
neutrino oscillates to either a $\nu_\mu$ or $\nu_\tau$ state, which do not
convert $n\leftrightarrow p$. A detailed, self-consistent calculation will
account for the phasings of various such mechanisms, which may be important at
the level of precision anticipated for in the next generation of cosmological
observations.

We emphasize that we couple neutrino-energy transport self-consistently and
concurrently to evaluation of the neutron-to-proton rates and nucleosynthesis
reaction network.  At each time step in \burst, the weak interaction
neutron-proton conversion rates ($n\leftrightarrow p$ rates),
\begin{align}
  \nu_e + n &\leftrightarrow p + e^-,\label{trans-eqn:np1}\\
  e^+ + n &\leftrightarrow p + \bnue,\label{trans-eqn:np2}\\
  n &\leftrightarrow p + e^- + \bnue,\label{trans-eqn:np3}
\end{align}
are determined using the evolved, non-equilibrium \nue and \bnue
spectra. The thermodynamics of the electromagnetic plasma is
coupled to the neutrino seas to account for heat flow
between the plasma and the neutrinos.  Non-equilibrium effects
generate entropy, increasing the total entropy of the plasma and the
neutrinos, through a timelike entropy-current flux.  Finally, we
integrate the neutrino occupation probabilities to determine the
energy density for calculating the Hubble expansion rate.  In this
way, self-consistency within the neutrino sector is maintained over
approximately $10^8$ Hubble times.
The overall architecture employed in \burst differs from the approaches used in
previous treatments (see Appendix \ref{trans-app:overview}).

The nuclear reaction network employed in the current code is based on
those of Refs.\ \cite{Wagoner:1969sy,SMK:1993bb} as augmented in
Ref.\ \cite{Fuller:2010nn}; details are discussed in
Ref.\ \cite{GFKP-5pts:2014mn}. Ongoing work is focused on incorporating
into the present approach a nuclear reaction network based on a
reaction formalism that respects unitarity.

The outline of this work is as follows.  In Sec.\ \ref{trans-sec:wd},
we present details of the transport code and weak-decoupling
calculations.  We investigate, in Sec.\ \ref{trans-ssec:wdcalcs}, the
contributions of the scattering processes to the out-of-equilibrium
neutrino spectra.  Section \ref{trans-sec:entropy} describes the
evolution of the entropy during the weak-decoupling process. Section
\ref{trans-sec:BBN} discusses primordial nucleosynthesis resulting
from the self-consistent coupling to the transport code. We conclude
in Sec.\ \ref{trans-sec:concl}.  Appendix \ref{trans-app:overview}
contains a summary of the calculations of different groups.
Appendices \ref{trans-app:nunu} and \ref{trans-app:other} describe the
analytical derivations of the collision terms.  We should emphasize
that the current manuscript represents a preliminary step toward the
objective of coupling neutrino kinetics to the nucleosynthesis
reaction network.  The proper treatment of neutrino flavor
oscillations and possible coherent effects requires a quantum kinetic
approach \cite{VFC:QKE}. Flavor oscillations have been
estimated\cite{Mangano:3.040} to change the production of \Heiv at
the 20\% level. The self-consistent approach that we consider here
might be expected to enhance this change; a detailed calculation is
required to estimate the actual effect. We detail further, ongoing
efforts in this work in the conclusion, Sec.\ \ref{trans-sec:concl}.
Throughout this paper we use natural units where $\hbar=c=k_B=1$.

In this manuscript we have provided a pedagogical presentation of
some familiar topics. This is done in the interest of giving a clear
presentation of our work and in the hopes of making our analytical and
numerical computations reproducible.

\begin{table*}
  \begin{tabular}{| c !{\vrule width 1.5 pt} c !{\vrule width 1.5 pt} c |}
    \hline
    $r$ & Process & $G_F^{-2}S_r\langle|\mathcal{M}_r|^2\rangle$ \\ 
    \midrule[1.5pt]
    1 & $\nu_i+\nu_i\leftrightarrow\nu_i+\nu_i$ & $2^6(P_1\cdot P_2)(P_3\cdot P_4)$\\ \hline
    2 & $\nu_i+\nu_j\leftrightarrow\nu_i+\nu_j$ & $2^5(P_1\cdot P_2)(P_3\cdot P_4)$\\ \midrule[1.5pt]
    3 & $\nu_i+\overline{\nu}_i\leftrightarrow\nu_i+\overline{\nu}_i$
    & $2^7(P_1\cdot P_4)(P_2\cdot P_3)$\\ \hline
    4 & $\nu_i+\overline{\nu}_j\leftrightarrow\nu_i+\overline{\nu}_j$
    & $2^5(P_1\cdot P_4)(P_2\cdot P_3)$\\ \hline
    5 & $\nu_i+\overline{\nu}_i\leftrightarrow\nu_j+\overline{\nu}_j$
    & $2^5(P_1\cdot P_4)(P_2\cdot P_3)$\\ \midrule[1.5pt]
    \multirow{3}{*}{6} & \multirow{3}{*}{$\nue+e^-\leftrightarrow e^-+\nue$} &
    $2^5[(2\sin^2\theta_W+1)^2(P_1\cdot Q_2)(Q_3\cdot P_4)$\\
    && $+4\sin^4\theta_W(P_1\cdot Q_3)(Q_2\cdot P_4)$\\
    && $-2\sin^2\theta_W(2\sin^2\theta_W+1)m_e^2(P_1\cdot P_4)]$\\ \hline
    \multirow{3}{*}{7} & \multirow{3}{*}{$\nu_{\mu(\tau)}+e^-\leftrightarrow e^-+\nu_{\mu(\tau)}$} &
    $2^5[(2\sin^2\theta_W-1)^2(P_1\cdot Q_2)(Q_3\cdot P_4)$\\
    && $+4\sin^4\theta_W(P_1\cdot Q_3)(Q_2\cdot P_4)$\\
    && $-2\sin^2\theta_W(2\sin^2\theta_W-1)m_e^2(P_1\cdot P_4)]$\\ \hline
    \multirow{3}{*}{8} & \multirow{3}{*}{$\nue+e^+\leftrightarrow e^++\nue$} &
    $2^5[(2\sin^2\theta_W+1)^2(P_1\cdot Q_3)(Q_2\cdot P_4)$\\
    && $+4\sin^4\theta_W(P_1\cdot Q_2)(Q_3\cdot P_4)$\\
    && $-2\sin^2\theta_W(2\sin^2\theta_W+1)m_e^2(P_1\cdot P_4)]$\\ \hline
    \multirow{3}{*}{9} & \multirow{3}{*}{$\nu_{\mu(\tau)}+e^+\leftrightarrow e^++\nu_{\mu(\tau)}$} &
    $2^5[(2\sin^2\theta_W-1)^2(P_1\cdot Q_3)(Q_2\cdot P_4)$\\
    && $+4\sin^4\theta_W(P_1\cdot Q_2)(Q_3\cdot P_4)$\\
    && $-2\sin^2\theta_W(2\sin^2\theta_W-1)m_e^2(P_1\cdot P_4)]$\\ \midrule[1.5pt]
    \multirow{3}{*}{10} & \multirow{3}{*}{$\nue+\bnue\leftrightarrow e^-+e^+$} &
    $2^5[(2\sin^2\theta_W+1)^2(P_1\cdot Q_4)(P_2\cdot Q_3)$\\
    && $+4\sin^4\theta_W(P_1\cdot Q_3)(P_2\cdot Q_4)$\\
    && $+2\sin^2\theta_W(2\sin^2\theta_W+1)m_e^2(P_1\cdot P_2)]$\\ \hline
    \multirow{3}{*}{11}
    & \multirow{3}{*}{$\nu_{\mu(\tau)}+\overline{\nu}_{\mu(\tau)}\leftrightarrow e^-+e^+$}
    & $2^5[(2\sin^2\theta_W-1)^2(P_1\cdot Q_4)(P_2\cdot Q_3)$\\
    && $+4\sin^4\theta_W(P_1\cdot Q_3)(P_2\cdot Q_4)$\\
    && $+2\sin^2\theta_W(2\sin^2\theta_W-1)m_e^2(P_1\cdot P_2)]$\\ \hline
  \end{tabular}
  \caption{\label{trans-tab:ssamps}Weak interaction processes relevant
  for neutrino weak decoupling. The left column labels the scattering,
  production, and annihilation processes in the middle column by an
  index $r$.  The right column gives the spin-averaged and summed
  square of the matrix element $\mathcal{M}_r$ for process $r$ with
  the Fermi constant and symmetry factor $S_r$ divided out. Indices
  $i$ and $j$ in the middle column for processes $r=1,\ldots,5$, which
  describe neutrino and antineutrino scattering, are distinct.
  Processes with an antineutrino scattering on a charged lepton,
  correspond to the parity-conjugate reactions of $r=6,\ldots,9$.
  Since they have identical matrix elements to these they are not
  shown in the table, although their effect is explicitly accounted for
  in antineutrino energy transport. $S_r$ is unity for all processes
  except $r=1$, where $S_1=1/2$.}
\end{table*}

\section{Neutrino weak decoupling calculations}
\label{trans-sec:wd}

Neutrinos decouple from the plasma, roughly speaking, when typical
rates of the weak processes, $\Gamma_w$ given in Table
\ref{trans-tab:ssamps}, fall below the Hubble rate: 
\begin{align} 
   \label{eqn:nwdc-approxtemp}
   \frac{\Gamma_w}{H} \lesssim \frac{G_F^2 T^5}{T^2/m_{\rm Pl}} 
   \simeq \left(\frac{T}{0.7\ {\rm MeV}}\right)^3.
\end{align} 
where $G_F = 1.166\times 10^{-11}$ MeV$^{-2}$ is the Fermi constant
and $m_{\rm Pl} = 1.221 \times 10^{22}$ MeV. By numerically evolving
the neutrino distributions for $\nu_e, \bar\nu_e,\nu_\mu, \bar\nu_\mu,
\nu_\tau$, and $\bar\nu_\tau$ we find, however, that the neutrinos
exchange entropy with the plasma until a temperature of nearly 100
keV, for many Hubble times beyond the estimate in
Eq.\ \eqref{eqn:nwdc-approxtemp} (see Fig.\
\ref{trans-fig:plot_entropy_vs_tcm}).  This is in part explained by
the fact that given the large entropy of the early universe, which is
carried by both photons/electrons/positrons and neutrinos, a
significant fraction of the neutrinos have energies larger than the
temperature. This effect is enhanced by plasma particles scattering
from the neutrinos, which preferentially up-scatter the neutrinos and
distort the high-momentum tails of the neutrino distributions. In this
section, we present the details of the numerical evaluation of the
collision integrals, the solution of the Boltzmann equation, and
performance statistics of the code, followed by details of the weak
decoupling calculations.

\subsection{Weak interaction processes}
\label{trans-ssec:weakints}
We discuss the weak interactions relevant for neutrino weak decoupling
here and their implementation in the collision integral $C$ in
the Boltzmann equation, Eq.\ \eqref{trans-eqn:boltz}.

Expressions for the neutral and charged current weak interaction
processes involving neutrinos, antineutrinos and the charged leptons
of the plasma are given in Table \ref{trans-tab:ssamps}. The table
gives the squared amplitudes $\langle |\mathcal{M}_r|^2\rangle$, where
$r$ labels two-body processes that are important during neutrino weak
decoupling \cite{1975ApJ...201..467T,1976ApJ...208L..19F}, averaged
over initial spin states and summed over final spins.  The initial
state particle four-momenta in Table \ref{trans-tab:ssamps} are given
particle numbers 1 and 2; final states are 3 and 4. That is:
\beq
  1 + 2 \leftrightarrow 3 + 4,
\eeq
where particle 1 is always a neutrino (or antineutrino).  We label
neutrino four-momenta as $P_i$ and charged lepton four-momenta as
$Q_i$.

The $\langle |\mathcal{M}_r|^2\rangle$ are different for
electron-flavor neutrinos compared to $\mu$ or $\tau$-flavor neutrinos
due to the charged-current interaction, which alters the factor
$2\sin^2\theta_W-1$ to $2\sin^2\theta_W+1$.\footnote{We note some
typographical differences between Table \ref{trans-tab:ssamps} and
Tables I and II in DHS. Row 10 here corresponds to Row 6 of Table I in
DHS. While the expression $G_F^{-2}S_6\langle|\mathcal{M}_6|^2\rangle$
is the same as that of DHS, the particle indexed 3 of our Row 10 is an
electron, and particle 3 of Row $6$ in Table I of DHS is a positron,
which should result in a different expression.  This discrepancy also
occurs between our Row 11 and Row 6 of Table (2) in DHS.  Our
expression for $r=10$, however, agrees with that of Row 7 of Table I
in Ref.\ \cite{1995PhRvD..52.1764H}.}  The Weinberg angle $\theta_W$ is
taken as $\sin^2\theta_W \approx 0.23$.  At the energy scales of
interest here the $\mu$ and $\tau$ neutrino species have the same
interactions.

\subsubsection{Collision integrals}
Given the amplitudes $\mathcal{M}_r$ of Table \ref{trans-tab:ssamps},
we may calculate the collision integral of Eq.\ \eqref{trans-eqn:boltz}:
\begin{align}
   \nonumber{C}^{(r)}_{\nu_1}[f_j] &= \frac{1}{2E_1}
   \int \frac{d^3p_2}{(2\pi)^3 2E_2}\frac{d^3p_3}{(2\pi)^3
   2E_3}\frac{d^3p_4}{(2\pi)^3 2E_4}\\
   &\times(2\pi)^4\delta^{(4)}(P_1+P_2-P_3-P_4)
   S_r\langle|\mathcal{M}_r|^2\rangle \nonumber \\
   &\times F_r(p_1,p_2,p_3,p_4),\label{trans-eqn:collint}
\end{align}
where $S_r$ is the symmetrization factor for identical particles, and
\begin{align} 
  \label{trans-eqn:capfdef}
   F_r(p_1,p_2,p_3,p_4) &= [1-f_1][1-f_2]f_3f_4 \nonumber\\
                         & -f_1f_2[1-f_3][1-f_4],\\
  \label{trans-eqn:capfdef12}
                         &= F^{(+)}_{r} - F^{(-)}_r.
\end{align}
Here we have suppressed time dependence and written the occupation
probability functions in abbreviated form.  For example, $f_1$ for
$r=1$ would read $f_{\nu_1}(p_1,t)$.
The quantities
$F_r^{(\pm)}$, corresponding to the first and second lines of
Eq.\ \eqref{trans-eqn:capfdef}, give the probability for scattering into
$(+)$ or out of $(-)$ the phase space volume for particle ``1''; they
include Pauli blocking factors $\sim (1-f_i)$.  The phase space
measure for particles $2,3$, and $4$, and the arguments of the
four-momentum conserving delta function
$\delta^{(4)}(p_1+p_2-p_3-p_4)$ and of $F_r$ are written schematically
with the dependence of $p_i$ on $r$, which can either be four-momentum $P_i$ or
$Q_i$, suppressed.
The factor $(2E_1)^{-1}$ ensures that an
integral over $d^3p_1/(2\pi)^3$ of the collision integral for $f_1$
vanishes in number-conserving processes; this is discussed in more
detail in Sec.\ref{trans-ssec:sumrules}.
All amplitudes in Table \ref{trans-tab:ssamps} are
proportional to $G_F$, the Fermi coupling constant. The square of the
Fermi coupling and a factor of $\tcm^5$ may be taken outside of the
collision integral [Eq.\ \eqref{trans-eqn:collint}] to give a
dimensionless expression with integration variable $\epsilon$, the
binning parameter for the occupation probabilities. The product
$G_F^2\,\tcm^5$ has dimensions of energy or inverse time, appropriate
to that for a rate.
The expression for the
collision integral appearing in Eq.\ \eqref{trans-eqn:boltz} is
\begin{align}
   \label{eqn:Cf}
   C_{\nu_i}[f_j] &= \sum_r C^{(r)}_{\nu_i}[f_j]
\end{align}
for processes $r$ that include $\nu_i$.

In general, for $2\to 2$ processes, Eq.\ \eqref{trans-eqn:collint} is a
nine-dimensional integral over the phase space of particles 2, 3, and
4. The four-momentum conserving $\delta$ function reduces the
collision integral to five dimensions.  Homogeneity and isotropy
further reduce Eq.\ \eqref{trans-eqn:collint} to a two-dimensional
expression in terms of single-particle energies of either species 2
and 3, or 2 and 4, or 3 and 4. The method of the reduction to
two-dimensions is distinct from and independent of that of DHS.
The reduction is tuned for the specific process in Table \ref{trans-tab:ssamps}
to ensure speed and accuracy in a parallel computation.  Appendix
\ref{trans-app:nunu} details our straightforward but lengthy method to obtain
the two-dimensional expression for the process in the first row of Table
\ref{trans-tab:ssamps}.  Appendix \ref{trans-app:other} gives the reduction
algorithm for collision integrals for the remaining processes of Table
\ref{trans-tab:ssamps}.  In both appendices, we relabel the indices of the
active particle species in Table \ref{trans-tab:ssamps} to simplify the
presentation.

\subsection{Numerical evaluation}
\label{trans-ssec:nonpertapproach}

In the interest of providing a complete description of the numerical
evaluation of the collision integrals of Eq.\ \eqref{trans-eqn:collint}
we describe here our choices for the energy ($\eps$) binning,
numerical quadrature, interpolation and extrapolation, and convergence
criteria.

\subsubsection{Binning} 
\label{sssec:bins} 

We employ a linear binning scheme for the occupation probabilities in
terms of the comoving invariant quantity $\epsilon=E_\nu/\tcm$.  The
interval from $\epsilon=0$ to $\epsilon=\epsmax$ is partitioned into
\nbins equal-width bins.  For a linear binning scheme, we use
$\nbins+1$ abscissas with the lowest abscissa at $\epsilon=0$.  The
\epsmax must be chosen large enough to support the $f_{\nu_i}$ and
$f_{\bar\nu_i}$.  We compare the numerically integrated equilibrium
energy spectrum to the analytical FD calculation at high temperature
and find agreement to a few parts in $10^6$.  

We have performed test calculations with values of \nbins from 100 to
1000.  The computing time has been verified empirically to scale as
$\nbins^3$.  Computation of the nuclear reaction network and
thermodynamic quantities associated with charged leptons and photons
incurs minimal computational overhead. Parallel code implementation of
the calculations results in reasonable wall-clock times $\sim$ days on
$\mathcal{O}(100)$ processors even with fine $\epsilon$ binning.
Typically, we find convergence for \nbins=100, as discussed later in
this section.

\subsubsection{Charged lepton quantities}
For the processes relevant to weak decoupling, the
occupation probabilities for the charged leptons are required. We
assume these are given by the FD equilibrium spectra with chemical
potential $\mu$ and temperature $T$:
\begin{align}
\label{trans-eqn:epmFD}
  f_{e^\pm}(E,T,\mp\mu) &= \frac{1}{\exp\left(E/T\pm\phie\right)+1},
\end{align}
where $\phie=\mu/T$ is the electron degeneracy parameter.  Here \phie
is determined by the requirement of charge neutrality in the
electron/positron/baryon plasma. We assume zero lepton number residing
in the neutrino seas\footnote{Our approach allows non-zero lepton
asymmetry. The assumption of zero neutrino lepton number is stipulated
for the present work and is in accord with the standard model.} and
neglect neutrino-nucleon charged-current transfer of electron lepton
number between the electrons/positrons and electron
neutrino/antineutrinos. This is plausible since the baryon-to-photon
ratio is small. Finite electron mass is taken into account in
Eq.\ \eqref{trans-eqn:epmFD} where $E=\sqrt{p^2+m_e^2}$.  We define, for
future use, the scaled mass $m_\eps$ as
\beq
\label{eqn:meps}
  m_\epsilon\equiv\frac{m_e}{\tcm}.
\eeq
We employ the comoving temperature, as its evolution
is simple.

\subsubsection{Numerical quadrature}
\label{sec:nquad}
Upon reduction of the three-body, nine-dimensional momentum integrals
as detailed in Appendix \ref{trans-app:nunu}, we may effect the
remaining momentum integrations, which are transformed to integrals
over $\eps$, via numerical quadrature. We refer to the integration
performed first (second) as ``inner'' (``outer''). We
neglect the neutrino rest mass and divide the energy (or,
equivalently, momentum) variable by \tcm to obtain the variables
$\eps_i$, where $i$ refers to either inner or outer integrations. If
the integral is over a charged-lepton kinematic variable, we use its
energy.  The squared-amplitude expressions require both
energies and three-momenta. We determine dimensionless momenta
as $p/\tcm = \sqrt{\epsilon^2 - m_\epsilon^2}$, where $m_\epsilon$
is given in Eq.\ \eqref{eqn:meps}.

Depending on the specific process in Table \ref{trans-tab:ssamps}, the
$\epsilon$ integral may be over a neutrino or a charged lepton.  For
the inner integral, irrespective of the species, the integration
method is a Gaussian quadrature
method \cite{Elhay:1987:AIF:35078.214351}. When the limits of the inner
integral are finite, we use Gauss-Legendre.  For finite intervals over
a range of $\eps$ larger than 200 and semi-infinite intervals, we use
Gauss-Laguerre.

When the outer integral is over an \epsval of a charged lepton, we use
either a Gauss-Legendre or Gauss-Laguerre method, depending on the
integration limits.  In the case that the outer integral is over a
neutrino energy, we use a five-point (Boole's)
rule \cite{Press:1993:NRF:563041} with abscissas aligned with the bin
points. This affords a slight improvement in performance by avoiding
interpolation for this integration of the occupation probabilities for
the neutrino energy of the outer integral.

\subsubsection{Interpolation and Extrapolation}
\label{trans-sssec:interp}

As detailed in Appendix \ref{trans-app:other}, we have the freedom to
choose which single-particle \epsvals to use in calculating the
collision integral.  The $2\to 2$ processes in Table
\ref{trans-tab:ssamps} have at least two neutrinos in the combined
initial and final states.  We use three of the four energy- or
momentum-conserving $\delta$ functions to eliminate an integral over
the phase space of one of the neutrino species.  This procedure
requires an interpolation over the \epsval of that species to
determine the occupation probability.  Processes that involve four
neutrinos or antineutrinos require an additional interpolation over
$\epsilon$ for the occupation probability of the inner integration
variable species.  The outer integration is either a Gaussian quadrature
method over a charged lepton, or a Boole's rule method over the bin
points.  In either case, no interpolation is required.  There is no
situation in which we need to interpolate the occupation probabilities
for the charged leptons since they are given by the equilibrium FD
expressions for electrons and positrons.

We use a fifth-order polynomial
interpolator \cite{Press:1993:NRF:563041} for the neutrino occupation
probabilities if the energy of the third or fourth neutrino does not
fall on an abscissa.  The accuracy of this interpolation is better
if we interpolate on the logarithms of the occupation probabilities,
as opposed to the occupation probabilities themselves. The domain of
integration is extended beyond \epsmax, to $\eps=300$, by
extrapolation. Beyond this point the occupation probability is taken
to be zero. None of our results are sensitive to these
extrapolations.

\subsubsection{Acceptance tolerance for rates}

When the occupation probabilities $f_i$ in
Eq.\ \eqref{trans-eqn:capfdef} are all equilibrium-distribution values, the
collision integral is zero, independent of the value of the squared
matrix elements. Numerical quadrature and interpolation, 
however, incur errors at the precision limitations of these methods
and the collision integrals attain small values when calculated under
equilibrium conditions. During the computation the need arises to set
the tolerance to accept a collision integral value as non-zero or,
conversely, to reject a value as the result of imprecision. To
accomplish this task, we use the net rate and forward-reverse-summed
(FRS) rate. The net rate is the value given by the collision integral
in Eq.\ \eqref{trans-eqn:collint}. The FRS rate corresponds to the sum
of contributions to the collision integral by substituting
$F^{(+)}_r+F^{(-)}_r$ for $F_r$ [Eq.\ \eqref{trans-eqn:capfdef}] into
Eq.\ \eqref{trans-eqn:collint}.

We calculate the net and FRS rates for each neutrino and antineutrino
species in each bin for all processes $r=1,\ldots,11$ (and the
antineutrino versions of interactions $r=6,\ldots,9$) in Table
\ref{trans-tab:ssamps} assuming thermal and chemical
equilibrium between the three flavors of neutrinos, antineutrinos,
positrons and electrons.  We sum over all of the processes to obtain
the collision integral for the net rate, and a modified collision
integral for the FRS rate.  For each neutrino species and each bin, we
calculate the precision ratio, defined as:
\begin{align}
   \label{trans-eqn:precratio}
   \mathcal{R}_{\nu_i}(\epsilon)\equiv
   \frac{\left|{C_{\nu_i}}[\feq_j(\epsilon)]\right|}
   {{C_{\nu_i}}[\feq_j(\epsilon)]_{\rm FRS}},
\end{align}
where $\feq_j(\epsilon)$ is the equilibrium FD occupation probability for
species $j$ at a given $\epsilon$ bin.  The FRS rate is numerically strictly
positive. The absolute value of the net rate is required to obtain a strictly
positive precision ratio since negative values can arise in and near
equilibrium due to finite numerical precision.
For diagnostic purposes only (i.e., not in our transport calculations) we take
$m_\epsilon=0$, meaning no temperature dependence in
Eq.\ \eqref{trans-eqn:precratio}.

In production runs of \burst, during weak decoupling, we calculate the
collision integrals for both the net and FRS rates at each time step.
We compare the ratio of values of the net and FRS rates for the
evolved, in general non-equilibrium distributions $f_i$, to those of
the equilibrium distributions [Eq.\ \eqref{trans-eqn:precratio}].
If the ratio in \eqref{trans-eqn:precratio} is larger than the 
tolerance threshold
\begin{align}
  \label{trans-eqn:precratio2}
  \left\{\frac{\left|{C}_{\nu_i}[f_j(\epsilon)]\right|}
  {{C}_{\nu_i}[f_j(\epsilon)]_{\rm FRS}}\right\} 
  \biggr/\mathcal{R}_{\nu_i}(\epsilon) > \nuratiotol,
\end{align}
the collision integral is accepted as non-zero and used in the evaluation of
the time derivative of the occupation probability $f_i(\epsilon)$.  If the
left-hand side of Eq.\ \eqref{trans-eqn:precratio2} is smaller than the
threshold, we set the collision integral to zero.  The precision ratio
$\mathcal{R}_{\nu_i}(\epsilon)$ never gets larger than a few parts in
$10^{12}$.

\subsection{Conservation sum rules}
\label{trans-ssec:sumrules}
We have tested the convergence of the numerical quadrature of the
collision integrals by studying number and energy sum rules.  Accurate
evaluation of the collision integral is necessary to maintain the
conservation of energy-momentum, particle number (for species with
conserved charges), and neutrino lepton number. These are discussed 
in the following two sections.

\subsubsection{Number and energy sum rules}

We define the total scaled errors in the number and energy densities
as:
\begin{align}
  \label{trans-eqn:sumrulenum}
  \delta\left(\cfrac{dn}{dt}\right)&=
  \frac{\sum\limits_{\nu}\displaystyle\int
  d\epsilon\,\epsilon^2\cfrac{df_{\nu}}{dt}\biggr|_{\rm net}}
  {\sum\limits_{\nu}\displaystyle\int
  d\epsilon\,\epsilon^2\cfrac{df_{\nu}}{dt}\biggr|_{\rm FRS}}, \\
  \label{trans-eqn:sumrulerho}
  \delta\left(\frac{d\rho}{dt}\right)&=
  \frac{\sum\limits_{\nu}\displaystyle\int
  d\epsilon\,\epsilon^3\frac{df_{\nu}}{dt}\biggr|_{\rm net}}
  {\sum\limits_{\nu}\displaystyle\int
  d\epsilon\,\epsilon^3\frac{df_{\nu}}{dt}\biggr|_{\rm FRS}},
\end{align}
respectively.  The summation over $\nu$ is for the three flavors of
neutrinos and antineutrinos and the denominators in these expressions
are strictly positive. We evaluate the sum rules including
contributions only from processes isolated within the neutrino seas,
{\em i.e.}, $r=1,2,\ldots,5$ in Table \ref{trans-tab:ssamps} to
gauge the effectiveness of the numerical evaluation in respecting
number and energy conservation. 
The spectra of the charged leptons are assumed
to be described by equilibrium distributions so scattering processes
involving electrons and positrons will not preserve the sum rules as
written in Eqs.\ \eqref{trans-eqn:sumrulenum} and
\eqref{trans-eqn:sumrulerho}. 

The neutrinos are assumed, in our computational approach, to be in thermal
equilibrium with the electrons and positrons until a temperature $\tin \gg 1$
MeV.  The comoving temperature and plasma temperature are equal for all
temperatures greater than the input temperature: $T=\tcm\ge\tin$.  At \tin, we
commence evaluation of the collision integrals and evolve the neutrino
occupation probabilities until a comoving temperature \tstop.  The computation
approach adopted in \burst utilizes an adaptive Cash-Karp
\cite{Press:1993:NRF:563041} time step.  It evolves observables at
$\sim3\times10^4$ steps on the interval defined by \tin and \tstop with a
fifth-order Runge-Kutta (RK5) algorithm.  All simulations in this paper have
$\epsmax = 20.0$, $\nbins=100$, $\tin=8\text{ MeV}$, and $\nuratiotol=30.0$.
The terminal temperature is $\tstop=15\text{ keV}$, corresponding to a plasma
temperature of $T\sim20\text{ keV}$.  In thermal equilibrium, the total scaled
errors are small but non-zero and evaluate to $\sim10^{-12}$ for both the
number and energy sum rules for 100 bins.

\begin{table*}
  \begin{tabular}{| c !{\vrule width 1.5 pt} c !{\vrule width 1.5 pt} c | c !{\vrule width 1.5 pt} c |}
    \hline
    Processes & $\tcm/T$ & $100\times\delta\rho_{\nue}$ &
    $100\times\delta\rho_{\num}$ & $\Delta\neff$\\ 
    \midrule[1.5pt]
    All & 0.7148 & 0.9282 & 0.3771 & 0.03397\\ \midrule[1.5pt]
    10, 11 & 0.7147 & 0.9383 & 0.2867 & 0.03063\\ \hline
    1, 2, 10, 11 & 0.7147 & 0.9268 & 0.2963 & 0.03078\\ \hline
    1, 2, 3, 4, 5, 10, 11 & 0.7147 & 0.8557 & 0.3465 & 0.03136\\ \midrule[1.5pt]
    6, 7, 8, 9 & 0.7140 & 0.1853 & 0.0639 & 0.00723\\ \hline
    1, 2, 6, 7, 8, 9 & 0.7140 & 0.1724 & 0.0778 & 0.00753\\ \hline
    1, 2, 3, 4, 5, 6, 7, 8, 9 & 0.7140 & 0.1559 & 0.0886 & 0.00763\\ \hline
  \end{tabular}
  \caption[Neutrino energy-transport runs]
  {\label{trans-tab:transsummary}Process-dependent changes in neutrino
     energy density properties.  For all runs $\epsmax=20.0$, $\nbins=100$,
     $\tin=8\,{\rm MeV}$, $\tstop=15\,{\rm keV}$, $\nuratiotol=30.0$.  The
     first column gives the processes used for a given run.  The second column
     is the ratio of comoving to plasma temperature.  For column two reference,
     $(4/11)^{1/3}=0.7138$.  Columns three and four are the relative changes of
     the \nue and \num energy densities.  The quantity $\Delta\neff$ is given
     by Eq.\ \eqref{trans-eqn:deltaneff}.  Round-off error of the neglected
     fifth significant digit in columns 2, 3, and 4 accounts for the one part
     in $10^4$ discrepancy with column 5.}
\end{table*}

We monitor the total scaled errors of Eqs.\ \eqref{trans-eqn:sumrulenum} and
\eqref{trans-eqn:sumrulerho} at each time step during our weak decoupling
calculations.  On average, we maintain accuracy to better than one part in
$10^6$ over the entire run.

\subsubsection{Neutrino lepton number conservation}
\label{sssec:nlnc}
Elastic processes satisfy
\begin{align}
   \label{eqn:lnc-el}
   \int d^3p\, C^{(r)}_{\nu_i}(p) &= 0,
\end{align}
since the processes $r=1,\ldots,4$ and $r=6,\ldots,9$ (and their
antineutrino counterparts) conserve neutrino (antineutrino) number.
The annihilation processes, $r=5,10$ and 11 satisfy, for example:
\begin{align}
   \label{eqn:lnc-inel}
   \int d^3p\, \left[
    C^{(\nu_e\bar\nu_e,\nu_\mu\bar\nu_\mu)}_{\nu_e}(p) 
  - C^{(\nu_e\bar\nu_e,\nu_\mu\bar\nu_\mu)}_{\bar\nu_e}(p) 
   \right]&= 0, \\
   \int d^3p\, \left[
    C^{(\nu_e\bar\nu_e,\nu_\mu\bar\nu_\mu)}_{\nu_\mu}(p) 
  - C^{(\nu_e\bar\nu_e,\nu_\mu\bar\nu_\mu)}_{\bar\nu_\mu}(p) 
   \right]&= 0.
\end{align}
Analogous relations hold for other annihilation processes that fall under
the reaction classes $r=5,10$ and 11.

We have confirmed that the neutrino lepton numbers are conserved at
the level of $\lesssim 10^{-14}$ for all values of the scale factor
$a(t)$.

\section{Results in the neutrino sector}
\label{trans-ssec:wdcalcs}

Our treatment of the Boltzmann-equation evolution of the neutrino
energy transport reveals novel features of the transport
characteristics of the active
neutrino sector. We focus first on these results, which are largely 
independent of the coupling to BBN through the nuclear reaction
network. The present calculations reveal, in particular, that the
history of $e^\pm$ annihilation to photons displays a rich set of
behaviors that has not been discussed before. We also look into the
role of QED radiative corrections. These results are in line with
previous work but they indicate that a more comprehensive treatment of the
plasma physics during the epochs we consider is warranted.

\subsection{Neutrino interactions and energy transport}
\label{ssec:niet}

Table \ref{trans-tab:transsummary} summarizes the neutrino energy
transport properties in the present calculations, which as mentioned
are carried out for computational parameters $\epsmax=20.0,
\nbins=100, \tin=8 \textnormal{ MeV}, \tstop=15 \textnormal{ keV}$,
and $\nuratiotol=30.0$. In this section, we focus on the first row of
the table, when all of the weak interactions of neutrinos (and the
antineutrino reactions corresponding to the parity conjugates of the
reactions $r=6,\ldots,9$) are computed. We discuss the results for
selective process evaluations corresponding to the remaining rows of
this table in the next section, Sec.\ \ref{ssec:neta}. We briefly
describe this table to orient the subsequent discussion.

The first column of Table \ref{trans-tab:transsummary} lists the
processes $r$ from Table \ref{trans-tab:ssamps} used for a given run.
The second column gives the ratio of the comoving to plasma
temperatures.  Columns three and four give the relative changes
of the \nue and \num energy densities, respectively,
with respect to
the equilibrium energy density:
\begin{align}
  \label{eqn:drho}
  \delta\rho_{\nu_i} &\equiv\frac{\rho_{\nu_i} - \rho_\nu^{\rm
  (eq)}}{\rho_\nu^{\rm (eq)}}, &
  \rho_\nu^{\rm (eq)} &=
  \frac{7}{8}\frac{\pi^2}{30}\tcm^4.
\end{align}
The
last column is the change in \neff.
\neff is defined through the energy density in ultra-relativistic
particles -- the radiation energy density $\rho_{\rm rad}(a)$ -- after
the epoch of photon decoupling as
\begin{align}
   \label{trans-eqn:neff}
   \rho_{\rm rad}(a_{\gamma d}) &=
   \left[2+\frac{7}{4}\left(\frac{4}{11}\right)^{4/3}\neff\right]
   \frac{\pi^2}{30}T_{\gamma d}^4,
\end{align}
where $a(t)$ is the scale factor at universal comoving time $t$,
$T(a)$ is the plasma or photon temperature and $T_{\gamma d} =
T(a_{\gamma d})$ is the photon temperature at the conclusion of the
epoch of photon decoupling.
We make the assumption that \tcmpl and $\delta\rho_{\nu_i}$ do not
change significantly for $10\,{\rm keV}\gtrsim T\gtrsim0.2\,{\rm eV}$.
Therefore, if we set the radiation energy density equal to the sum of
the photon and neutrino densities in Eq.\ \eqref{trans-eqn:neff}, we can
determine \neff from \tcmpl and $\delta\rho_{\nu_i}$:
\begin{align}
  \label{trans-eqn:neff2}
  \neff &= \left[\frac{(\tcm/T_{\gamma d})}{(4/11)^{1/3}}\right]^4
  \nonumber \\
  &\times
  \left[(1+\delta\rho_{\nue}(a_{\gamma d})) 
  + 2(1+\delta\rho_{\num}(a_{\gamma d}))\right].
\end{align}
In writing Eq.\ \eqref{trans-eqn:neff2}, we have assumed that
antineutrinos have the same relative change in energy density as
neutrinos.  The change in \neff is given as:
\begin{align}
  \label{trans-eqn:deltaneff}
  \Delta\neff &= \neff - 3,
\end{align}
where \neff is given by Eq.\ \eqref{trans-eqn:neff2}.
It is clear from this table that the dominant contribution to the
parameter $\Delta\neff$ is due to annihilation processes $r=10$ \& 11.
Additionally, for $\Delta\neff \sim 0.05$, the value typically quoted
in the literature \cite{neff:3.046}, the effect of charged lepton
scattering is small but not negligible.

Another feature apparent in Table \ref{trans-tab:transsummary} is
non-linearity in the combination of processes. Adding, for example,
the values of $\Delta\neff$ for Table \ref{trans-tab:transsummary}
rows 4 and 5, which sums all of the processes $r=1,\ldots,11$ (and the
implied charged-lepton antineutrino scattering processes) with
$\Delta\neff=0.039$ is not equivalent to the Table
\ref{trans-tab:transsummary} first row with $\Delta\neff=0.034$.
Finally, the ratio of the comoving temperature to the plasma
temperature $\tcm/T$ is largely set by the annihilation processes.  We
note, however, that this does not uniquely determine $\Delta\neff$ as
Eq.\ \eqref{trans-eqn:neff} implies.

\begin{figure}
  \includegraphics[width=\columnwidth]{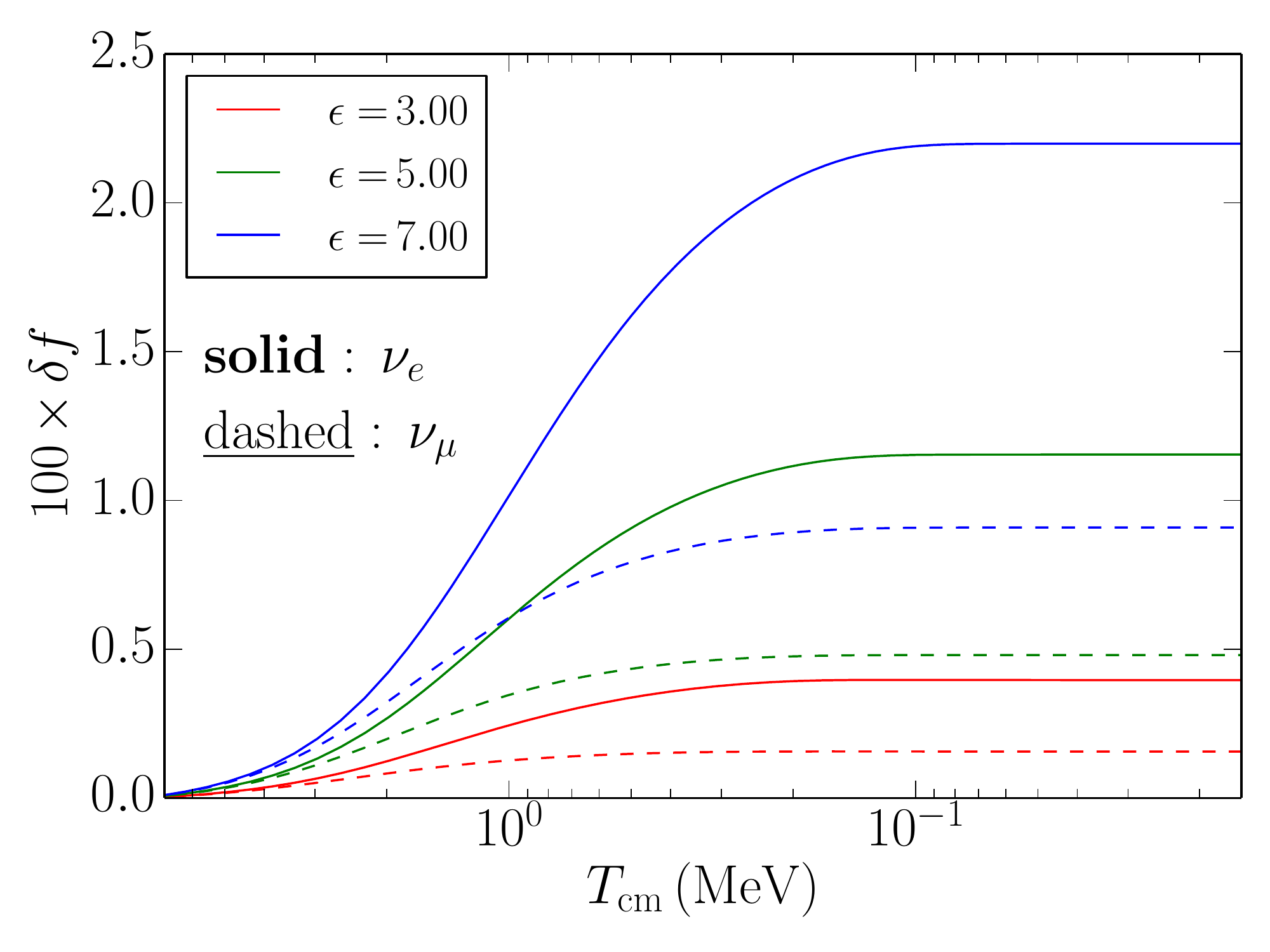}
  \caption[Relative change in occupation probability as a function of
  \tcm]{\label{trans-fig:plot_doccprob_tcm}(Color online) The relative
  change, as in Eq.\ \eqref{trans-eqn:dfandfeq}, in the occupation
  probability as a function of the comoving temperature \tcm.  Three
  values of $\epsilon$ are evaluated at $\epsilon=3,5$ and $7$.  The
  solid lines are for electron-flavor neutrinos, and the dashed lines
  are for muon-flavor neutrinos.  The larger $\delta f$
  correspond to larger $\epsilon$ values.}
\end{figure}

\begin{figure}
  \includegraphics[width=\columnwidth]{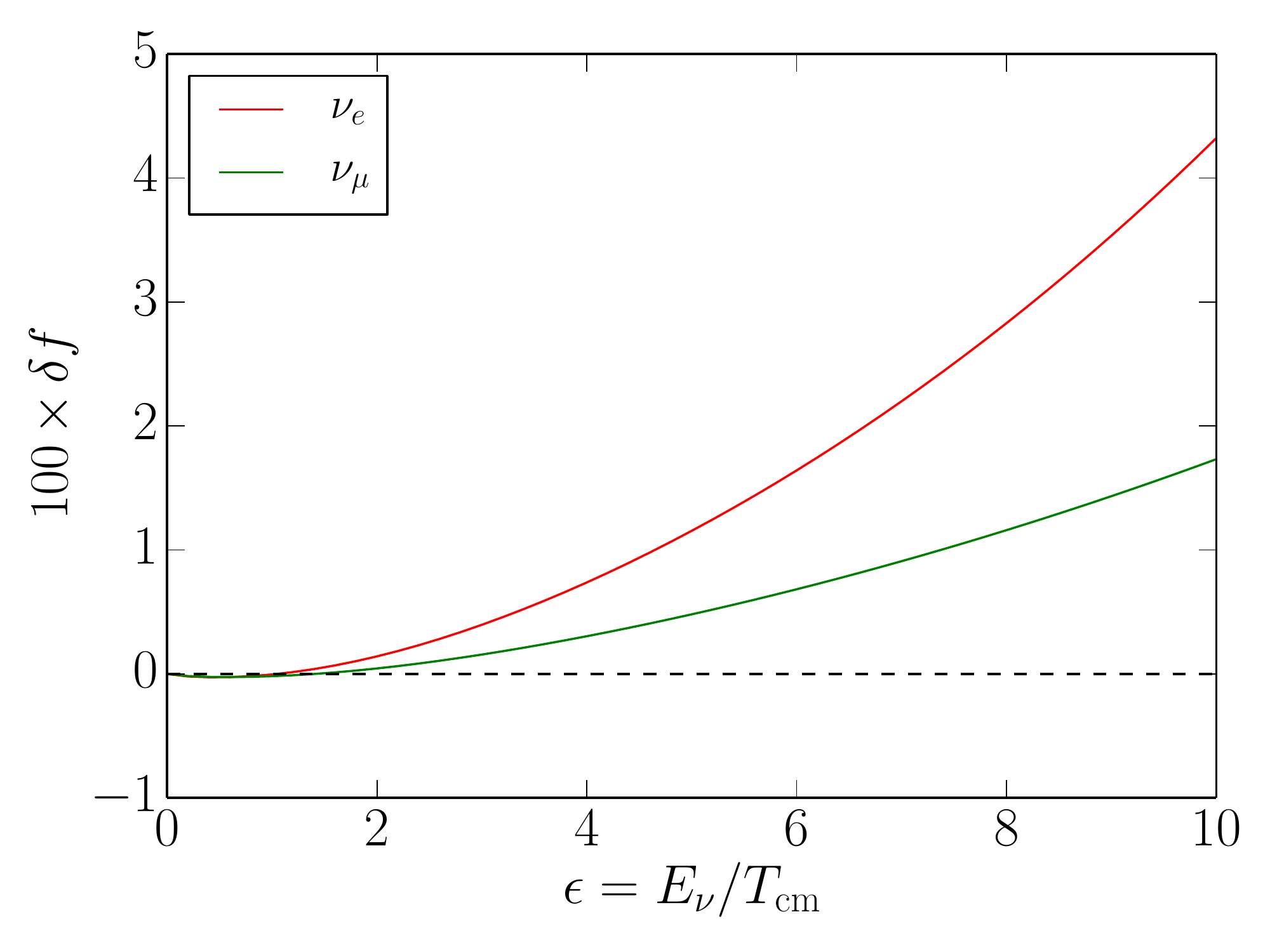}
  \caption[Relative change in occupation probability versus
  $\epsilon$]{\label{trans-fig:plot_doccprob_eps}(Color online) The
  relative change, as in Eq.\ \eqref{trans-eqn:dfandfeq}, in the occupation
  probability as a function of $\epsilon$ for $\tcm=1$ keV. The larger
  change is the electron-flavor neutrinos, over the muon-flavor
  neutrinos. The antineutrino evolution is nearly identical to the
  neutrino evolution for all flavors.}
\end{figure}

\begin{figure}
  \includegraphics[width=\columnwidth]{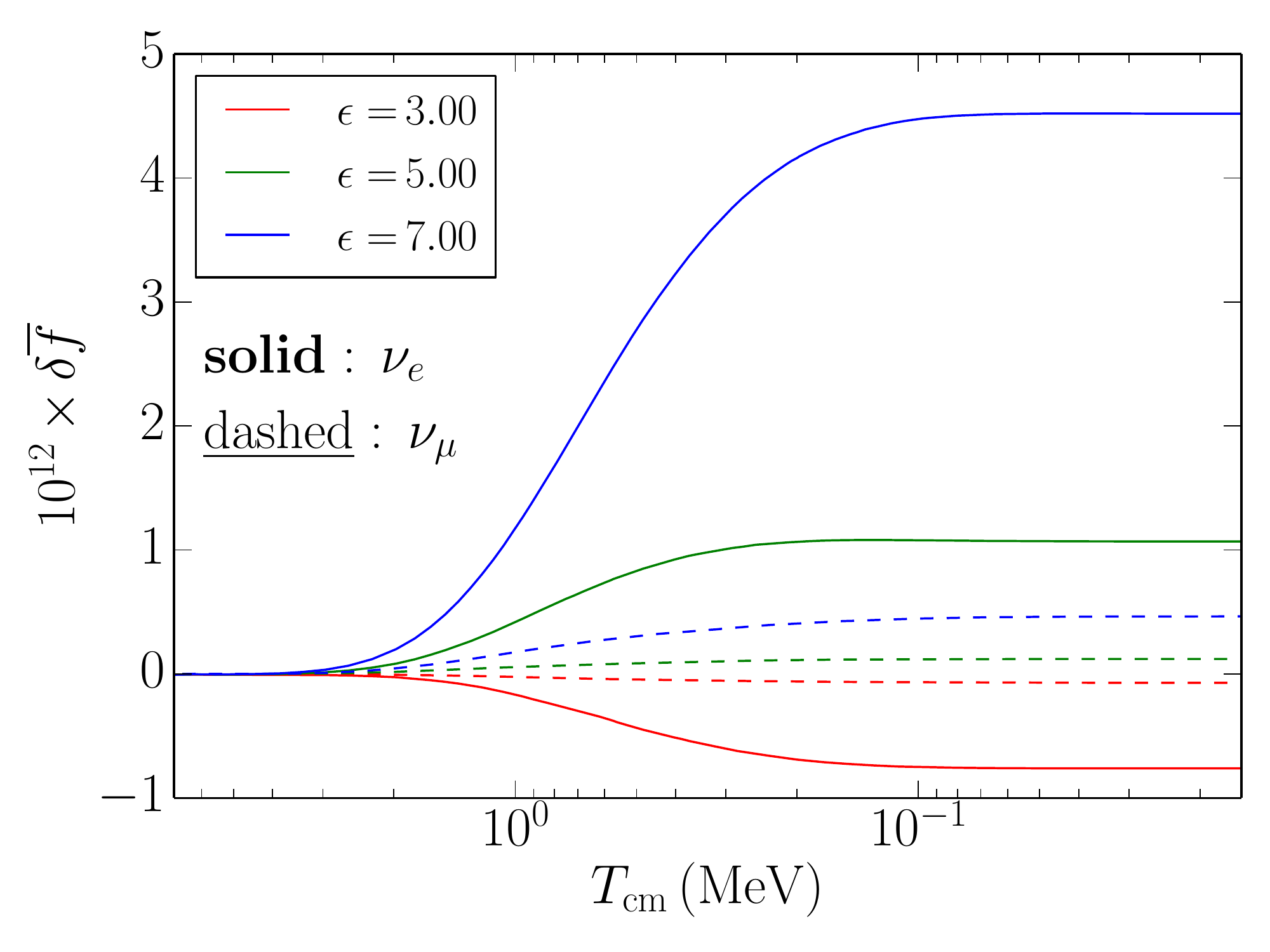}
  \caption[Difference between neutrino and antineutrino spectra as a
  function of \tcm]{\label{trans-fig:plot_doccbar_vs_tcm}(Color
  online) The difference in relative changes in the occupation
  probabilities of $\nu$ and $\overline{\nu}$ [Eq.\ \eqref{eqn:dfbar}]
  as a function of comoving temperature \tcm.  Three values of
  $\epsilon$ are plotted at $\epsilon=3,5$ and $7$.  The solid lines
  are for electron-flavor neutrinos, and the dashed lines are for
  muon-flavor neutrinos.  The \nue experience a larger change than the
  \num.}
\end{figure}

\begin{figure}[h]
   \includegraphics[width=\columnwidth]{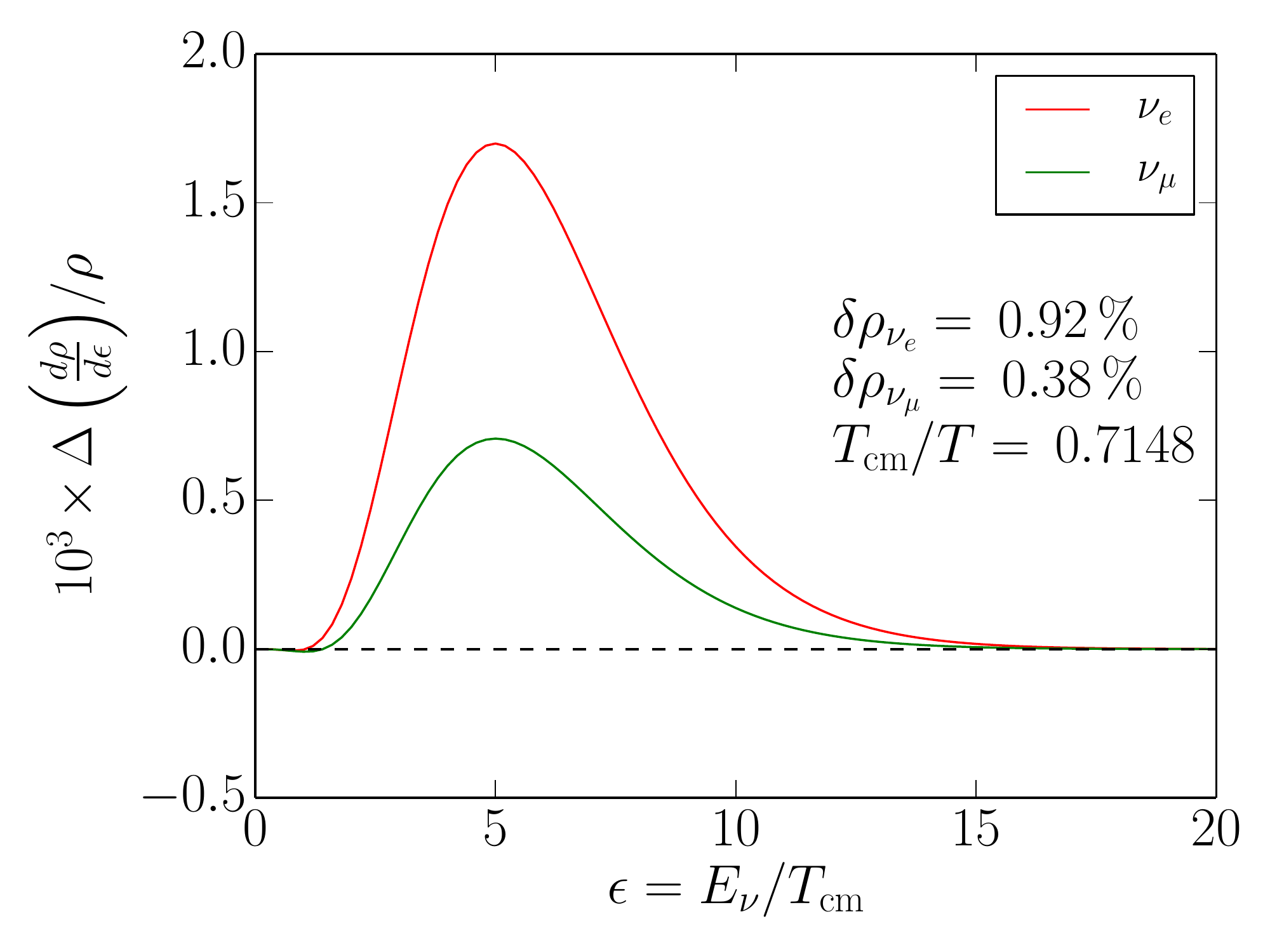}
   \caption[Normalized change in differential energy density versus
   $\epsilon$]{\label{trans-fig:plot_drho_vs_eps}(Color online) The
   normalized change in the differential energy density
   [Eq.\ \eqref{eqn:ded}] as a function
   of $\epsilon$.  The electron neutrinos exhibit a larger change
   compared to the muon neutrinos. The antineutrino evolution is
   nearly identical to the neutrino evolution for all flavors.}
\end{figure}

\Cref{trans-fig:plot_doccprob_tcm,trans-fig:plot_doccbar_vs_tcm,trans-fig:plot_doccprob_eps,trans-fig:plot_drho_vs_eps}
show relative changes in the neutrino spectra for a calculation with transport
versus a no-transport calculation.  All processes are active in the transport
calculation, i.e.\ row 1 of Table \ref{trans-tab:transsummary}. The no
transport calculation maintains FD-like distributions at temperature parameter
\tcm.
We compare our present results, in
detail, to DHS and Ref.\ \cite{2000NuPhB.590..539E}. To this end, we
first define several quantities to facilitate this comparison and then
turn to a detailed discussion of each of these figures.

We define $\delta f$ at a given time
$t$ and $\epsilon$ to be the relative change in the occupation
probabilities with respect to the FD occupation probability:
\begin{align} 
\label{trans-eqn:dfandfeq} 
\delta f &\equiv
\frac{f(\epsilon,t)-\feq(\epsilon)}{\feq(\epsilon)} 
\end{align} 
where
\begin{align} 
\feq(\epsilon) &= \frac{1}{e^\epsilon + 1}.  
\end{align}
We note that \feq does not depend explicitly on time or temperature.
Figures \ref{trans-fig:plot_doccprob_tcm} and
\ref{trans-fig:plot_doccprob_eps} show $\delta f$ as a function,
respectively, of \tcm for $\epsilon=3,5$ and 7 and as a function of
$\epsilon$ at a comoving temperature $\tcm=1\text{ keV}$.  Figure
\ref{trans-fig:plot_doccbar_vs_tcm} displays the difference in
the relative change for neutrinos and antineutrinos:
\begin{align}
\label{eqn:dfbar}
  \delta\overline{f}\equiv \delta f_\nu - \delta f_{\bar{\nu}} =
  \frac{f_\nu - f_{\bar{\nu}}}{\feq}.
\end{align}
Figure \ref{trans-fig:plot_drho_vs_eps} shows the normalized change in
the differential energy density:
\begin{align}
  \frac{\displaystyle\Delta\left(\frac{d\rho}{d\epsilon}\right)}{\rho}&=
  \frac{\displaystyle\left[\frac{\epsilon^3}{2\pi^2}
  f(\epsilon)-\frac{\epsilon^3}{2\pi^2}\feq(\epsilon)\right]}
  {\displaystyle\frac{1}{2\pi^2}\int dx\,x^3\feq(x)}\\
\label{eqn:ded}
  &=\frac{120}{7\pi^4}\,\epsilon^3[f(\epsilon)-\feq(\epsilon)].
\end{align}
The antineutrino behavior is nearly identical to the neutrino
behavior for all flavors.  

For each \epsval in Fig.\ \ref{trans-fig:plot_doccprob_tcm} the
relative change in the electron-flavor (\nue) is larger than the
relative change in the muon-flavor (\num) neutrino sea.  The
annihilation and scattering rates with electrons and positrons are
faster due to the contribution of the charged-current diagrams for
\nue, which are absent for \num, as noted in
Ref.\ \cite{2000NuPhB.590..539E}.  In addition to the larger affect on
the \nue spectra, the charged-current processes keep the \nue in
thermal contact with the charged leptons longer than \num.  This is
apparent from Fig.\ \ref{trans-fig:plot_doccprob_tcm} where freeze-out
corresponds to the point where the derivative of the curves goes to
zero.  The \num freeze-out occurs at an earlier epoch than the \nue
freeze-out. Additionally, freeze-out occurs later for larger
$\epsilon$ values, as noted in DHS. These results are generally
consistent with DHS and Ref.\ \cite{2000NuPhB.590..539E}.  For example,
the $\epsilon=5$, \nue curve rises at a more rapid rate than, and
crosses, the $\epsilon=7$, \num curve. Figure (4) of
Ref.\ \cite{2000NuPhB.590..539E} also exhibits this crossing between the
$\epsilon=5$, \nue curve and the $\epsilon=7$, \num curve. Comparing
Fig.\ \ref{trans-fig:plot_doccprob_tcm} with Figs.(3a) and (3b) of
DHS, confirms the similar behavior of \burst and DHS.

Figure \ref{trans-fig:plot_doccprob_eps}, plotted at a temperature of
$\tcm=1\text{ keV}$ well after weak decoupling, shows that the \nue
have a larger distortion than the \num and that this effect is
enhanced at large $\epsilon$.  An interesting feature of Fig.\
\ref{trans-fig:plot_doccprob_eps} is the negative relative change for
$\epsilon\lesssim1$.  It appears to occur in Fig.\ (5) of both DHS and
Ref.\ \cite{2000NuPhB.590..539E} but is not explicitly mentioned in
either reference. We investigate this phenomenon in more detail in the
subsections below. In addition, our relative changes are in good
agreement with those of DHS for both \nue and \num.

In Fig.\ \ref{trans-fig:plot_doccbar_vs_tcm}, we exhibit the difference
in the relative change of \nue and \bnue, and also \num and \bnum.
The electron-flavor shows an enhanced effect over the muon-flavor for
all
\epsvals.  For both flavors, at \epsvals$=5$ and $7$ in Fig.\
\ref{trans-fig:plot_doccbar_vs_tcm}, the relative changes are
positive.  The negative differences for $\epsilon=3$ indicate there is
an abundance of antineutrinos over neutrinos, independent of flavor.
The differences between neutrino and antineutrino distributions for
both $e$ and $\mu$ are small for all epsilon values considered here.
This raises, however, the important issue of how neutrino flavor
evolves under the full quantum kinetic
evolution \cite{VFC:QKE}.

Figure \ref{trans-fig:plot_drho_vs_eps} shows where the largest change
in the energy-density spectrum occurs.  Figure
\ref{trans-fig:plot_drho_vs_eps} is approximately equivalent to Fig.\ 
\ref{trans-fig:plot_doccprob_eps} multiplied by $\epsilon^3\feq$.  The
peak of the normalized change in the differential energy density is
located at $\epsilon\sim5$, for both \nue and \num.  Figure (6) of
Ref.\ \cite{2000NuPhB.590..539E} also shows a peak at an
$\epsilon\sim5$.  Although Fig.\ \ref{trans-fig:plot_doccprob_eps}
shows that the deviation from equilibrium of the occupation
probabilities increases for increasing \epsvals, the probability is
small enough in the high-$\epsilon$ bins that the large changes from
equilibrium have little effect on the total energy density.

Integrating the neutrino energy distributions in Fig.\
\ref{trans-fig:plot_drho_vs_eps}, we find relative changes in the energy
density of: $\delta\rho_{\nue} = 0.0092$, and $\delta\rho_{\num} = 0.0038$.
The temperature ratio is given in the first row of Table
\ref{trans-tab:transsummary} as $\tcm/T=0.7148$. Using
Eq.\ \eqref{trans-eqn:deltaneff}, we find $\Delta\neff=0.034$.  The quantities
$\delta\rho_{\nue}$, $\delta\rho_{\num}$, \tcmpl, and \neff all agree closely
with both Ref.\ \cite{2000NuPhB.590..539E} and DHS.

\begin{figure}[h]
   \includegraphics[width=\columnwidth]{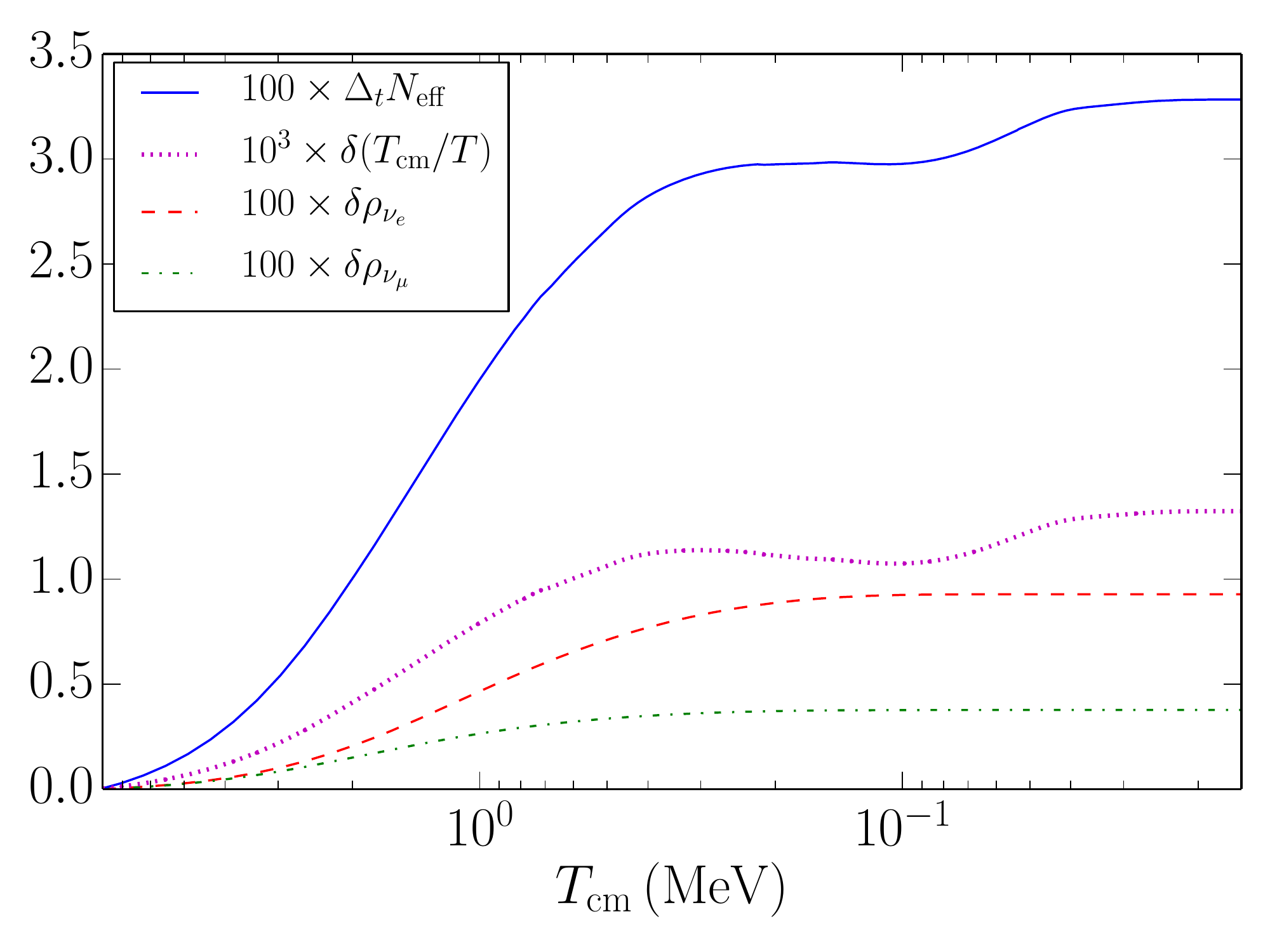}
   \caption[Changes in energy densities, \neff, and \tcmpl versus \tcm.]
   {\label{trans-fig:plot_drho_vs_tcm}(Color online) Quantities related to
      energy density and temperature are plotted against the comoving
      temperature parameter.
      The blue solid curve shows the change in \neff using
      Eq.\ \eqref{trans-eqn:neffevolve}.
      The red dashed curve shows the relative change in the
      energy density of \nue.
      The green dash-dot curve shows the relative change in the
      energy density of \num.
      The magenta dotted curve shows the relative change in \tcmpl using
      Eq.\ \eqref{trans-eqn:dtcmpl}. At a given \tcm, $\delta\tcmpl>0$ is equivalent to
      a lower plasma temperature in the transport case compared to no transport.}
\end{figure}

Figure \ref{trans-fig:plot_drho_vs_tcm} shows how the energy densities,
\neff, and \tcmpl evolve with \tcm until they reach their asymptotic
values.  The $\delta\rho_{\nue}$ and
$\delta\rho_{\num}$ are computed from Eq.\ \eqref{eqn:drho}
and the relative
change in \tcmpl is computed by comparing the evolution of the
temperature with transport $(\tcmpl)_\textnormal{all}$ and without
$(\tcmpl)_\textnormal{none}$:
\begin{align}
  \delta(\tcmpl) &= \frac{(\tcmpl)_\textnormal{all}
  - (\tcmpl)_\textnormal{none}}{(\tcmpl)_\textnormal{none}}.
  \label{trans-eqn:dtcmpl}
\end{align}
Finally, to calculate the time evolution change in \neff, we use
\begin{align}
   \Delta_t\neff & \equiv 
   [1+\delta(\tcmpl)]^4
   \nonumber \\
   &\times [(1+\delta\rho_{\nue})+2(1+\delta\rho_{\num})]- 3,
   \label{trans-eqn:neffevolve}
\end{align}
where the subscript $t$ denotes time dependence, in contrast to the asymptotic
limit of Eq.\ \eqref{trans-eqn:deltaneff}. As may be seen in Fig.\
\ref{trans-fig:plot_drho_vs_tcm}, \neff does not converge to 3.034, the value
consistent with DHS. The reason its asymptotic value is instead 3.033 is due to
the fact that the run with no transport has $(\tcmpl)_\textnormal{none}\ne
(4/11)^{1/3}$ in the asymptotic limit.
If we assume the neutrinos are in thermal equilibrium for $T>\tin$, the
temperature ratio incurs a modification from the finite electron rest mass as:
\begin{align}
\label{trans-eqn:tcmplfinitem} 
\left(\frac{\tcm}{T}\right)_\textnormal{none} 
&= \left(\frac{4}{11}\right)^{1/3}
\left(1+\frac{5}{22\pi^2}z^2\right),
\end{align}
to second order in $z\equiv m_e/\tin$.  Setting $\tin=8\,{\rm MeV}$,
we find an altered \tcmpl gives $\Delta\neff=0.001$.

The evolution of $\delta (\tcm/T)$ in Fig.\ \ref{trans-fig:plot_drho_vs_tcm}
displays interesting features that are driven by the specifics of the loss of
entropy in the plasma from the annihilation of electrons and positrons to
neutrinos and the transfer of entropy from electrons/positrons to photons
through annihilation (see Sec.\ \ref{trans-sec:entropy} for a detailed
discussion of entropy). The annihilation of electrons and positrons into
neutrinos can be seen in the rise of the $\delta \rho_\nu$ curves in Fig.\
\ref{trans-fig:plot_drho_vs_tcm}.  For $\tcm \gtrsim 200\,{\rm keV}$, entropy
is lost from the plasma into the neutrino seas resulting in a lower plasma
temperature for the transport case (where entropy is lost) versus the
no-transport case (where entropy is not lost).  The increase in $\delta
(\tcm/T)$ for $\tcm \gtrsim 400\,{\rm keV}$ is caused by this entropy loss.

To analyze the entropy transfer from the electron/positron components to the
photons, we need the total number densities of electrons and positrons
\beq\label{eqn:ne}
  n_{e^\pm}(T,\mp\mu) = 2\int\! \frac{d^3p}{(2\pi)^3} f_{e^\pm}(E,T,\mp\mu),
\eeq
which, in local thermodynamic equilibrium, are solely functions of the plasma
temperature and the electron chemical potential.
Fig.\ \ref{trans-fig:plot_drho_vs_tcm} shows a different phasing of scale
factor and temperature for the two cases. At a given \tcm, the plasma
temperature is always lower in the transport case versus the no-transport case.

\begin{figure}
  \includegraphics[width=\columnwidth]{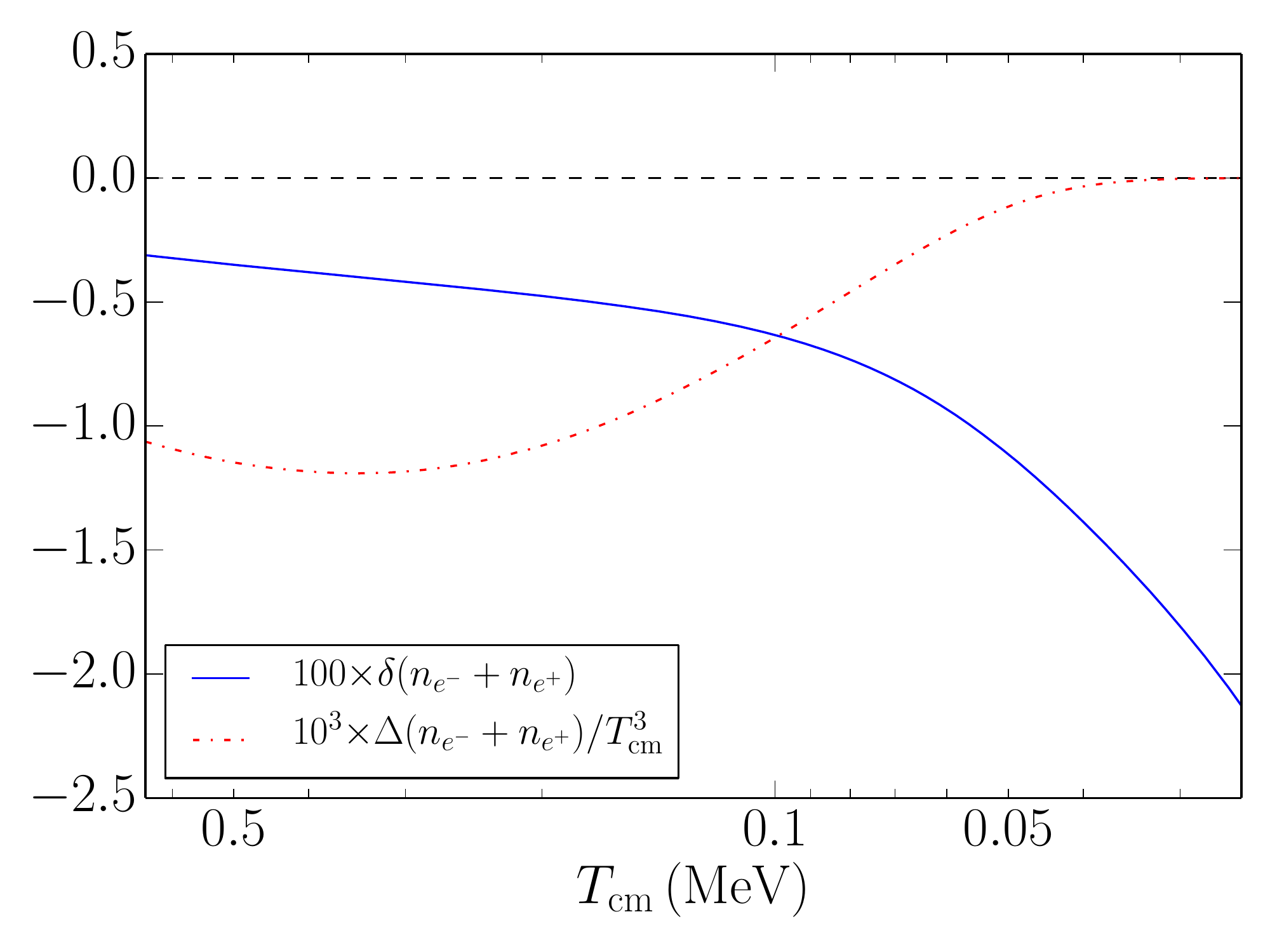}
  \caption{\label{trans-fig:ndens_vs_tcm}(Color online)
  Quantities related to charged lepton number density are plotted against the
  comoving temperature parameter.
  The blue solid curve is the relative change in the sum of positron and
  electron number densities as calculated in Eq.\ \eqref{eqn:delnecl}.
  The red dash-dot curve is the absolute change in the sum of positron and
  electron number densities, divided by $\tcm^3$.}
\end{figure}

Using a notation similar to that of Eq.\ \eqref{trans-eqn:dtcmpl}, we define the
absolute change in the number density of charged leptons as 
\beq\label{trans-eqn:absdelne}
  \Delta(n_{e^-} + n_{e^+})\equiv
  (n_{e^-}+n_{e^+})_{\rm all} -(n_{e^-}+n_{e^+})_{\rm none},
\eeq
and the relative change in the number as
\begin{equation}\label{eqn:delnecl}
  \delta (n_{e^-}+n_{e^+}) \equiv \frac{
  \Delta(n_{e^-} + n_{e^+})}
  {(n_{e^-}+n_{e^+})_{\rm none}}.
\end{equation}
The quantity $(n_{e^-} + n_{e^+}) / \tcm^3$ is proportional to the total number
of electrons and positrons in a comoving volume.  The dot-dashed curve in Fig.\
\ref{trans-fig:ndens_vs_tcm} shows this quantity, while the solid curve in
Fig.\ \ref{trans-fig:ndens_vs_tcm} shows the relative change in the number of
charged leptons.  The absolute change in the total number of charged leptons in
a comoving volume is negative. This implies that there are fewer charged
leptons and hence a lower plasma temperature in the transport case than in the
no-transport case at a given \tcm.  The slope of the absolute change represents
the different annihilation rates.  The negative slope (at $\tcm \gtrsim
400\,{\rm keV}$) indicates that the annihilation rate is greater in the
transport case than in no-transport, while the opposite is true for $\tcm
\lesssim 400\,{\rm keV}$.

The rate at which entropy is transferred from the electrons and positrons to
the photons is proportional to the annihilation rate divided by the plasma
temperature.  For $100\,{\rm keV} \lesssim \tcm \lesssim 400\,{\rm keV}$, the
competition between a larger annihilation rate in the no-transport case and
the lower plasma temperature in the transport case results in a slight decrease
in $\delta (\tcmpl)$.  Finally for $\tcm \lesssim 100\,{\rm keV}$, the greater
annihilation rate in no-transport results in an increasing $\delta(\tcm/T)$
until virtually no electrons and positrons remain and $\delta(\tcm/T)$ reaches
its asymptotic value.

We turn now to the study of the individual and joint contributions of
the annihilation and elastic processes.

\subsection{Neutrino energy transport analysis}
\label{ssec:neta}

We return to the discussion of the weak interaction processes in Table
\ref{trans-tab:ssamps} and their effect on neutrino observables
that probe the changes to their energy density -- the comoving-plasma
temperature ratio, energy density changes, and $\Delta\neff$. The
component contributions are collected in the following subsections in
terms of the annihilation and elastic channels.

\subsubsection{Annihilation Channel}
\label{sssec:ac}

\begin{figure}[h]
  \includegraphics[width=\columnwidth]{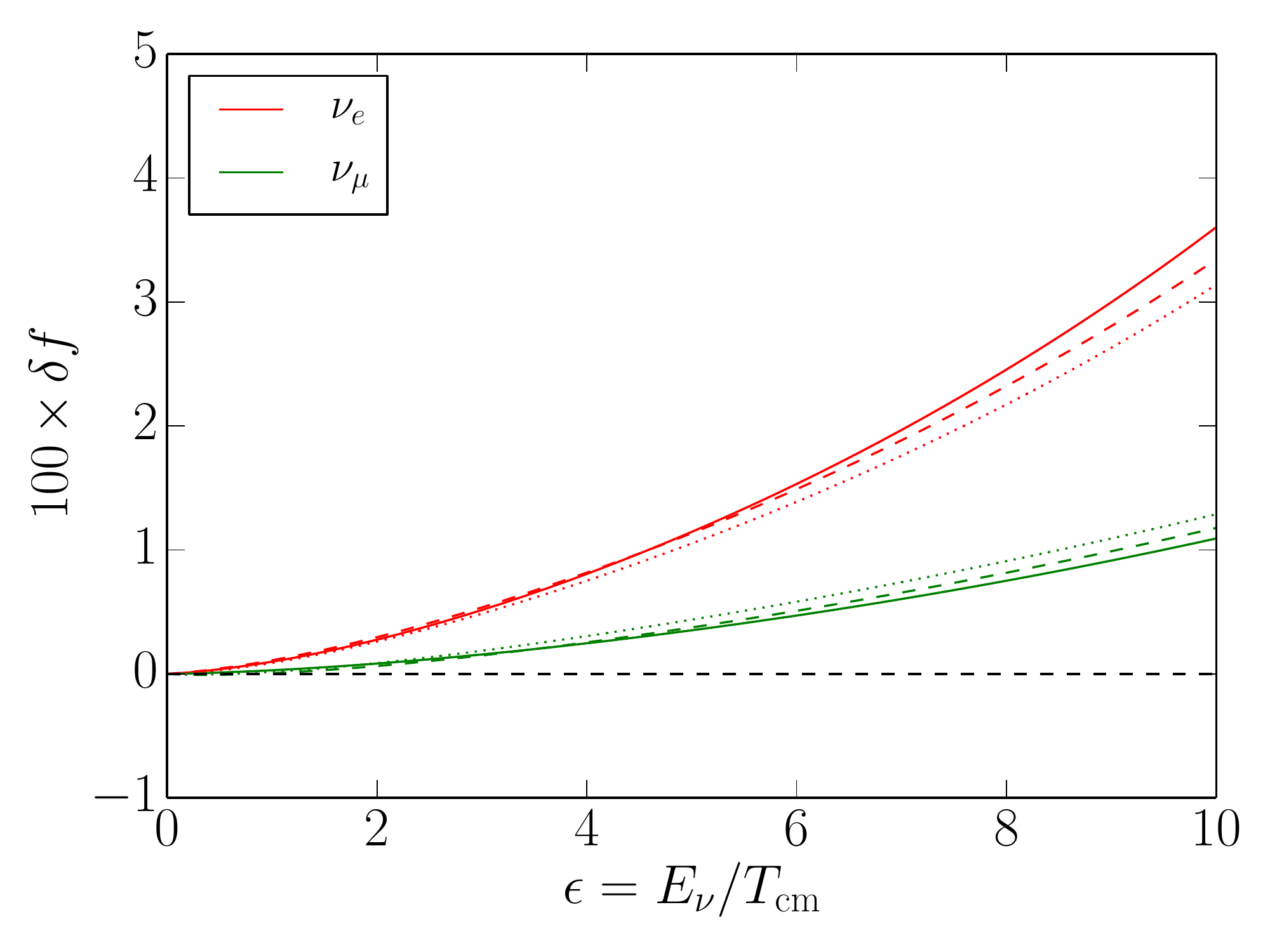}
  \caption[Relative change in occupation probabilities versus
  $\epsilon$ for annihilation combinations]
  {\label{trans-fig:plot_docc_vs_eps_ac}(Color online) The change in
  the neutrino occupation probabilities relative to the equilibrium
  distributions as a function of $\epsilon$. Electron neutrinos
  exhibit a larger change compared to the muon neutrinos.  Solid lines
  correspond to processes $r=10$ and 11. Dashed lines correspond to
  processes $r=1,2,10$ and 11. Dotted lines correspond to
  processes $r=1,\ldots,5,10$ and $11$.}
\end{figure}

We focus here on annihilation-channel effects, which for the present
purposes are defined according to \Cref{trans-tab:ssamps} as processes
$r=10$ and 11.  Figure \ref{trans-fig:plot_docc_vs_eps_ac} shows the
relative changes in the occupation probabilities for the annihilation
channels as a function of $\epsilon$.  The solid curves are when all
weak interaction transport processes are neglected, except for
annihilation of a neutrino and antineutrino into an electron-positron
pair. The dashed curves are the same as the solid curves with the
addition of processes $r=1$ and 2. The dotted curves further include
processes $r=3, 4$ and 5. These processes do not exchange population
among individual \epsvals for a given flavor. Instead, there is an
equilibration between the \nue and \num flavors. This is clearly seen
in the figure since the difference between the dashed and dotted
curves decreases relative to the solid, annihilation curves.

\subsubsection{Elastic Scattering Channel}
\label{sssec:esc}

\begin{figure}[h]
  \includegraphics[width=\columnwidth]{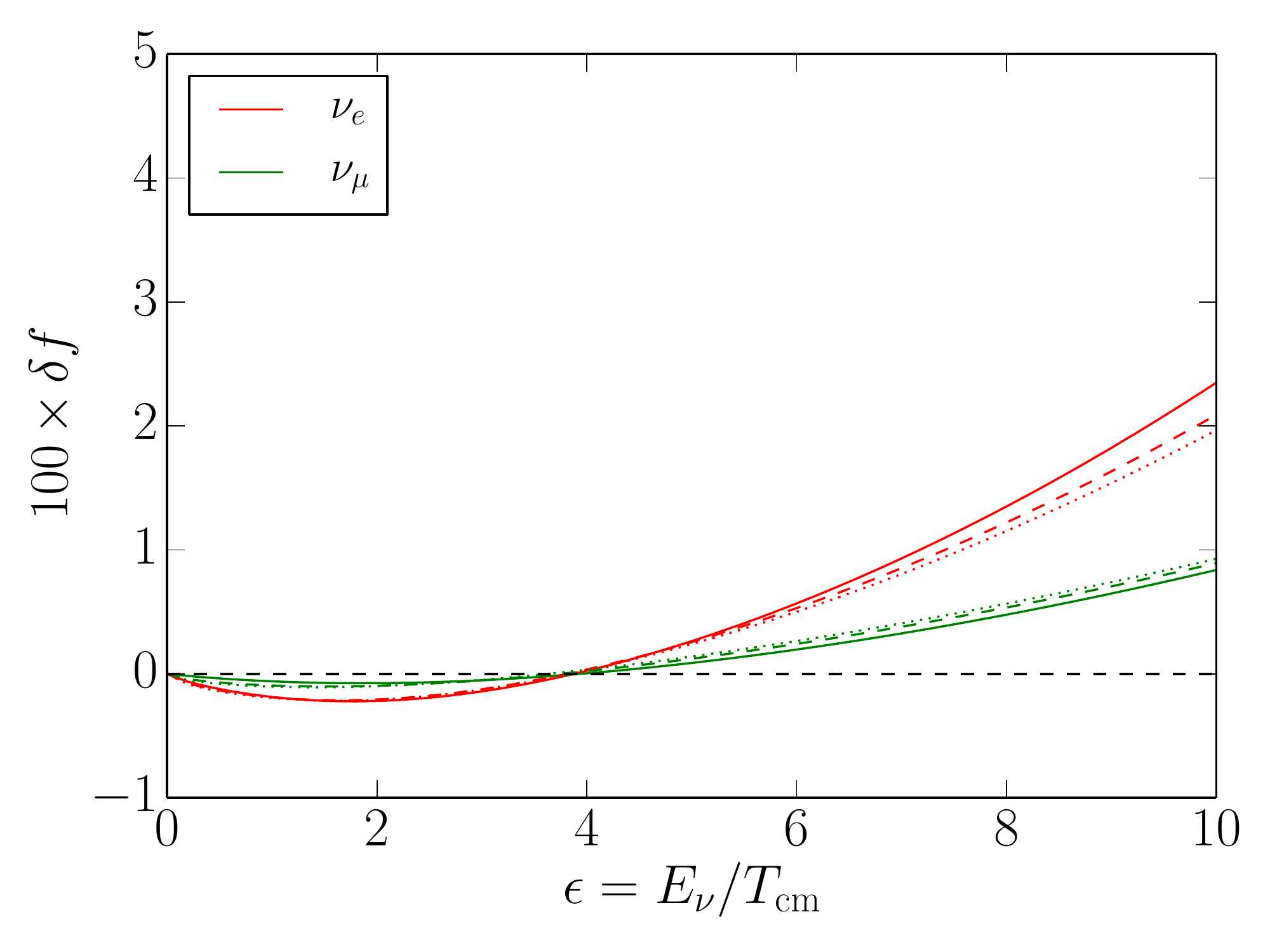}
  \caption[Relative change in occupation probabilities versus
  $\epsilon$ for elastic scattering combinations]
  {\label{trans-fig:plot_docc_vs_eps_nc}(Color online) The change in
  the neutrino occupation probabilities relative to the equilibrium
  distributions as a function of $\epsilon$.  The electron neutrinos
  again exhibit larger changes relative to the muon neutrinos.  Solid
  lines correspond to processes $6,\ldots,9$. Dashed lines correspond
  to processes $1,2,6, \ldots,9$. Dotted lines correspond to processes
  $r=1,\ldots,9$.}
\end{figure}

Figure \ref{trans-fig:plot_docc_vs_eps_nc} shows changes relative to
equilibrium for the combinations of processes involving elastic scattering of
neutrinos on the charged leptons. It is distinguished by its behavior at low
$\epsilon$, where the change relative to equilibrium goes negative for
$\epsilon\lesssim4$, a feature not found in Fig.\
\ref{trans-fig:plot_docc_vs_eps_ac}.  Neutrinos, whose number are conserved in
processes $r=6,\ldots,9$, upscatter and populate the larger epsilon bins. \neff
increases, and conversely, the energy in the plasma decreases. Processes
$1,\ldots,5$ behave much the same way as they do for the annihilation
combinations. These processes act to equilibrate the occupation probabilities
among flavors for a given bin. Perhaps surprisingly, the five neutrino-only
processes do not appear to equilibrate the bins for a given flavor; the
addition of the processes in rows $r=1,2$ do not change the intersection with
the horizontal axis at $\eps\simeq4$.  The further additions of the processes
in rows $r=3,4,5$ also preserve the intersection point with the horizontal
axis. Note that in both Figs.\ref{trans-fig:plot_docc_vs_eps_ac} and
\ref{trans-fig:plot_docc_vs_eps_nc} there is a larger divergence between the
solid, dashed, and dotted lines with increasing $\epsilon$-value for the \nue
as compared to the \num.  This is because the population is transferred from \nue
into \emph{both} \num and \nut.  

When adding the annihilation and elastic scattering channels together,
as in Fig.\ \ref{trans-fig:plot_doccprob_eps}, the annihilation
channels are able to repopulate the low \epsvals states.  Annihilation
erases much of the deficit caused by elastic scattering,
although Fig.\ \ref{trans-fig:plot_doccprob_eps} shows that annihilation
cannot entirely erase the deficit for $\epsilon\lesssim1$.

\subsection{Finite temperature QED radiative corrections}
\label{trans-ssec:QED}

Our calculations show a new sensitivity to finite temperature QED radiative
corrections.  The feedback between neutrino energy transport and plasma
conditions is especially sensitive to electron-positron pair number density as
these are key targets for neutrino scattering.  Previous
works \cite{1982PhRvD..26.2694D,1982NuPhB.209..372C,1994PhRvD..49..611H,1999PhRvD..59j3502L}
include the effects of finite temperature QED radiative corrections.  The
corrections have been calculated for the electron mass and wave function
renormalization, the electron-photon vertex and infrared photon emission and
absorption. These corrections have an effect on a variety of quantities
including the dispersion relation of the electron mass, weak interaction rates,
and the equation of state.

In the present study, we include radiative corrections to the
self-interaction energy for electrons, positrons, and photon
dispersion relations. We follow the approach employed in
Ref.\ \cite{Mangano:3.040}.
This formulation is adopted primarily for comparison with previous results.  We
note, however, that the feedback between neutrino energy distribution evolution
and plasma conditions can depend sensitively on the electron-positron pair
density and that, in turn, can depend on these corrections.  This highlights
the need for a more complete treatment of finite temperature plasma effects.
In any case, finite temperature QED radiative corrections by themselves have a
small effect on, say, the relative \Heiv abundance change, which is at the
level of $10^{-4}$, smaller than the neutrino transport effects that we
are primarily concerned with in this work.

We apply finite temperature QED radiative corrections to
contributions to the thermodynamic quantities $\rho$, $p$, $d\rho/dT$,
$d\rho/d\phi_e$, $d(n_{e^-} - n_{e^+})/dT$, and $d(n_{e^-} -
n_{e^+})/d\phi_e$.
For the collision integrals of Sec.\
\ref{trans-ssec:weakints}, we take the vacuum value of electron rest
mass.  In a preview of Sec.\ref{trans-sec:BBN}, the weak interaction
terms involving neutrinos and free nucleons utilize the electron rest
mass at its vacuum value and do not include the higher-order effects
detailed in
Refs.\ \cite{1982PhRvD..26.2694D,1982NuPhB.209..372C,1999PhRvD..59j3502L}.
Our $n\leftrightarrow p$ rates do not take into account the renormalization of
the electron rest mass in any BBN computations. We have computed the effects of
non-zero electron degeneracy $\phi_e\ne 0$ and found them to be negligible, so
we take $\phi_e=0$ in these radiative corrections.  We maintain $\phie\ne0$ in
calculations of neutrino transport, weak rates, and overall charge neutrality
with baryons.

Following Refs.\ \cite{1994PhRvD..49..611H,Mangano:3.040} we take the shift in
the electron rest mass $\delta m_e$ to be
\begin{align}
  \delta m_e^2(p,T) &= \frac{2\pi\alpha T^2}{3}
  +\frac{4\alpha}{\pi}\int\limits_0^{\infty}dk\frac{k^2}{E_k}
  \frac{1}{e^{E_k/T}+1}\nonumber\\
  &-\frac{2m_e^2\alpha}
  {\pi p}\int\limits_0^\infty dk\, \frac{k}{E_k}
  \log\left| \frac{p+k}{p-k}\right| \frac{1}{e^{E_k/T}+1},
\end{align} 
where $\alpha=e^2/(4\pi)$, $E_k=\sqrt{k^2+m_e^2}$, $T$ is again the plasma
temperature. To be consistent
with the procedure adopted in Ref.\ \cite{Mangano:3.040}, we ignore the last, momentum $p$ dependent term.  According to Appendix
B of Ref.\ \cite{1994PhRvD..49..611H} this relation is valid for $T\ll m_e$. We
note that the range of temperatures relevant for BBN include temperatures which
do not satisfy this condition.  It is nevertheless applied in the interest of
comparison with previous results.  Likewise for comparison purposes, the change
in the photon mass \cite{1997PhRvD..56.5123F} is taken as
\begin{align}
\delta m_\gamma^2 =
\frac{8\alpha}{\pi}\int\limits_0^{\infty}dk\frac{k^2}{E_k}
\frac{1}{e^{E_k/T}+1}.
\end{align}

We compute radiative corrections to the thermodynamic quantities by numerical
integration of appropriately weighted distribution functions, including the
dispersion relations with terms $\delta m_e(T)$ and $\delta m_\gamma(T)$.  By
contrast, Ref.\ \cite{Mangano:3.040} applies these same corrections but with a
perturbative approach to the calculation of the thermodynamic quantities.
Consequently, we obtain different asymptotic values for certain cosmological
quantities from Ref.\ \cite{Mangano:3.040}.  To wit, without the inclusion of
neutrino transport, we obtain a value of $\tcm/T=0.7150$, implying
$\neff=3.020$.  In the presence of neutrino transport, our values for the
relevant neutrino parameters are:
\begin{align}
  \tcm/T&=0.7159,\\
  \delta\rho_{\nu_e}&=8.908\times10^{-3},\\
  \delta\rho_{\num}&=3.537\times10^{-3},\\
  \neff&=3.052.
\end{align}
Ref.\ \cite{neff:3.046} employed a binned spectrum to investigate neutrino
oscillations.  The authors included the QED effects and found $\neff=3.046$.
Our value of \neff is reasonably close, although it does differ from
Ref.\ \cite{neff:3.046} by $\sim12\%$. This difference may be due to differing
methods of numerical evaluation. We are primarily concerned with changes in the
primordial abundances relative to our baseline values which stem from
self-consistent neutrino transport/BBN effects. Note that these
transport-induced changes are an order-of-magnitude larger than the QED
finite-temperature effects. Future work will focus on these subdominant
contributions.

\section{Entropy transfer and generation} \label{trans-sec:entropy}

The textbook
treatment \cite{1990eaun.book.....K,Dodelson:2003mc,2008cosm.book.....W} of entropy
exchange in the early universe takes into account entropy flow among the
various components of the cosmic fluid but assumes that entropy generation
is negligible. According to the Boltzmann H theorem, however, entropy
increases whenever non-equilibrium kinetics obtain.  Homogeneity and isotropy
of the Friedmann-Lema\^{i}tre-Robertson-Walker metric preclude a spacelike heat
flow. The entropy in a comoving volume can change, however, if there is a
timelike heat flow which respects the overall symmetry of homogeneity and
isotropy on any spacelike surface $t={\rm constant.}$

Of course, the weak decoupling of neutrinos from the plasma prior to and during
BBN is a classic example of a non-equilibrium process and we therefore expect
the total entropy to increase with increasing time/scale factor or decreasing
comoving temperature.  We have calculated the total entropy of the neutrino
plus other plasma constituents in the early universe and find that it varies at
the sub-percent level.  Nevertheless, there are entropy flows between the
photon/electron/positron plasma and the decoupling neutrinos which are
considerably larger than this and which alter nucleosynthesis relative to a
no-transport case.

\begin{figure*}
  \includegraphics[width=0.75\textwidth]{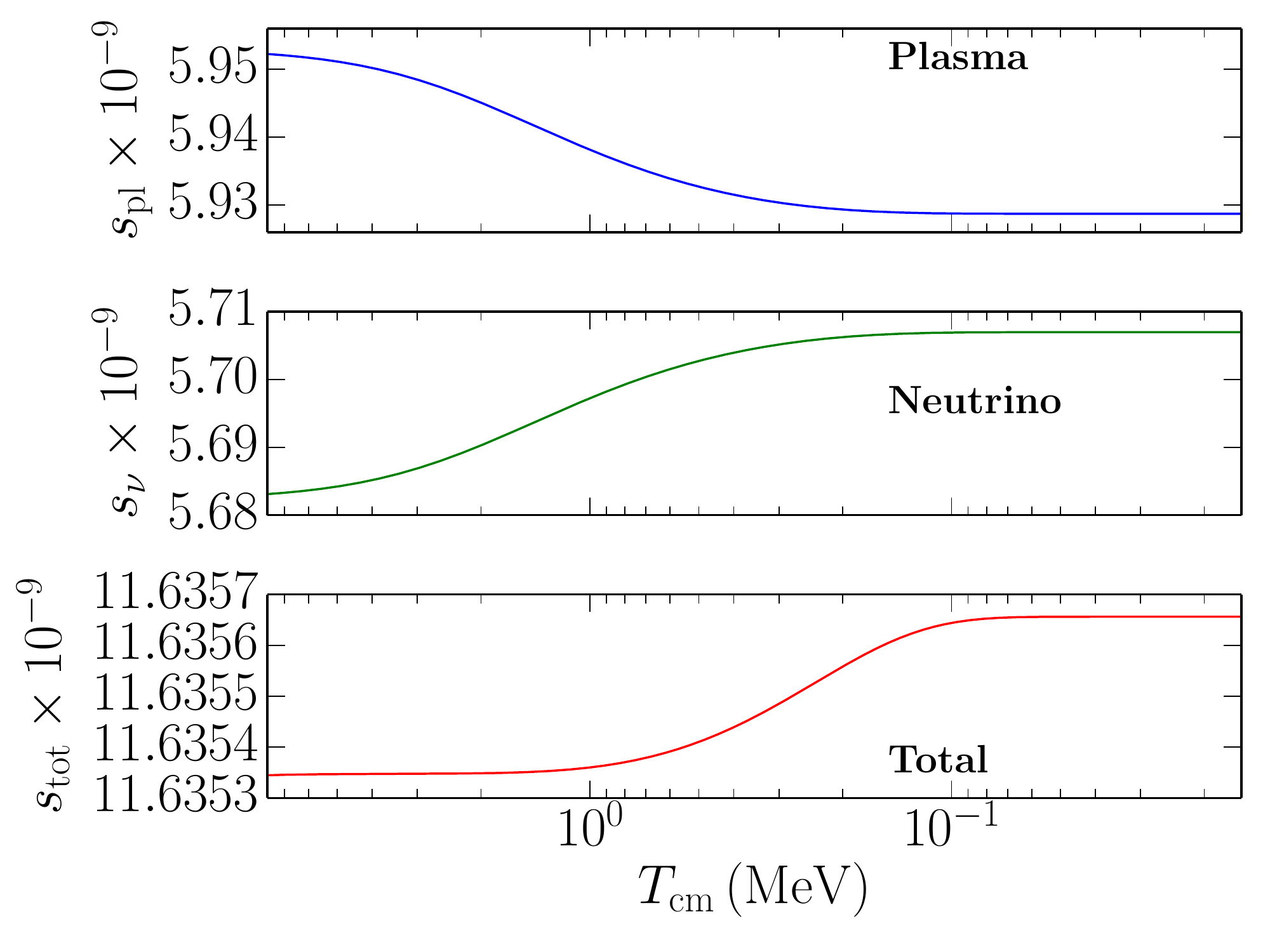}
  \caption[Entropy versus \tcm]{\label{trans-fig:plot_entropy_vs_tcm}
     (Color online) The entropy-per-baryon for three sectors as
     functions of comoving temperature.  The top panel (blue line) is
     the evolution of the entropy per baryon in the plasma, $s_{\rm
     pl}$.  The middle panel (green line) is the evolution of the
     entropy per baryon in the neutrino sector, $s_{\nu}$.  The lower
     panel (red line) is the evolution of the total entropy per
     baryon, $s_{\rm tot}$.}
\end{figure*}

The entropy may be calculated generally -- in either equilibrium or
non-equilibrium states -- as
\begin{align}
   \label{trans-eqn:general_S}
   S_i=-\int\frac{d^3x\,d^3p}{(2\pi)^3}[f_i\ln f_i+(1-f_i)\ln(1-f_i)],
\end{align}
for a species $i$. For species in equilibrium the above reduces to
the familiar thermodynamic relation for the entropy per baryon $s$:
\begin{align}
   \label{trans-eqn:equil_s}
   s\equiv\frac{S}{n_bV} =
   \frac{1}{n_b}\frac{\rho+P-\sum\limits_i\mu_i n_i}{T},
\end{align}
for baryon number density, $n_b$, energy density $\rho$, pressure $P$,
chemical potential $\mu_i$ and number density $n_i$ for species $i$.  

Assuming homogeneity and isotropy, Eq.\ \eqref{trans-eqn:general_S}
leads to the entropy per baryon for general, non-equilibrium,
states:
\begin{align}
   \label{trans-eqn:nonequil_s}
   s_i=-\frac{\tcm^3}{2\pi^2n_b}
   \int\limits_{0}^{\infty}d\epsilon\,\epsilon^2
   [f_i\ln f_i+(1-f_i)\ln(1-f_i)],
\end{align}
where the occupation probabilities $f_i$ are taken to be functions of
$\epsilon$ and time.  As described earlier, \tcm in
Eq.\ \eqref{trans-eqn:nonequil_s} is a proxy for inverse scale factor,
$a^{-1}(t)$. Since the comoving baryon number $n_ba^3$ is covariantly
conserved, we have
\beq
  \frac{\tcm^3}{n_b}={\rm const}.
\eeq

Taking the time derivative of Eq.\ \eqref{trans-eqn:nonequil_s}, we determine the
change in the entropy of species $i$ by using the Boltzmann equation and
general collision integral, $C_i[f_j]$,
\begin{align}
   \frac{ds_i}{dt} = -\frac{\tcm^3}{2\pi^2n_b}
   \int\limits_{0}^{\infty}d\epsilon\,
   \epsilon^2{C}_i[f_j]\ln\left(\frac{f_i}{1-f_i}\right).
\end{align}
The Boltzmann H theorem implies that the sum of the constituent entropies must
be non-negative but the derivative of a given species, of course, has arbitrary
sign.

We write the total entropy change in the
neutrino sector as a summation over the individual species:
\beq\label{trans-eqn:dsnudt}
  \frac{ds_\nu}{dt} =
  -\frac{\tcm^3}{2\pi^2n_b}\sum\limits_{i=1}^{6}
  \int\limits_{0}^{\infty}d\epsilon\,
  \epsilon^2{C_{\nu_i}}[f_j]
  \ln\left(\frac{f_{\nu_i}}{1-f_{\nu_i}}\right).
\eeq
Assuming equilibrium distributions for photons, electrons, and
positrons, and ignoring the negligible contribution from baryons, we
compute the change in entropy of the plasma employing equilibrium
thermodynamics, taking account of energy conservation. We assume that
cooling of the plasma occurs only due to interactions between
neutrinos and charged leptons in scattering and annihilation
processes. Heating due to nucleosynthesis, primarily from the release
of binding energy of $^4$He, is neglected since the relative
contribution of the binding energy heat to the plasma is $\sim
10^{-9}$. This gives the change in entropy of the plasma due to
heating of the neutrinos:
\begin{align}
   \label{trans-eqn:dspldt}
   \frac{d\spl}{dt} &= \frac{1}{n_bT}\frac{dq}{dt}
   =-\frac{\tcm^4}{2\pi^2n_bT}
   \sum_{i=1}^6\int\limits_{0}^{\infty}d\epsilon\,
   \epsilon^3 \, C_{\nu_i} [f_j],
\end{align}
where $q$ is the energy flux per unit volume and the sum is over
neutrinos $\nu_i$. The minus sign is required to define heat flow
$q>0$ out of the plasma. The time derivative of the total entropy per
baryon is the sum of Eqs.\ \eqref{trans-eqn:dsnudt} and
\eqref{trans-eqn:dspldt}:
\begin{align}
  \frac{d\stot}{dt} &=
  -\frac{\tcm^3}{2\pi^2n_b}\sum\limits_{i=1}^{6}
  \int\limits_{0}^{\infty}d\epsilon\,
  \epsilon^2{C_{\nu_i}}[f_j]\nonumber \\
  &\times \left[\epsilon\frac{\tcm}{T}
  +\ln\left(\frac{f_i}{1-f_i}\right)\right],
\end{align}
which must be positive to satisfy the
H theorem \cite{1988kteu.book.....B}.

We show the change in the entropy-per-baryon components in Fig.\
\ref{trans-fig:plot_entropy_vs_tcm}.  The blue curve of the top panel
shows the entropy-per-baryon of the plasma as a function of comoving
temperature. As expected, the plasma loses entropy as it heats and
decouples from the neutrinos. The green curve of the middle panel
gives the entropy-per-baryon in the neutrino seas as a function of comoving temperature. It
is increasing due to heating from the plasma, also as expected. The
red curve in the lower panel is the sum of $s_{\rm pl}$ and $s_{\nu}$.
As expected, it is a monotonically increasing
function of time.
We discuss the role of the entropy flows as shown in Fig.\
\ref{trans-fig:plot_entropy_vs_tcm} in more detail below.

The lower panel of Fig.\ \ref{trans-fig:plot_entropy_vs_tcm} shows
that the epoch of weak decoupling occurs over $\sim10^3$ Hubble times.
Starting at the left side of the figure at $\tcm=8$ MeV we see that
entropy\footnote{We refer to ``entropy'' from here on, though it is to
be understood that this is the entropy-per-baryon.} is already being
exchanged between the plasma and the neutrinos, although at a low
rate. Until the comoving temperature reaches about 2 MeV, this entropy
exchange between the components of the cosmic fluid occurs {\em in
equilibrium} since the total entropy (bottom panel) is constant. Near
1 MeV, the total entropy begins to deviate from its high-temperature,
equilibrium value. In the region of temperatures from 8 MeV $> \tcm >
2$ MeV, the rates of {\em equilibrium} entropy exchange are
increasing. The component entropies (top and middle panels), near
$\tcm \approx 1$ MeV, reach a point of inflection and, concomitantly,
the total entropy begins to increase, deviating significantly from its
high-temperature (low) value. More than half of the entropy
transferred to the neutrinos from the plasma is complete by this
temperature. As the comoving temperature continues to drop, going
below 1 MeV, heating becomes more effective at changing the total
entropy. During this epoch of {\em entropy production}, the entropy
generated by weak-interaction driven kinetic processes is larger than
the entropy lost from the neutrinos.  We note that, contrary to the
order-of-magnitude estimates of Eq.\ \eqref{eqn:nwdc-approxtemp}, the
process of weak decoupling, measured as the point at which the
derivatives drop below some near-zero value, lasts until $\tcm \simeq
90-100$ keV, well into the epoch of BBN.

Table \ref{trans-tab:entropysummary} shows the initial, final, and relative
changes in the entropy for the same runs as those performed in Section
\ref{trans-sec:wd}. The first row (``None''), corresponds to the ``standard''
cosmology without transport. The second row (``All'') corresponds to the curves
in Fig.\ \ref{trans-fig:plot_entropy_vs_tcm}.  Component contributions to the
collision integrals, corresponding to various $r$ in Table
\ref{trans-tab:ssamps}, are given in the remaining rows. Here it is apparent
that the dominant process contributing to entropy generation are due to the
annihilation processes $r=10$ and 11.

\begin{table*}
  \begin{tabular}{| c !{\vrule width 1.5 pt} c !{\vrule width 1.5 pt} c !{\vrule width 1.5 pt} c |}
    \hline
    Processes & $10^{-9}\times\spl^{(i)}$ & $10^{-9}\times\spl^{(f)}$ &
    $(\spl^{(i)} - \spl^{(f)})/\spl^{(f)}$\\ 
    \midrule[1.5pt]
    None & 5.929 & 5.929 & 0\\ \midrule[1.5pt]
    All & 5.952 & 5.929 & $3.977\times10^{-3}$\\ \midrule[1.5pt]
    10, 11 & 5.950 & 5.929 & $3.574\times10^{-3}$\\ \hline
    1, 2, 10, 11 & 5.950 & 5.929 & $3.574\times10^{-3}$\\ \hline
    1, 2, 3, 4, 5, 10, 11 & 5.950 & 5.928 & $3.663\times10^{-3}$\\ \midrule[1.5pt]
    6, 7, 8, 9 & 5.933 & 5.929 & $7.426\times10^{-4}$\\ \hline
    1, 2, 6, 7, 8, 9  & 5.933 & 5.928 & $7.798\times10^{-4}$\\ \hline
    1, 2, 3, 4, 5, 6, 7, 8, 9 & 5.933 & 5.928 & $7.798\times10^{-4}$\\ \hline
  \end{tabular}
  \caption[Entropy
  runs]{\label{trans-tab:entropysummary}Process-dependent changes in
     the plasma entropy.  For all runs $\epsmax=20.0$, $\nbins=100$,
     $\tin=8\,{\rm MeV}$, $\tstop=15\,{\rm keV}$, $\nuratiotol=30.0$.
     Column one gives the processes used for a given run similar to
     Table \ref{trans-tab:transsummary}.  The second column is the
     initial entropy-per-baryon in the plasma at \tin.  Column three
     is final \spl at \tstop.  Column four is the relative change
  between columns two and three.}
\end{table*}

\section{Weak Freeze-Out and Nucleosynthesis}
\label{trans-sec:BBN}


In this section we examine how the charged current weak reactions involving
nucleons and the strong and electromagnetic nuclear reactions are affected by
the evolving neutrino and plasma components. As outlined above, the
scattering-driven non-equilibrium evolution of the neutrino energy distribution
functions through the weak decoupling epoch is nonlinearly coupled to the
plasma thermodynamic conditions.
The plasma of photons, electrons, and positrons is maintained in thermal
equilibrium by electromagnetic interactions whose rates are much faster than the
Hubble rate for all epochs under present consideration. The rapid fall-off in
the weak interactions of neutrinos with neutrons and protons, however, result
in the weak freeze-out of the \np ratio where chemical equilibrium is no longer
maintained even though thermal equilibrium still obtains.  Systems where
equilibrium is maintained instantaneously are, at any given time, insensitive
to the previous history of the system. Quantities characterizing systems which
are out of equilibrium, on the other hand, {\it can be sensitive} to previous
history. In fact, since the \np ratio and nuclear reactions are not in chemical
equilibrium, the out-of-equilibrium neutrino energy distributions alter BBN
abundance yields over the no-transport case.

As discussed in Appendix \ref{trans-app:overview}, DHS
(Ref.\ \cite{Dolgov:1997ne}) has taken into account effects of neutrino transport
during weak decoupling on energy density, the weak interactions and the plasma
temperature derivative. The work of DHS, which is most similar to our present
treatment, however, employs a perturbative approach for nucleosynthesis. There,
the primordial nucleosynthesis was ``post-processed'' by using the results from
DHS' prior solution of the coupled set of neutrino Boltzmann equations.
Our treatment concurrently solves the Boltzmann equations for the neutrino
occupation probabilities and the light nuclide abundances or mass fractions,
given by:
\beq
  Y_i\equiv\frac{n_i}{n_b}\quad{\rm and}\quad
  X_i\equiv A_iY_i,
\eeq
where for a given species $i$: $n_i$ is the number density, $A_i$ is the atomic
mass number, $Y_i$ is the abundance, and $X_i$ is the mass fraction.  The
quantity $n_b$ is the baryon number density.  This fully coupled,
self-consistent approach results in a significant enhancement of effects that
change the light element abundances from the treatments without transport or
with transport included perturbatively, as we detail in this section.

In both our transport and no-transport BBN calculations we employ the value of
the baryon-to-photon ratio from Ref.\ \cite{PlanckXVI:2014}, corresponding to
$\bardens=\Omega_b h^2=0.022068$.  This also corresponds to the final entropy
per baryon in the plasma of $\spl=5.929\times10^9$ units of Boltzmann's
constant.  We emphasize that these are the \emph{final} values of these
quantities after all transport and entropy generating reactions have ceased,
i.e., as measured at the CMB decoupling epoch.  A standard BBN, baseline
calculation assuming constant comoving entropy, but not including QED and other
corrections, yields the following values for the primordial mass fraction of
\Heiv, relative abundances of deuterium, \Heiii and \Livii (with respect to
hydrogen):
\begin{align}
  Y^{\rm (N)}_P \equiv X_{^4{\rm He}}
  &= 0.2438,\label{trans-eqn:ypnot}\\
  \left({\rm D/H}\right)^{\rm (N)} \equiv Y_{\rm D}/Y_{\rm H}
  &= 2.627\times10^{-5},\label{trans-eqn:dhnot}\\
  \left(^3{\rm He/H}\right)^{\rm (N)} &= 1.049\times10^{-5},\label{trans-eqn:3henot}\\
  \left(^7{\rm Li/H}\right)^{\rm (N)} &= 4.277\times10^{-10}.\label{trans-eqn:7linot}
\end{align}
We refer to the abundances in this baseline computation as (N).  The standard
BBN calculation and associated reaction network employed here is detailed in
Refs.\ \cite{Wagoner:1966pv, Wagoner:1969sy, GFKP-5pts:2014mn}. We emphasize
that the (N) baseline computation does not include Coulomb corrections (CC),
zero-temperature radiative corrections (0T), and transport-induced corrections
(Trans).  As an alternative baseline we consider the inclusion of Coulomb
corrections (given by Eq.\ (5b) in Ref.\ \cite{1980ApJS...42..447F}) to the
reactions Eqs.\ \eqref{trans-eqn:np1} and \eqref{trans-eqn:np3} (on page
\pageref{trans-eqn:np1}); and zero-temperature radiative corrections (Eq.\
(2.14) in Ref.\ \cite{1982PhRvD..26.2694D}) to reactions \eqref{trans-eqn:np1},
\eqref{trans-eqn:np2}, and \eqref{trans-eqn:np3}.  See Ref.\
\cite{2010PhRvD..81f5027S} for a detailed discussion of the Coulomb corrections
to BBN.  The QED corrections discussed in Sec.\ref{trans-ssec:QED} are excluded
for this baseline.  The helium-4 (hereafter shortened to helium) mass fraction
and relative abundances for this baseline are
\begin{align}
  Y^{\rm (Q)}_P &= 0.2478,\label{trans-eqn:ypcc0t}\\
  \left({\rm D/H}\right)^{\rm (Q)} &= 2.650\times10^{-5},\\
  \left(^3{\rm He/H}\right)^{\rm (Q)} &= 1.052\times10^{-5},\\
  \left(^7{\rm Li/H}\right)^{\rm (Q)} &= 4.317\times10^{-10}.\label{trans-eqn:7licc0t}
\end{align}
We refer to the abundances in this baseline computation as (Q).  The (Q)
baseline allows us to compare to other nucleosynthesis codes.  To wit, in the
(Q) baseline we obtain for the primordial helium mass fraction $Y_P = 0.2478$,
which is within $\sim 0.1\%$ of the value from the {\sc parthenope} code
\cite{2008CoPhC.178..956P} of $0.24725$ \cite{Nollett:2014bb}.  Table
\ref{trans-tab:highorderbbn} shows the effect on the abundances for these
cases.  


We use a semi-implicit Heun's method to integrate the BBN nuclear reaction
network \cite{letsgoeu2} from $t=t_n$ to $t=t_{n+1}$.  To calculate the
abundance derivatives we need the abundance values themselves, $Y_j$, and a
set of thermodynamic/transport quantities, namely $T$, \tcm, \phie, $\rho_b$,
and the \nue, \bnue occupation probabilities.  We integrate the RK5 method by
partitioning the time interval $\Delta t=t_{n+1}-t_n$ into six subintervals
(see Ref.\ \cite{Press:1993:NRF:563041} for details on the fifth-order
Runge-Kutta method with a Cash-Karp time step).  We step through each
subinterval and evolve the above set of thermodynamic/transport quantities (and
other quantities as well) but not the $Y_j$.  We extrapolate the small
nucleosynthesis contributions to the derivative of \phie (from alterations of
the \np ratio) and to the plasma-temperature derivative (from the release of
nuclear binding energy and the \np ratio) for each of the subintervals in the
RK5 method.  The baryon-to-photon ratio is small enough that the extrapolation
does not produce substantial error in either the gross thermodynamics of the
plasma or the Boltzmann neutrino-energy transport network (see Sec.\
\ref{trans-sec:entropy}).  We store within memory the set of
thermodynamic/transport quantities needed for the reaction network at two
specific subintervals while integrating the RK5 method: the first subinterval
(corresponding to the start of the time interval, $t=t_n$); and the fifth
subinterval (corresponding to the end of the time interval, $t=t_n+\Delta t$).
Once the RK5 terminates, we check for numerical convergence.  If the
convergence criteria failed, we repeat the RK5 calculation (beginning at
$t=t_n$) with a smaller time step.  If the convergence criteria succeeded, we
accept the thermodynamic/transport quantities at $t_n+\Delta t=t_{n+1}$ and
proceed to integrate the nuclear reaction network with Heun's method to obtain only
the $Y_j$ at $t_{n+1}$.  Heun's method requires an initial evaluation at
the start of the interval and a second evaluation at the end of the interval.
We recall the set of thermodynamic/transport quantities stored in memory to use
in the integration of the nuclear reaction network.  Specifically, for the
first computation we recall $T(t_n)$, $\tcm(t_n)$, $\phie(t_n)$, $\rho_b(t_n)$,
$f_{\nue}(\epsilon,t_n)$, $f_{\bnue}(\epsilon,t_n)$, and the \emph{current}
values of the abundances $Y_j(t_n)$ to calculate a first set of abundance
derivatives.  This is accomplished by utilizing the Jacobian of a linearized
Boltzmann equation for nuclear reactions and subsequently diagonalizing a
matrix (see Refs.\ \cite{Wagoner:1969sy} and \cite{letsgoeu2} for details on
this procedure).  Using the time step value $\Delta t$ and the first set of
abundance derivatives, we estimate the new values of the abundances,
$\widetilde{Y}_j(t_n+\Delta t)$.  At this stage in Heun's method, the
$\widetilde{Y}_j(t_n+\Delta t)$ are only \emph{estimates} of the abundances at
$t_n+\Delta t$; they are not the calculated abundances, i.e., the
$Y_j(t_{n+1})$.  Next, we calculate a second set of abundance derivatives again
using the Jacobian of the linearized Boltzmann equation.  This new set of
derivatives requires the second set of thermodynamic/transport quantities,
i.e., $T(t_{n+1})$, $\tcm(t_{n+1})$, etc., and the previous estimates of the
abundances, namely the $\widetilde{Y}_j(t_n+\Delta t)$.  Finally, we average
the two sets of abundance derivatives and arrive at a derivative for each
nuclide.  We use this derivative and the time step to calculate the new value
of the abundances $Y_j(t_{n+1})$.  After we integrate the nuclear reaction
network and have obtained the $Y_j$, we proceed to the next time point
$t_{n+1}$ and repeat the process.

With the two baseline calculations in hand, we are in a position to study the
effect that weak interaction processes and neutrino transport has on the
primordial abundances {\em relative to these baseline cases}.  This comparison
is done in Tables \ref{trans-tab:bbnsummary} and \ref{trans-tab:highorderbbn}.

In both tables \ref{trans-tab:bbnsummary} and \ref{trans-tab:highorderbbn} we
show the change in a nuclide, $\delta Y$, relative to the (N) baseline
case as
\beq
  \delta Y\equiv\frac{Y^{\rm (proc)}- Y^{\rm (N)}}{Y^{\rm (N)}},
\eeq
where $Y^{\rm (proc)}$ is the quantity of interest for the specific set of
processes. $Y^{\rm (N)}$ is the quantity of interest for the case of no
transport and no higher-order corrections to the $n\leftrightarrow p$ rates.
i.e.\ our (N) baseline value labeled ``None'' in row 1 of Tables
\ref{trans-tab:bbnsummary} and \ref{trans-tab:highorderbbn}.

\begin{figure*}
  \includegraphics[width=0.75\textwidth]{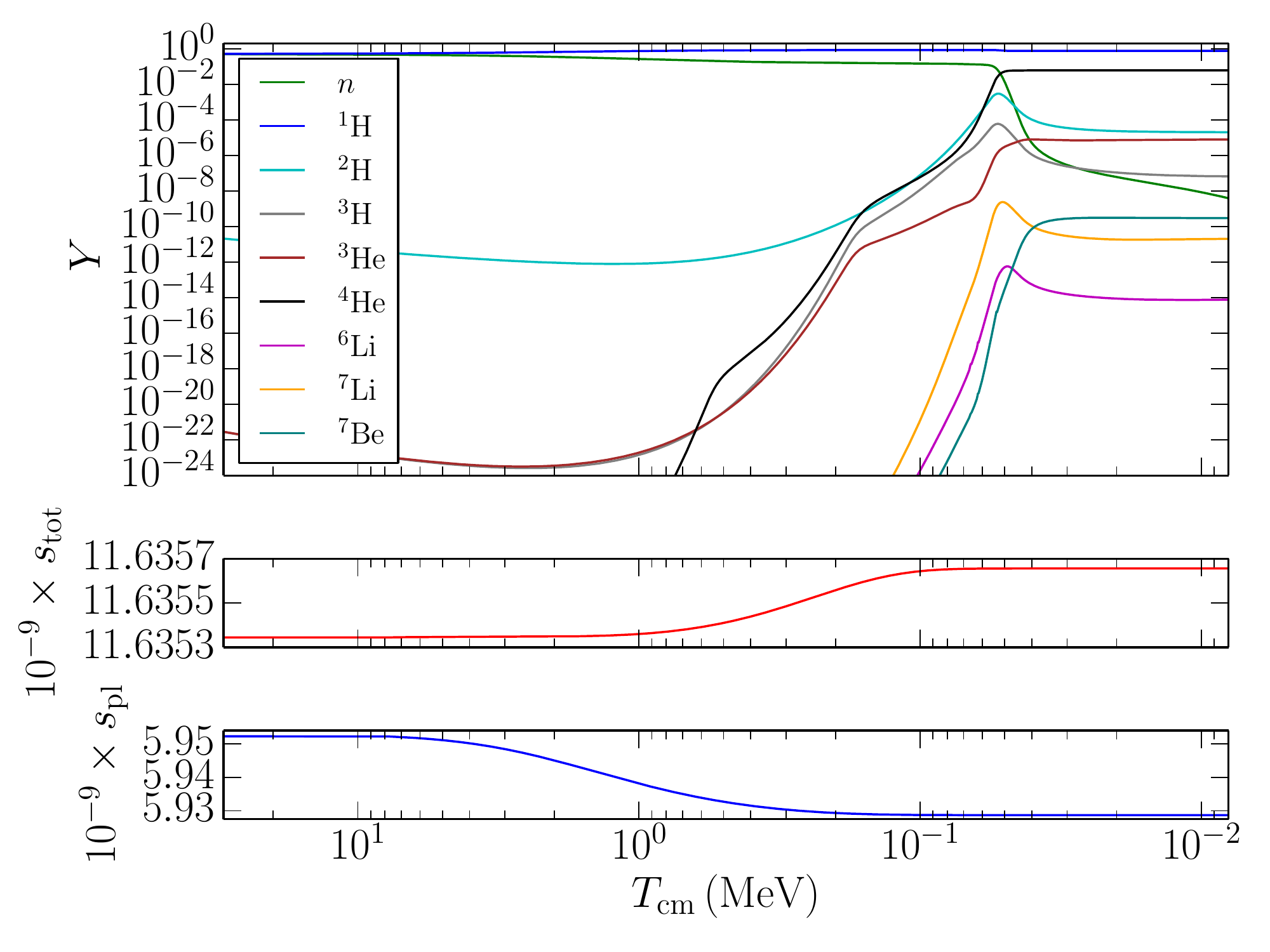}
  \caption[Evolution of abundances and entropy versus the comoving
  temperature
  parameter]{\label{trans-fig:plot_abunds_total_s_vs_tcm}(Color
  online)
     [Top panel] Evolution of nuclear abundances as a function of
     comoving temperature parameter. [Middle panel] The evolution of
  the total entropy as a function of scale factor. [Lower panel]
  Evolution of entropy carried by the photon/electron/positron
  plasma. Coulomb and zero-temperature radiative corrections
  are not included in this plot.}
\end{figure*}

\begin{table*}
   \begin{adjustbox}{width=\textwidth}
  \begin{tabular}{| c !{\vrule width 1.5 pt} c | c !{\vrule width 1.5 pt} c | c 
  !{\vrule width 1.5 pt} c | c !{\vrule width 1.5 pt} c | c |}
    \hline
    Processes & $Y_P$ & $\delta Y_P$ & $10^5\times{\rm D/H}$ &
    $\delta({\rm D/H})$ & $10^5\times\,^3{\rm He/H}$ &
    $\delta(^3{\rm He/H})$ & $10^{10}\times\,^7{\rm Li/H}$ & $\delta(^7{\rm Li/H})$\\ 
    \midrule[1.5pt]
    None & 0.2438 & 0 & 2.627 & 0 & 1.049 & 0 & 4.277 & 0\\ \midrule[1.5pt]
    All & 0.2440 & $4.636\times10^{-4}$ & 2.636 & $3.686\times10^{-3}$ &
    1.050 & $1.209\times10^{-3}$ & 4.260 & $-3.916\times10^{-3}$\\ \midrule[1.5pt]
    10, 11 & 0.2439 & $2.124\times10^{-4}$ & 2.635 & $3.202\times10^{-3}$ &
    1.050 & $1.048\times10^{-3}$ & 4.262 & $-3.650\times10^{-3}$\\ \hline
    1, 2, 10, 11 & 0.2439 & $1.515\times10^{-4}$ & 2.635 & $3.155\times10^{-3}$ &
    1.050 & $1.032\times10^{-3}$ & 4.261 & $-3.672\times10^{-3}$\\ \hline
    1, 2, 3, 4, 5, 10, 11 & 0.2439 & $2.415\times10^{-4}$ &
    2.635 & $3.148\times10^{-3}$ & 1.050 & $1.029\times10^{-3}$ &
    4.262 & $-3.543\times10^{-3}$\\ \midrule[1.5pt]
    6, 7, 8, 9 & 0.2440 & $6.730\times10^{-4}$ & 2.629 & $1.002\times10^{-3}$ &
    1.049 & $3.348\times10^{-4}$ & 4.276 & $-3.536\times10^{-4}$\\ \hline
    1, 2, 6, 7, 8, 9 & 0.2440 & $5.455\times10^{-4}$ & 2.629 & $9.034\times10^{-4}$ &
    1.049 & $3.001\times10^{-4}$ & 4.275 & $-3.972\times10^{-4}$\\ \hline
    1, 2, 3, 4, 5, 6, 7, 8, 9  & 0.2440 & $5.533\times10^{-4}$ & 2.629 & $8.981\times10^{-4}$ &
    1.049 & $2.981\times10^{-4}$ & 4.276 & $-3.797\times10^{-4}$\\ \hline
  \end{tabular}
  \end{adjustbox}
  \caption[BBN runs]{\label{trans-tab:bbnsummary}Process-dependent
     changes in the BBN abundances.  For all runs $\epsmax=20.0$,
     $\nbins=100$, $\tin=8\,{\rm MeV}$, $\tstop=15\,{\rm keV}$,
     $\nuratiotol=30.0$.  The first column gives the processes used for a
     given run similar to Table \ref{trans-tab:transsummary}.  Column
     two is the primordial mass fraction of $^4$He and column three is
     the relative change from the (N) baseline case with no neutrino transport,
     i.e.\ the first row.  Column four is the relative abundance of D
     and column five the relative change.  Column six is the relative
     abundance of $^3$He and column seven the relative change.  Column
     eight is the relative abundance of $^7$Li and column nine the
  relative change.}
\end{table*}

\begin{table*}
  \begin{tabular}{| c !{\vrule width 1.5 pt} c | c !{\vrule width 1.5 pt} c | c 
  !{\vrule width 1.5 pt} c | c !{\vrule width 1.5 pt} c | c |}
    \hline
    Processes & $Y_P$ & $\delta Y_P$ & $10^5\times{\rm D/H}$ &
    $\delta({\rm D/H})$ & $10^5\times\,^3{\rm He/H}$ &
    $\delta(^3{\rm He/H})$ & $10^{10}\times\,^7{\rm Li/H}$
    & $\delta(^7{\rm Li/H})$\\ 
    \midrule[1.5pt]
    None & 0.2438 & 0 & 2.627 & 0 & 1.049 & 0 & 4.277 & 0\\ \midrule[1.5pt]
    CC & 0.2474 & $1.463\times10^{-2}$ & 2.647 & $7.898\times10^{-3}$ &
    1.052 & $2.737\times10^{-3}$ & 4.317 & $9.344\times10^{-3}$\\ \hline
    0T & 0.2442 & $1.454\times10^{-3}$ & 2.629 & $7.816\times10^{-4}$ &
    1.049 & 0.0 & 4.281 & $9.365\times10^{-4}$\\ \hline
    CC, 0T & 0.2478 & $1.613\times10^{-2}$ & 2.650 & $8.719\times10^{-3}$ &
    1.052 & $3.021\times10^{-3}$ & 4.321 & $1.030\times10^{-2}$\\
    \midrule[1.5pt]
    Trans & 0.2440 & $4.636\times10^{-4}$ & 2.636 & $3.686\times10^{-3}$ &
    1.050 & $1.209\times10^{-3}$ & 4.260 & $-3.916\times10^{-3}$\\
    \midrule[1.5pt]
    CC, 0T, Trans & 0.2479 & $1.644\times10^{-2}$ & 2.659 & $1.236\times10^{-2}$ &
    1.053 & $4.209\times10^{-3}$ & 4.304 & $6.231\times10^{-3}$ \\ \hline
  \end{tabular}
  \caption[BBN runs with higher order corrections]
  {\label{trans-tab:highorderbbn}Changes in primordial abundances in
  BBN for Coulomb and radiative corrections. The first column gives
  the processes used for a given run. Rows correspond to various
  corrections as: ``CC'' for Coulomb corrections; ``0T'' for
  zero-temperature radiative corrections; ``Trans''\ for neutrino
  transport calculation with computational parameters as given in Table
  \ref{trans-tab:bbnsummary}.
  The notation for the relative changes is the same as in Table
  \ref{trans-tab:bbnsummary}.  Row 4 is our (Q) baseline.}
\end{table*}

Table \ref{trans-tab:bbnsummary} gives the primordial mass fractions or
relative abundances when various processes in Table \ref{trans-tab:ssamps} are
included or neglected.  This table shows that the transport calculations
produce a $5\times10^{-4}$ increase in the expected $^4$He yield compared to
$Y^{\rm (N)}_P$.  Table \ref{trans-tab:highorderbbn} gives corresponding
changes between the (N) baseline case and the cases with Coulomb,
zero-temperature radiative, or transport-induced corrections.  We compare the
helium yield in row 4 [the (Q) baseline] with the helium yield in the last row
(labeled CC, 0T, Trans).  The no-transport value in this baseline case is
0.2478, as mentioned above, while the same weak rate physics but with transport
gives 0.2479, a roughly $3\times10^{-4}$ increase in the helium yield.  This is
similar to the comparison above with the cases without Coulomb and
zero-temperature radiative corrections, showing that the transport-induced
alterations in light element abundance yields are somewhat robust to how this
set of corrections to the $n\leftrightarrow p$ rates is treated.  The increase
in $Y_P$ is in rough agreement with DHS irrespective of the baseline.
Our hypothesis was that a high-energy enhancement of the \nue occupation
probability would lead to a smaller \np ratio and subsequent decrease in $Y_P$.
We have found the opposite behavior.
Comparing the cases with and without transport, we find two competing processes
affecting the helium abundance.  With transport there is an enhanced population
of \nue and \bnue relative to FD equilibrium, and this results in an enhanced
neutron destruction in the channel $\nue+n\rightarrow p+e^-$.  (The
$\bnue+p\rightarrow n+e^+$ channel is hindered by a threshold energy.) A
decrease in the neutron number leads to a decrease in helium.  Second, a larger
energy density in the neutrino sector yields a faster expansion rate, and this
means a larger neutron number during weak freeze-out, which would produce a
higher helium yield.  Tables \ref{trans-tab:bbnsummary} and
\ref{trans-tab:highorderbbn} show that the net change in helium with these two
effects is nearly a wash, with the faster expansion rate being the more
dominant process and a very small increase in helium (0.2478 to 0.2479 in
Table\ref{trans-tab:highorderbbn}).  

We have investigated our theory by the following numerical test.  We run our
code with all of the neutrino transport processes activated to allow the
neutrino occupation probabilities to go out of FD equilibrium. We follow the
flow of entropy out of the plasma and calculate the Hubble expansion rate, but
we do not use the modified occupation probabilities in calculating the
neutron-to-proton rates.  Instead, we simply use FD occupation probabilities
when calculating the weak interaction rates.  This program ensures that we have
the same thermodynamics and phasing of $T$ with \tcm, and therefore tests how
effective the high-energy tail of the $\nu_e$ distribution is at lowering the
helium abundance.  The results of the test are a slight increase of helium to
0.2480, over the 0.2479 value in row 6 of table \ref{trans-tab:highorderbbn}.
The increase in the $\nu_e$ occupation probability has a very slight overall
leverage on the helium abundance.  For this test, the changes in deuterium,
helium-3, and lithium-7 are even smaller.

Neutrino transport alters the deuterium abundance computed from the baselines
in a significant and interesting way.  Table \ref{trans-tab:bbnsummary} (second
row) shows an increase of about $0.4\%$ in the predicted BBN D/H value relative
to the (N) baseline.  Table \ref{trans-tab:highorderbbn} gives the
corresponding changes relative to our (Q) baseline case, and again shows a
comparable fractional increase in the deuterium yield.  This is a change which is
comparable to the level of BBN nuclear physics input uncertainties (i.e., in
the ${\rm D}(p,\gamma)\,^3{\rm He}$) cross sections) \cite{Adelberger:2011bp}
and these might be improved upon by {\em ab initio} many-body
calculations \cite{Marcucci:2015dp}.  Moreover, our calculated increase is not
far from the speculated precision in the primordial D/H abundance attainable
with thirty-meter-class telescopes and observations of isotope-shifted Lyman
series hydrogen absorption lines in nearly pristine hydrogen clouds seen along
lines of sight to high-redshift
quasars \cite{2000PhR...333..409T,2003ApJS..149....1K,Pettini:2012yd,Cooke:2014do}.

Tables \ref{trans-tab:bbnsummary} and \ref{trans-tab:highorderbbn} also show
the changes in the lithium-7 yield with and without transport.  The changes in
this case are roughly $0.3\%$ relative to either baseline calculation.  This
reduction is more than two orders of magnitude below that needed to address the
factor of 3 or 4 overprediction of the primordial \Livii\ abundance that
constitutes the ``lithium problem.'' It has been argued that there is no
nuclear physics fix for this problem (see for example Refs.\
\cite{Fuller:2010nn,2009PhRvD..79j5001S}).  BSM physics, like out of
equilibrium decays of massive particles (see Refs.
\cite{FKK:2011di,2015PhRvD..91h3505N,DarkRadScherrer:2012}), or massive
particle decay and post-BBN cascade nucleosynthesis \cite{1994ApJ...423...50J}
may be required if the observationally inferred lithium abundance is indeed
primordial \cite{2000ApJ...530L..57R}.

Figure \ref{trans-fig:plot_abunds_total_s_vs_tcm} shows the transport-coupled
BBN light element abundance histories as a function of decreasing \tcm or,
equivalently, increasing time.  Note that the entropy in the plasma \spl is
decreasing, as entropy flows into the decoupling neutrino seas, primarily in
the early phases of BBN. This is during the epoch when many but not all nuclear
species are maintained in NSE by relatively large nuclear reaction rates.  In
NSE the nuclear abundances are set by the \np ratio, the nuclear binding
energies, and the relevant entropy-per-baryon, namely \spl.

The nonlinear effect of neutrino transport on \neff, discussed in
Section \ref{trans-ssec:wdcalcs}, is also observed in the primordial
abundances. It is associated with the change in the phasing of the
time development of the entropy, plasma temperature, and $n/p$ ratio
relative to the no-transport, constant comoving entropy case.
The transport-induced higher expansion rate implies that the plasma temperature
will decrease at a more rapid rate after the alpha-particle-formation epoch
($\tcm\sim 50\,{\rm keV}$ in Fig.\ \ref{trans-fig:plot_abunds_total_s_vs_tcm}).
After the majority of neutrons are isolated within alpha particles, deuterium
begins to decrease until it freezes-out at $\tcm\sim 20\,{\rm keV}$.  There is
less time for the temperature-sensitive deuterium destruction reactions to
operate.  At the deuterium peak (coincident with the alpha-particle formation
epoch), the most effective deuterium destruction channels are the purely strong
interactions, namely $^3{\rm He}(d,p)\alpha$ and $t(d,n)\alpha$, where we have
used shorthand nuclear notation for deuterium ($d$), tritium ($t$), and $^4$He
($\alpha$).  The former reaction is hindered relative to the latter because the
former has a larger Coulomb barrier than the latter. This is evidenced by the
parallel tracks of deuterium and tritium for $50\,{\rm keV}>\tcm>20\,{\rm
keV}$.  However, deuterium is also efficiently destroyed by the electromagnetic
channels: $d(p,\gamma)^3{\rm He}$ and $d(\gamma,p)n$.  In addition, the
electromagnetic reactions are also sensitive to temperature.  Note that the
former reaction produces $^3{\rm He}$ and there is a slight increase of $^3{\rm
He}$ on the lower temperature side of the deuterium peak.  Due to the increased
energy density from neutrino transport, both the strong and electromagnetic
reactions have less time to destroy deuterium compared to the no-transport
case.  The result is a higher deuterium yield.

The neutrino transport calculations alter the neutrino energy
distribution functions and thus the charged-current weak interaction
histories for the $n/p$ ratio, which change BBN abundance yields over
those in Eqs.\ \eqref{trans-eqn:ypnot}--\eqref{trans-eqn:7linot} and
Eqs.\ \eqref{trans-eqn:ypcc0t}--\eqref{trans-eqn:7licc0t}.  The most
important charged-current processes at late times ($\tcm\sim$ several
hundred keV) are those without thresholds, i.e., $\nu_e+n\rightarrow p
+ e^-$ and $e^+ + n\rightarrow p +\bnue$
\cite{2005PhRvD..72f3004A,2010PhRvD..81f5027S}.
Helium is sensitive to $n/p$,
which is altered by a competition between the effect of high-energy
electron-flavor neutrinos that convert neutrons to protons and the
effect of the increased energy density present in neutrinos and
anti-neutrinos of all flavors that cause an increase in the cosmic
expansion rate. The deuterium yield also appears to be sensitive to
the freeze-out of the nuclear reactions due to the increased expansion
rate.

\begin{figure}[h]
  \includegraphics[width=\columnwidth]{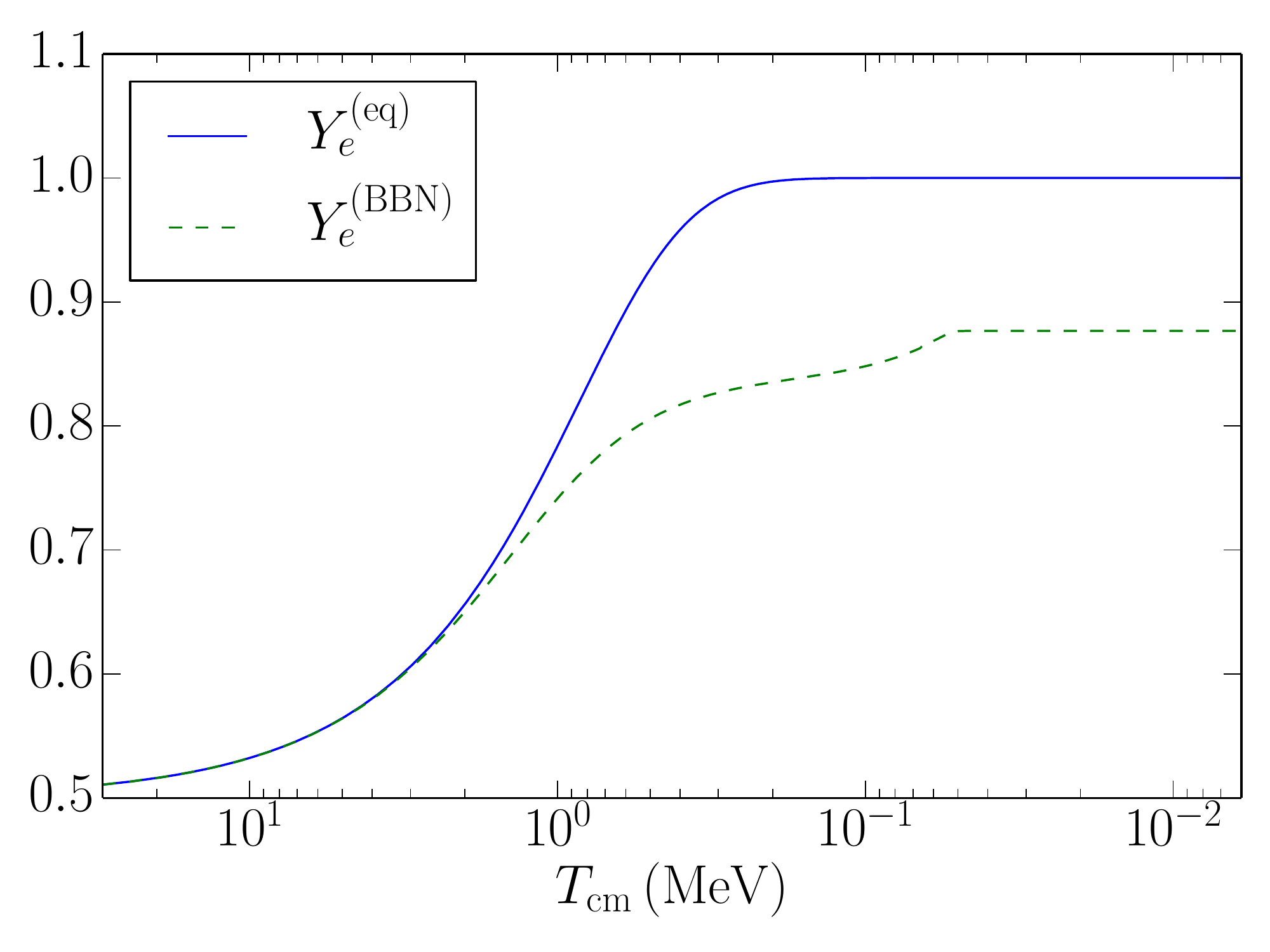}
  \caption[Evolution of electron fraction versus the comoving
  temperature parameter]{\label{trans-fig:plot_Y_e_vs_tcm} Evolution
  of the electron fraction $Y_e$ as a function of comoving temperature
  parameter \tcm. Coulomb and zero-temperature radiative corrections
  are not included in this plot.}
\end{figure}

Figure \ref{trans-fig:plot_Y_e_vs_tcm} shows a plot of how the electron
fraction $Y_e$ evolves with comoving temperature.  $Y_e$ is given in terms of
the $n/p$ ratio by $Y_e=1/(1+n/p)$.  The evolution of the $n/p$ ratio is given
by:
\begin{align}
\label{trans-eqn:dnpdt}
  \frac{d}{dt}(n/p) &= (1 + n/p)\left(\lambda_p - \lambda_n\,n/p\right),
\end{align}
where $\lambda_p$ and $\lambda_n$ are the total weak charged-current proton and
neutron destruction rates, respectively.  Figure
\ref{trans-fig:plot_Y_e_vs_tcm} gives the actual electron fraction, $Y_e^{\rm
(BBN)}$, determined from Eq.\ \eqref{trans-eqn:dnpdt} with the
transport-calculated neutrino energy distributions and the nuclear reaction
network.  In addition, Fig.\
\ref{trans-fig:plot_Y_e_vs_tcm} shows the electron fraction assuming weak
equilibrium throughout the range of temperature considered
\begin{align}
   \label{eqn:Ye_eq}
  Y_e^{\rm (eq)} = \frac{1}{1+e^{-\delta m_{np}/T + \phie - \xi_{\nu_e}}},
\end{align}
where $\delta m_{np} \equiv m_n - m_p$, the differences of the neutron and
proton rest masses, respectively, \phie is the electron degeneracy, and
$\xi_{\nu_e}$ is the \nue/\bnue degeneracy parameter.  In the equilibrium plot
in Fig.\ \ref{trans-fig:plot_Y_e_vs_tcm}, we take $\phie=\xi_{\nue}=0$.

\section{Conclusion}
\label{trans-sec:concl}

We have performed a fully self-consistent, and simultaneous, calculation of the
evolution of the neutrino energy distribution functions and all strong,
electromagnetic and weak nuclear reactions during the epochs in the early
universe where neutrinos decouple and primordial nuclear abundances are set. A
key result of this calculation is to show that there is nonlinear feedback
between the time/scale factor evolution of the neutrino sector and the
corresponding evolution of the photon/electron/positron/baryon plasma. The
neutrino energy transport part of this calculation yields essentially the same
final result as previous treatments, at least in terms of final, post BBN,
relic neutrino energy distributions, where the final entropy-per-baryon (or
baryon-to-photon ratio) matches the value of this quantity as inferred from CMB
measurements at the photon decoupling epoch. However, our calculation reveals
that the history of, and phasing of the neutrino and plasma components, and
associated entropy flow and generation, is altered when nonlinear feedback is
included.  Systems in instantaneous equilibrium are, of course, blind to the
history of conditions and system parameters.  This is not the case for
non-equilibrium systems. Indeed, our calculations show that out-of-equilibrium
components in the early universe can be sensitive to the history of how
neutrino energy distributions and plasma thermodynamic conditions got to their
final states.  These non-equilibrium components include the entire weakly
interacting system of neutrinos, electrons, positrons, neutrons, and protons,
as well as important segments of the strong and electromagnetic nuclear
reaction network.  In fact, our calculations show changes in the BBN light
element abundance yields relative to baseline BBN calculations with no neutrino
energy transport.  These changes stem, in part, from feedback between the
non-equilibrium sectors.  Appendix \ref{trans-app:overview} gives an account of
previous important work in this area which we have built on.

In Sec.\ \ref{trans-sec:wd} we describe in detail the new computational tool,
\burst, that we developed to do the coupled, simultaneous modeling of all
standard-model early universe components.  We have tested the performance and
accuracy of the code, as discussed in Sec.\ \ref{trans-ssec:nonpertapproach}.
Evaluation of integrals of the collision terms entering the Boltzmann equation
assure the conservation of neutrino lepton number at the level of 1 part in
$10^{14}$.  Efficient numerical methods have been developed for execution on
parallel platforms.  In Appendices \ref{trans-app:nunu} and
\ref{trans-app:other} we provide a detailed exposition of the neutrino
scattering processes and corresponding kernels and integrals used in these
parallel computations.

\begin{figure*}
  \includegraphics[width=0.75\textwidth]{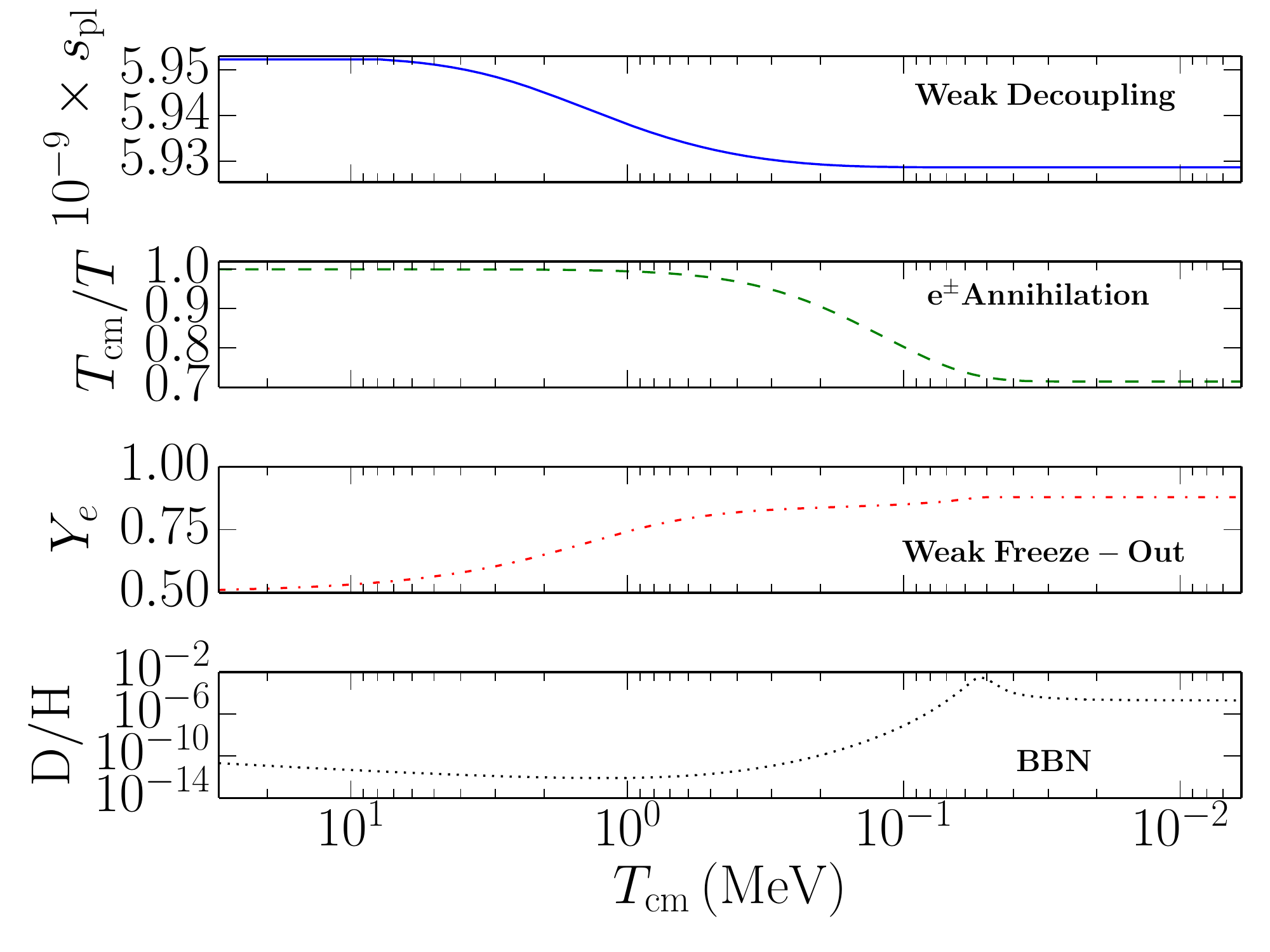}
  \caption[Entropy, temperature ratio, electron fraction, and D/H
  versus \tcm]{\label{trans-fig:plot_epochs_vs_tcm}(Color online)
  The entropy,
     temperature ratio, electron fraction, and relative abundance of
     deuterium as functions of comoving temperature.  The blue solid
     line is the evolution of the entropy per baryon, $\spl$, in the
     plasma as a function of comoving temperature, \tcm.  The green
     dashed line is the evolution of the ratio of comoving temperature
     to plasma temperature, $\tcm/T$ as a function of \tcm.  The red
     dash-dot line is the evolution of the electron fraction, $Y_e$,
     as a function of \tcm.  The black dotted line is the evolution of
     the relative abundance of deuterium, D/H, as a function of \tcm.}
\end{figure*}

Section \ref{trans-ssec:wdcalcs} details the results specific to the neutrino
sector when the charged lepton, neutrino and photon components are evolved.
Our calculations reveal a rich history of timelike entropy flow between the
components of the early universe.  If we use \neff to parameterize the increase
in the energy density of neutrinos, we find $\neff=3.034$ when we include only
transport processes, and $\neff=3.052$ when we include transport and QED
effects.  Furthermore, we have uncovered a novel late-time rise in relativistic
energy density ($\Delta_t\neff$, shown in Fig.\
\ref{trans-fig:plot_drho_vs_tcm}) when the entropy in $e^\pm$ pairs is
transferred to photons.

We have dissected the contributions of each neutrino scattering and reaction
channel to distortions in neutrino energy spectra, along with the concomitant
effects on the entropy flows.  A key conclusion of our work is that the changes
in cosmological quantities stemming from various processes do not add
incoherently and must be followed in a full non-linear, coupled treatment.

Our work may suggest a new way in which in-medium corrections to
electron/positron rest masses and other plasma corrections are important. It is
clear from our calculations that the timelike flow of entropy between the
components in the early universe medium, the evolution of the neutron-to-proton
ratio and nuclear abundances, and the phasing of these processes relative to
scale factor and plasma temperature, must be adequately modeled if we hope to
predict light element BBN abundance yields to better than $\sim 1\%$ precision.
The key processes facilitating entropy transfer, and determining the associated
phasing, are those involving neutrino-electron/positron
scattering/annihilation.  Consequently, the number density history of
$e^\pm$-pairs can be important. We presented calculations with and without
neutrino energy transport, both of which included QED
corrections to the thermodynamics of the plasma. From these calculations we can
see that important aspects of entropy flow and neutron-to-proton evolution and
phasing are being set at temperatures less than twice the electron rest mass.
In this regime, the $e^\pm$-pair density can be nearly exponentially sensitive
to the {\it in-medium} electron rest mass. This suggests that finite
temperature corrections take on a new importance in calculating neutrino
transport and BBN.

Transport-induced changes in the \nprates, as detailed in Sec.\
\ref{trans-sec:BBN}, appear to be independent of the particular implementation
of Coulomb \cite{2010PhRvD..81f5027S} and zero-temperature radiative
corrections.  However, other effects on the \nprates, such as
finite-temperature radiative corrections, in-medium renormalization of the
electron and positron rest masses, or inclusion of in-medium nucleo rest mass
corrections and nuclear recoil \cite{1999PhRvD..59j3502L} may indeed alter the
primordial abundances in a nonlinear way when neutrino transport is included.

The transport collision-term calculations detailed in Sec.\
\ref{trans-ssec:wdcalcs} are an incremental step in describing the neutrino
evolution of the early universe. We do not include neutrino oscillations in
this paper (see \cite{neff:3.046} for an analysis of this problem with neutrino
oscillations). However, we plan to expand the current Boltzmann-solver code
into a quantum-kinetic approach for handling neutrino flavor density matrices,
i.e. along the lines of the QKEs in Ref.\ \cite{VFC:QKE}.  If the QKEs lead to
an equilibration between \nue and \num (as highlighted in
\Cref{trans-fig:plot_doccprob_tcm,trans-fig:plot_doccbar_vs_tcm,trans-fig:plot_doccprob_eps,trans-fig:plot_drho_vs_eps}),
we would expect a decrease in \neff due to the lack of a charged-current
diagram in the \num scattering processes.  The equilibration has two indirect
effects on nucleosynthesis.  A decrease in \neff is equivalent to less
radiation energy density which would imply a smaller Hubble expansion rate.
The slower expansion rate delays weak freeze-out and would yield a smaller \np
(and smaller primordial helium abundance) compared to the case we explore here:
energy transport without oscillations.  In addition to modifying the energy
density, neutrino oscillations will decrease the population of the \nue,
causing an earlier epoch of weak freeze-out and subsequent increase of \np.  It
would appear, from our calculations in this paper, that $Y_P$ is more sensitive
to the change in expansion rate than the distortion of the \nue spectrum.
Furthermore, our best prediction of \neff is $\neff=3.052$, which is $\sim0.01$
larger than the Ref.\ \cite{neff:3.046} value of 3.046.  The principal
difference between our treatment and that of Ref.\ \cite{neff:3.046} is our
exclusion of neutrino oscillations.  The difference may be a signature of
oscillations.  However, the QKE problem is inherently non-linear and requires a
sophisticated calculation to verify quantitative and qualitative predictions.

We have demonstrated, in Sec.\ \ref{trans-sec:entropy} that the
textbook treatment of neutrino weak decoupling from the
plasma, which assumes covariant conservation of entropy,
is largely satisfied. Small numerical deviations from comoving entropy
conservation, as shown in Figs.\ \ref{trans-fig:plot_entropy_vs_tcm},
\ref{trans-fig:plot_abunds_total_s_vs_tcm}, and
\ref{trans-fig:plot_epochs_vs_tcm} have a significant and potentially
measurable effect on the light primordial nuclide abundances at the
level of $\sim 0.5\%$. We discussed the two regimes of {\em
equilibrium} entropy exchange and entropy generation. Further, we
have shown that the
weak-decoupling and BBN epochs overlap in time significantly and 
suggest that they should be regarded as the single epoch termed the
``weak-decoupling-nucleosynthesis'' epoch.


As presented in Section V, the neutrino transport-altered history of weak
decoupling and weak freeze-out results in potentially significant changes in
predicted BBN abundance yields. As Table IV shows, these include an increase of
$0.4\%$ in the BBN-predicted D/H yield relative to baseline no-transport
calculations. In addition, Table V shows the changes in the primordial light
nuclide abundances when selections of radiative and Coulomb corrections are
taken into account.  These variations are similar in size to the uncertainties
associated with observational uncertainties and those due to nuclear physics
uncertainties. The anticipated precision of extremely large, thirty-meter-class
telescopes will hopefully approach this precision within a decade.

Figure \ref{trans-fig:plot_epochs_vs_tcm} shows the temperature range
spanning the weak-decoupling, $e^\pm$ annihilation, weak
freeze-out, and BBN epochs.  The curve in the first panel is the
evolution of the entropy in the photon/electron/positron/baryon
plasma, \spl, with respect to \tcm and illuminates the physics of the
weak-decoupling epoch.  The curve in the second panel is the
evolution of the temperature ratio $\tcm/T$ with respect to \tcm
through the $e^\pm$ annihilation epoch.  The curve in the third panel
is the electron fraction $Y_e$
and shows the evolution of the weak
freeze-out epoch.  The last curve in the fourth panel is the relative
abundance of deuterium D/H characteristic of the BBN epoch.


Our calculations show that careful, high precision modeling of the complicated,
nonlinear interplay of out-of-equilibrium constituents in the early universe is
required if the goal is to predict BBN light element abundance yields and relic
neutrino properties to better than one percent precision. In fact, the expected
high precision data from Stage-IV CMB experiments and from thirty meter class
telescopes sets up a unique opportunity to probe cosmology and BSM physics
operating in the early universe \cite{Cooke:2014do}.  Moreover, new laboratory
data on neutrino properties, e.g., the neutrino mass hierarchy or neutrino rest
mass constraints \cite{2013arXiv1310.4340D}, may be forthcoming, and these may
impact the evolution of the neutrino and nuclear components in the early
universe.  High-precision BBN/neutrino calculations likely will be an important
cornerstone of this enterprise.

\acknowledgments
We thank
Fred Adams,
Eve Armstrong,
Kam Arnold,
Daniel Blaschke,
Lowell Brown,
John Carlstrom,
John Cherry,
Vincenzo Cirigliano,
Scott Dodelson,
Lauren Gilbert,
Luke Johns,
Brian Keating,
Lloyd Knox,
Adrian Lee,
Eric Michelsen,
Ken Nollett,
Amol Patwardhan,
Shashank Shalgar,
Meir Shimon,
Gary Steigman,
Mike Turner,
and Nicole Vassh
for useful conversations with respect to cosmology, neutrino physics, nuclear
physics, plasma physics, Fortran 90, and parallel computing.  We acknowledge
the Integrated Computing Network at Los Alamos National Laboratory for
supercomputer time.  This work was supported in part by NSF grant PHY-1307372
at UC San Diego, by the Los Alamos National Laboratory Institute for
Geophysics, Space Sciences and Signatures Subcontract No.\ 257842, and the
National Nuclear Security Administration of the U.S.\ Department of Energy at
Los Alamos National Laboratory under Contract No.\ DE-AC52-06NA25396.

\appendix

\section{Overview of past approaches}
\label{trans-app:overview}
The departure from equilibrium in weak decoupling and its subsequent
effects on nucleosynthesis has been studied by many groups.  Early
work \cite{1982PhRvD..26.2694D,1982NuPhB.209..372C,
   1989ApJ...336..539H,1991PhRvD..44..393R}
considered the neutrinos to have a thermal distribution after weak decoupling
and calculated the consequences of
$e^\pm$ annihilation on the temperature of this thermal
distribution.
The focus of Refs.\ \cite{1982PhRvD..26.2694D,1982NuPhB.209..372C} was
to examine the effect of finite-temperature radiative corrections to
the neutron-proton conversion rates to study
changes in helium production.  The corrections change the neutrino
spectra and thus increase the ratio \tcmpl.
Refs.\ \cite{1989ApJ...336..539H,1991PhRvD..44..393R} treated the
neutrinos and the electrons/positrons as two relativistic gases with
different temperatures and calculated the relaxation time over which
these two components of the universe would return toward thermal
equilibrium.

Recent work has instead shifted toward solving the coupled Boltzmann
equations, that deal directly with the neutrino distribution
functions.  One approach has been to treat neutrino scattering
processes and neutrino production in electron-positron annihilation
approximately, which treats the out-of-equilibrium effects as a
perturbation on the FD spectrum of neutrinos. 
Ref.\ \cite{1992PhRvD..46.3372D} solves the Boltzmann equations using
Maxwell-Boltzmann statistics, yielding both a perturbation in the
neutrino temperature and a first-order perturbation to the FD neutrino
spectrum.  Refs.\ \cite{2000NuPhB.590..539E,Mangano:3.040} adopt a
perturbative approach in introducing orthogonal polynomials that use
the FD occupation probability as a weight function.

Other approaches (and the approach we employ in this work) discretize
the neutrino distribution function on a comoving invariant momentum,
$p/T_{\rm cm}$.  The neutrino distribution function is binned,
creating a coupled set of Boltzmann equations for the evolution of
each bin to be solved numerically.  These approaches do not use the
Maxwell-Boltzmann statistics approximation used in past works (cf.,
Ref.\ \cite{1992PhRvD..46.3372D}), instead using the full FD blocking
factors in the collision integrals.  Ref.\ \cite{1995PhRvD..52.1764H}
introduced a general neutrino distribution function and showed values
consistent to those in Ref.\ \cite{1992PhRvD..46.3372D} when using the
Maxwell-Boltzmann approximation.
Ref.\ \cite{Gnedin:1998ne} used a pseudo-logarithmic binning scheme
and employed a unique numerical scheme which did not require the
calculation of the full Jacobian matrix.

Ref.\ \cite{Dolgov:1997ne} (DHS) uses 100 linearly-spaced bins
spanning $0 \leq p/T_{\rm cm} \leq 20$ (200 bins are also used, but
acceptable convergence is found with 100 bins).  In their seminal
work, DHS runs a number of convergence tests and finds that the
convergence of the results is most sensitive to the number of time
steps (steps in scale factor) taken from a prescribed initial to final
epoch as compared to number of bins, initial epoch, binned spectrum
vs. a perturbation function of the spectrum, or ODE evolution
algorithm:  simple (presumable Eulerian) time evolution vs.
Bulirsch-Stoer routine.  An addendum \cite{1999NuPhB.543..269D}
improved the accuracy of their code and found no change in the result,
$\neff = 3.034$.

The above works focused on out-of-equilibrium processes that augment
the neutrino seas relative to a scenario where there is a sharp
decoupling of neutrinos from the plasma.  Finite-temperature quantum
electrodynamic (QED) radiative corrections provide
$\mathcal{O}(\alpha)$ corrections to the dispersion relations of
electrons, positrons and photons which, in turn, affect the entropy
transferred to the photon-baryon plasma when the electrons and
positrons annihilate.  Ref.\ \cite{1994PhRvD..49..611H} calculated the
QED correction to the dispersion relation for electrons to first order
in the fine-structure constant, $\alpha$.  Using the altered
dispersion relation, Ref.\ \cite{1999PhRvD..59j3502L} employed entropy
conservation to find the corrections to the temperature ratio
$\tcm/T$.  Ref.\ \cite{1997PhRvD..56.5123F} introduced a QED
correction to the dispersion relation for photons, giving photons an
effective non-zero mass.  Ref.\ \cite{Mangano:3.040} extended their
method to perturbatively solve the Boltzmann equation while
simultaneously including the QED corrections to the dispersion
relations for positrons/electrons \cite{1994PhRvD..49..611H} and
photons \cite{1997PhRvD..56.5123F}.  Ref.\ \cite{2015NuPhB.890..481B}
assumed that the neutrino spectra were close to thermal equilibrium,
but needn't be in chemical equilibrium, so they used orthogonal
polynomials based on a weight function mirroring a FD spectrum with
non-unit fugacity.

Lastly, Ref.\ \cite{neff:3.046} combined a linear binning of the
neutrino distribution spectrum and the QED corrections to obtain
$\neff = 3.046$, a result that is routinely cited in the literature.
Another aspect of this work was the inclusion of neutrino
oscillations, which changed the energy densities of the individual
neutrino species, but did not overall change \neff.  Table
\ref{table:past-works} summarizes
the approaches of past work in this area.  While not all works cite a
value of \neff, we use Eqs.\ \eqref{trans-eqn:neff} and
\eqref{trans-eqn:neff2} to translate from the published calculations
to \neff.

\begin{table*}
\begin{tabular}{| l !{\vrule width 1.5 pt} l !{\vrule width 1.5 pt} @{\quad}l |}
\hline 
Refs. & Notes & \neff \\
\midrule[1.5pt]
\cite{1982PhRvD..26.2694D,1982NuPhB.209..372C} & 
  \begin{minipage}[t]{1.5\columnwidth}\begin{itemize}
    \item finite-temperature radiative corrects to neutron-to-proton rates
    \item average cross sections to estimate neutrino production
          during $e^\pm$ annihilation
  \end{itemize} \end{minipage}&
  3.020 \\ \hline
\cite{1989ApJ...336..539H,1991PhRvD..44..393R} &
   \begin{minipage}[t]{1.5\columnwidth} \begin{itemize}
     \item  relaxation-time formalism to calculate changes in neutrino temperature post-weak-decoupling
   \end{itemize}\end{minipage} &
  \begin{tabular}[t]{@{}l@{}}3.024 \cite{1989ApJ...336..539H}  \\ 3.022 \cite{1991PhRvD..44..393R} \end{tabular} \\ \hline
\cite{1992PhRvD..46.3372D} & 
   \begin{minipage}[t]{1.5\columnwidth} \begin{itemize}
     \item  coupled set of Boltzmann equations in weak decoupling
     \item  Maxwell-Boltzmann statistics
     \item solves for a change in the neutrino temperature and a first-order change to the neutrino distribution functions
   \end{itemize}\end{minipage} &
   3.022 \\ \hline
\cite{2000NuPhB.590..539E,Mangano:3.040} &
   \begin{minipage}[t]{1.5\columnwidth} \begin{itemize}
     \item  perturbative approach solving Boltzmann equations using a series of orthogonal polynomials to describe perturbations from a FD neutrino spectrum
   \end{itemize}\end{minipage} &
   3.035 \\ \hline
\cite{1995PhRvD..52.1764H} &
   \begin{minipage}[t]{1.5\columnwidth} \begin{itemize}
     \item  coupled set of Boltzmann equations in weak decoupling using FD statistics (cf., Ref \cite{1992PhRvD..46.3372D})
     \item solves for a change in the neutrino temperature and a general neutrino distribution function
   \end{itemize}\end{minipage} &
   \begin{minipage}[t]{0.09\columnwidth}
   3.017\newline
   or 3.027 
   \end{minipage}\\ \hline
\cite{Gnedin:1998ne} &
   \begin{minipage}[t]{1.5\columnwidth} \begin{itemize}
     \item solves coupled Boltzmann equations by binning the neutrino distribution function
     \item pseudo-logarithmic binning scheme:  40 linearly-spaced bins per decade ranging from $10^{-5.5} \leq p/T_{\rm cm} \leq 10^{1.7}$
     \item employ unique numerical scheme that does not require calculation of the full Jacobian matrix -- more efficient than standard adaptive RK5 scheme by a factor of 20-60
   \end{itemize}\end{minipage} &
   3.022 \\ \hline
DHS \cite{Dolgov:1997ne,1999NuPhB.543..269D} &
   \begin{minipage}[t]{1.5\columnwidth} \begin{itemize}
     \item  solves coupled Boltzmann equations by binning the neutrino distribution function
     \item 100 linearly-spaced bins between 0 $\leq p/T_{\rm cm} \leq 20$
     \item includes convergence studies regarding binning of neutrino spectrum and ODE solver
   \end{itemize}\end{minipage} &
   3.034 \\
\midrule[1.5pt]
\cite{1994PhRvD..49..611H,1999PhRvD..59j3502L,1997PhRvD..56.5123F} &
   \begin{minipage}[t]{1.5\columnwidth} \begin{itemize}
     \item  introduces QED corrections to electron and photon dispersion relations 
     \item no Boltzmann evolution, just conservation of comoving entropy
   \end{itemize}\end{minipage} &
   3.011 \cite{1999PhRvD..59j3502L}\\ \hline
\cite{Mangano:3.040} &
   \begin{minipage}[t]{1.5\columnwidth} \begin{itemize}
     \item  includes QED corrections to the perturbative approach with orthogonal polynomials described above for Ref.\ \cite{Mangano:3.040}
   \end{itemize}\end{minipage} &
   3.0395 \\ \hline
\cite{2015NuPhB.890..481B} &
   \begin{minipage}[t]{1.5\columnwidth} \begin{itemize}
     \item  includes QED corrections to the perturbative approach with orthogonal polynomials 
     \item assumes neutrino spectra in thermal equilibrium, but not necessarily in chemical equilibrium
     \item uses a different set of orthogonal polynomials as compared with Ref.\ \cite{Mangano:3.040}
   \end{itemize}\end{minipage} &
   3.044 \\ \hline
\cite{neff:3.046} &
   \begin{minipage}[t]{1.5\columnwidth} \begin{itemize}
     \item  includes QED corrections along with solving Boltzmann equations by binning the neutrino distribution function
     \item improved numerical technique as compared to Ref.\ \cite{Mangano:3.040}
   \end{itemize}\end{minipage} &
   3.046 \\ \hline
\end{tabular} 
\caption{Summary of previous work.  For all works that do not
explicitly report a value of \neff (Refs.\
\cite{1982PhRvD..26.2694D,1982NuPhB.209..372C,
1989ApJ...336..539H,1991PhRvD..44..393R,1992PhRvD..46.3372D,1995PhRvD..52.1764H}),
Eq.\ \eqref{trans-eqn:neff} or \eqref{trans-eqn:neff2} is used to
estimate a value of \neff from parameters reported. For
Ref.\ \cite{1995PhRvD..52.1764H}, we include two estimates of \neff: (1)
the relative changes in energy density implicitly contain the change
in temperature (3.017), and (2) the energy densities do not contain
the change in temperature (3.027).  \label{table:past-works}}
\end{table*}


\section{Neutrino--neutrino scattering}\label{trans-app:nunu}
This appendix details the reduction of the collision integral for
neutrino--neutrino elastic scattering.  We will make the approximation
that neutrinos are massless for all three flavors.  We start with the
summed--squared amplitude for neutrinos scattering on other neutrinos
with identical flavor (Row 1 of Table \ref{trans-tab:ssamps}):
\begin{align}
  &\nu(1) + \nu(2) \leftrightarrow\nu(3) + \nu(4),\\
  &\langle|\mathcal{M}_1|^2\rangle = 2^7G_F^2(P_1\cdot P_2)(P_3\cdot P_4).
\end{align}
Conservation of four-momentum implies $P_1\cdot P_2=P_3\cdot P_4$.  
The collision integral is \cite{1990eaun.book.....K}:
\begin{widetext}
\begin{align}
  I\equiv C_{\nu_1}[f_j] = \frac{1}{2p_1}&
  \int\frac{d^3p_2}{(2\pi)^3 2p_2}\frac{d^3p_3}{(2\pi)^3 2p_3}\frac{d^3p_4}{(2\pi)^3 2p_4}
  (2\pi)^4\delta^{(4)}(P_1+P_2-P_3-P_4)
  2^7S_1G_F^2(P_1\cdot P_2)^2 F(p_1,p_2,p_3,p_4).
\end{align}
Using the three-momentum part of $\delta^{(4)}$, we eliminate the integral over
$d^3p_4$:
\begin{align}
  I = \frac{1}{2p_1}\frac{(2\pi)^42^7S_1G_F^2}{2^3(2\pi)^9}
  \int\frac{d^3p_2}{p_2}(P_1\cdot P_2)^2\int\frac{d^3p_3}{p_3p_4}\delta(p_1+p_2-p_3-p_4)
  F(p_1,p_2,p_3,p_4)|_{p_4=|\mathbf{p}_1+\mathbf{p}_2-\mathbf{p}_3|},
\end{align}
\end{widetext}
where $\mathbf{p}_i$ is the three-momentum of the $i^{\rm th}$ particle, and
$p_4$ is no longer an integration variable, but instead related to the other
integration and free variables through:
\beq
  p_4^2 = |\mathbf{p}_1+\mathbf{p}_2|^2 + p_3^2 -2|\mathbf{p}_1+\mathbf{p}_2|p_3\cos\theta_3,
\eeq
where we have defined the integration variable $\theta_3$ to be the angle
between $\mathbf{p}_1+\mathbf{p}_2$ and $\mathbf{p}_3$.  To simplify $\int
d^3p_3$, we first consider $\int d\theta_3$ and use the following
u-substitution:
\begin{align}
  &u^2 = p_4^2,\\
  \implies&2p_4du = -2|\mathbf{p}_1+\mathbf{p}_2|p_3d(\cos\theta_3).
\end{align}
The new expression for the collision term is:
\begin{widetext}
\begin{align}
  I &= \frac{2^3S_1G_F^2}{(2\pi)^5p_1}
  \int\frac{d^3p_2}{p_2}(P_1\cdot P_2)^2\int d\phi_3\int dp_3p_3
  \int\limits_{-1}^1 d(\cos\theta_3)\frac{1}{p_4}\delta(p_1+p_2-p_3-p_4)F(p_1,p_2,p_3,p_4)\\
  &=\frac{2^3S_1G_F^2}{(2\pi)^5p_1}\int
  \frac{d^3p_2}{p_2}(P_1\cdot P_2)^2(2\pi)\int dp_3p_3
  \int\limits_{u(-1)}^{u(1)}\left(-\frac{du\,p_4}{|\mathbf{p}_1+\mathbf{p}_2|p_3}\right)
  \frac{1}{p_4}\delta(p_1+p_2-p_3-p_4)F(p_1,p_2,p_3,p_4)\\
  &\equiv\frac{2^3S_1G_F^2}{(2\pi)^4p_1}
  \int\frac{d^3p_2}{p_2}\frac{(P_1\cdot P_2)^2}{|\mathbf{p}_1+\mathbf{p}_2|}\int dp_3
  \int\limits_a^b du\,\delta(p_1+p_2-p_3-u)F(p_1,p_2,p_3,u),\label{trans-app:collterm1}
\end{align}
where the limits of integration on $\int du$ are:
\begin{align}
  a = u(1) &= (|\mathbf{p}_1+\mathbf{p}_2|^2 + p_3^2 -2|\mathbf{p}_1+\mathbf{p}_2|p_3)^{1/2}
  =||\mathbf{p}_1+\mathbf{p}_2| - p_3|,\\
  b = u(-1) &= (|\mathbf{p}_1+\mathbf{p}_2|^2 + p_3^2 +2|\mathbf{p}_1+\mathbf{p}_2|p_3)^{1/2}
  =|\mathbf{p}_1+\mathbf{p}_2| + p_3.
\end{align}
\end{widetext}
For $\int du$ to be non-zero, the argument of the delta function must vanish
within the integrable domain of $\int du$, i.e.\:
\beq\label{trans-app:ineq1}
  a<p_1+p_2-p_3<b.
\eeq
We will solve inequality \eqref{trans-app:ineq1} for $p_3$, and modify the
limits of $\int dp_3$ to ensure non-zero $\int du$.  We will consider two
cases: $p_3<|\mathbf{p}_1+\mathbf{p}_2|$ and $p_3>|\mathbf{p}_1+\mathbf{p}_2|$.

\noindent{\it Case 1}: $p_3<|\mathbf{p}_1+\mathbf{p}_2|$.  The first inequality
of Eq.\ \eqref{trans-app:ineq1} reads:
\begin{align}
  |\mathbf{p}_1+\mathbf{p}_2| - p_3 &< p_1 + p_2 - p_3,\\
  \implies|\mathbf{p}_1+\mathbf{p}_2|&<p_1 + p_2,
\end{align}
reproducing the triangle inequality which is always true.  Therefore, this
inequality provides no new constraints.  The second inequality of
Eq.\ \eqref{trans-app:ineq1} reads:
\begin{align}
  p_1 + p_2 - p_3&<|\mathbf{p}_1+\mathbf{p}_2|+p_3\\
  \implies p_3&>\frac{1}{2}(p_1 + p_2  - |\mathbf{p}_1+\mathbf{p}_2|)\equiv\pmin.
\end{align}
The possibility exists that $\pmin>|\mathbf{p}_1+\mathbf{p}_2|$
depending on the angle between $\mathbf{p}_1$ and $\mathbf{p}_2$.  If
this possibility were true, then the collision integral would vanish.
Thus, the portion of the collision integral relevant to this case is:
\begin{align}
I_1&=\frac{2^3S_1G_F^2}{(2\pi)^4p_1}
  \int\frac{d^3p_2}{p_2}\frac{(P_1\cdot P_2)^2}{|\mathbf{p}_1+\mathbf{p}_2|}
  \nonumber \\
  &\times \int\limits_{\pmin}^{\pmed} dp_3\,F(p_1,p_2,p_3,p_1+p_2-p_3),
  \end{align}
where:
\beq
  \pmed = \max(\pmin,|\mathbf{p}_1+\mathbf{p}_2|).
\eeq

\noindent{\it Case 2}: $p_3>|\mathbf{p}_1+\mathbf{p}_2|$.  The first inequality
of Eq.\ \eqref{trans-app:ineq1} reads:
\begin{align}
  p_3 - |\mathbf{p}_1+\mathbf{p}_2|&< p_1 + p_2 - p_3\\
  \implies p_3&<\frac{1}{2}(p_1 + p_2  + |\mathbf{p}_1+\mathbf{p}_2|)\equiv\pmax.
\end{align}
\pmax is always greater than $|\mathbf{p}_1+\mathbf{p}_2|$.  The second
inequality is independent of the specific case, so $p_3>\pmin$.  In this case,
the possibility arises that $\pmin>|\mathbf{p}_1+\mathbf{p}_2|$ again, so the
portion of the collision integral relevant to this case is:
\begin{align}
   I_2 &=\frac{2^3S_1G_F^2}{(2\pi)^4p_1}
  \int\frac{d^3p_2}{p_2}\frac{(P_1\cdot
  P_2)^2}{|\mathbf{p}_1+\mathbf{p}_2|} \nonumber \\
  &\times \int\limits_{\pmed}^{\pmax} dp_3\,F(p_1,p_2,p_3,p_1+p_2-p_3),
  \end{align}

We add $I_1$ to $I_2$ to calculate the total collision integral $I$.  This
requires the use of \pmin and \pmax (but not \pmed) to set the limits of $\int
dp_3$ and ensure that $\int du$ is non-zero.  We write
Eq.\ \eqref{trans-app:collterm1} as:
\begin{align}\label{trans-app:collterm2}
   I &=\frac{2^3S_1G_F^2}{(2\pi)^4p_1}
  \int\frac{d^3p_2}{p_2}\frac{(P_1\cdot P_2)^2}{|\mathbf{p}_1+\mathbf{p}_2|}
  \nonumber \\
  &\times
  \int\limits_{\pmin}^{\pmax} dp_3\,F(p_1,p_2,p_3,p_1+p_2-p_3).
  \end{align}
To simplify $\int d^3p_2$, we define $\theta_2$ to be the angle between
$\mathbf{p}_1$ and $\mathbf{p}_2$.  Eq.\ \eqref{trans-app:collterm2} becomes:
\begin{widetext}
\begin{align}
  I &=\frac{2^3S_1G_F^2}{(2\pi)^4p_1}
  \int d\phi_2\int dp_2\,p_2\int d(\cos\theta_2)\frac{(P_1\cdot P_2)^2}{|\mathbf{p}_1+\mathbf{p}_2|}
  \int\limits_{\pmin}^{\pmax} dp_3\,F(p_1,p_2,p_3,p_1+p_2-p_3)\\
  &=\frac{2^3S_1G_F^2}{(2\pi)^4p_1}(2\pi)\int dp_2\,p_2
  \int d(\cos\theta_2)\frac{p_1^2p_2^2(1-\cos\theta_2)^2}{(p_1^2+p_2^2+2p_1p_2\cos\theta_2)^{1/2}}
  \int\limits_{\pmin}^{\pmax} dp_3\,F(p_1,p_2,p_3,p_1+p_2-p_3)\\
  &=\frac{2^3S_1G_F^2p_1}{(2\pi)^3}\int\limits_0^{\infty} dp_2\,p_2^3
  \int\limits_{-1}^1 d(\cos\theta_2)
  \frac{(1-\cos\theta_2)^2}{(p_1^2+p_2^2+2p_1p_2\cos\theta_2)^{1/2}}
  \int\limits_{\pmin}^{\pmax} dp_3\,F(p_1,p_2,p_3,p_1+p_2-p_3).\label{trans-app:collterm3}
\end{align}
It will behoove us to make a change of variables on $\int d(\cos\theta_2)$.
Define $y$ such that:
\begin{align}
  y^2 &= p_1^2+p_2^2+2p_1p_2\cos\theta_2,\\
  \implies\cos\theta_2 &= \frac{y^2 - p_1^2-p_2^2}{2p_1p_2},\\
  \implies d(\cos\theta_2) &= \frac{y\,dy}{p_1p_2}.
\end{align}
We write Eq.\ \eqref{trans-app:collterm3} as:
\begin{align}
  I&=\frac{2^3S_1G_F^2p_1}{(2\pi)^3}\int\limits_0^{\infty} dp_2\,p_2^3
  \int\limits_{|p_1-p_2|}^{p_1+p_2} \frac{dy}{p_1p_2}
  \left(1-\frac{y^2 - p_1^2-p_2^2}{2p_1p_2}\right)^2
  \int\limits_{\pmin}^{\pmax} dp_3\,F(p_1,p_2,p_3,p_1+p_2-p_3)\\
  &=\frac{2S_1G_F^2}{(2\pi)^3p_1^2}\int\limits_0^{\infty} dp_2
  \int\limits_{|p_1-p_2|}^{p_1+p_2} dy\,[(p_1+p_2)^2-y^2]^2
  \int\limits_{\pmin}^{\pmax} dp_3\,F(p_1,p_2,p_3,p_1+p_2-p_3).\label{trans-app:collterm4}
\end{align}
\end{widetext}
Notice that the only term in the integrand of $\int dp_3$ is the occupation
probability product and difference $F$.  This term is independent of any
angles, and thus independent of the integration variable $y$.  However, the
limits of $\int dp_3$ do depend on $y$.  We define the step functions $H$ as:
\beq
  H(x) = \begin{cases}
    1\text{ if }x>0\\
    0\text{ if }x<0
  \end{cases}.
\eeq
We can rewrite $\int dp_3$ with step functions so that
Eq.\ \eqref{trans-app:collterm4} becomes:
\begin{widetext}
\begin{align}
  I &= \frac{2S_1G_F^2}{(2\pi)^3p_1^2}\int\limits_0^{\infty} dp_2
  \int\limits_{|p_1-p_2|}^{p_1+p_2} dy\,[(p_1+p_2)^2-y^2]^2
  \int\limits_0^{p_1+p_2} dp_3\,F(p_1,p_2,p_3,p_1+p_2-p_3)
  H(p_3-\pmin)H(\pmax-p_3)\\
  &= \frac{2S_1G_F^2}{(2\pi)^3p_1^2}\int\limits_0^{\infty} dp_2
  \int\limits_0^{p_1+p_2} dp_3\,F(p_1,p_2,p_3,p_1+p_2-p_3)
  \int\limits_{|p_1-p_2|}^{p_1+p_2} dy\,[(p_1+p_2)^2-y^2]^2
  H(p_3-\pmin)H(\pmax-p_3).\label{trans-app:collterm5}
\end{align}
\end{widetext}
For Eq.\ \eqref{trans-app:collterm5} to be non-zero, the $H$ functions must both
have positive arguments.  For $H(p_3-\pmin)$:
\begin{align}
  &p_3-\pmin>0,\\
  \implies&p_3 - \frac{1}{2}(p_1 + p_2  - y)>0,\\
  \implies&y>p_1+p_2-2p_3.\label{trans-app:cond1}
\end{align}
For $H(\pmax-p_3)$:
\begin{align}
  &\pmax-p_3>0,\\
  \implies&y>2p_3-p_1-p_2.\label{trans-app:cond2}
\end{align}
Conditions \eqref{trans-app:cond1} and \eqref{trans-app:cond2} imply
$y>|p_1+p_2-2p_3|$.  $p_3$ is bounded above by $p_1+p_2$, so the step functions
do not modify the upper limit of $\int dy$.  Let us define $x_0$ as the lower
limit of $\int dy$.  The expression for $x_0$ is:
\beq
  x_0 = \max(|p_1+p_2-2p_3|,|p_1-p_2|).
\eeq
If $p_1>p_2$, then:
\beq
  x_0=\begin{cases}
    p_1+p_2-2p_3&\text{if }p_3<p_2\\
    p_1-p_2&\text{if }p_2<p_3<p_1\\
    2p_3-p_1-p_2&\text{if }p_3>p_1
  \end{cases}.
\eeq
If $p_1<p_2$, then:
\beq
  x_0=\begin{cases}
    p_1+p_2-2p_3&\text{if }p_3<p_1\\
    p_2-p_1&\text{if }p_1<p_3<p_2\\
    2p_3-p_1-p_2&\text{if }p_3>p_2
  \end{cases}.  
\eeq

Equation \eqref{trans-app:collterm5} becomes:
\begin{widetext}
\begin{align}
   \label{trans-app:collterm6}
   I&=\frac{2S_1G_F^2}{(2\pi)^3p_1^2}
   \left\{\int\limits_0^{p_1} dp_2
      \left[\vphantom{\int\limits_0^{p_2}}\int\limits_0^{p_2} dp_3\,F
	 \int\limits_{p_1+p_2-2p_3}^{p_1+p_2} dy\,[(p_1+p_2)^2-y^2]^2
	 +\int\limits_{p_2}^{p_1} dp_3\,F
	 \int\limits_{p_1-p_2}^{p_1+p_2} dy\,[(p_1+p_2)^2-y^2]^2\right.\right.\nonumber\\
	 &+\left.\left.\int\limits_{p_1}^{p_1+p_2} dp_3\,F
      \int\limits_{2p_3-p_1-p_2}^{p_1+p_2} dy\,[(p_1+p_2)^2-y^2]^2\right]
   +\int\limits_{p_1}^{\infty}dp_2....\right\}
\end{align}
where $\int_{p_1}^{\infty}dp_2...$ is similar to $\int_0^{p_1}dp_2$ except
$p_1$ and $p_2$ are permuted in the arguments.  We have dropped the arguments
of $F$ for ease in notation.  Each $\int dy$ in Eq.\ \eqref{trans-app:collterm6}
is analytic:
\begin{align}
  J_1(p_1,p_2,p_3) &\equiv
  \int\limits_{p_1+p_2-2p_3}^{p_1+p_2} dy\,[(p_1+p_2)^2-y^2]^2
  =\frac{16}{15}p_3^3[10(p_1+p_2)^2-15(p_1+p_2)p_3+6p_3^2],\\
  \label{trans-app_nunu:j2}J_2(p_1,p_2) &\equiv
  \int\limits_{p_1-p_2}^{p_1+p_2} dy\,[(p_1+p_2)^2-y^2]^2
  =\frac{16}{15}p_2^3[10p_1^2+5p_1p_2+p_2^2],\\
  J_3(p_1,p_2,p_3) &\equiv
  \int\limits_{2p_3-p_1-p_2}^{p_1+p_2} dy\,[(p_1+p_2)^2-y^2]^2
  =\frac{16}{15}[(p_1+p_2)^5-10(p_1+p_2)^2p_3^3+15(p_1+p_2)p_3^4-6p_3^5].
\end{align}
We are assuming that particle 1 and 2 are in the same flavor state.  In this
case, the symmetrization factor is $S_1=1/2$, and Eq.\ \eqref{trans-app:collterm6}
becomes:
\begin{alignat}{3}
  I=\frac{G_F^2}{(2\pi)^3p_1^2}\left\{\vphantom{\int\limits_0^{p_1}}\right.
  &{}&&\int\limits_0^{p_1} dp_2
  \left[\int\limits_0^{p_2}dp_3\,FJ_1(p_1,p_2,p_3)
  +\int\limits_{p_2}^{p_1}dp_3\,FJ_2(p_1,p_2)
  +\int\limits_{p_1}^{p_1+p_2}dp_3\,FJ_3(p_1,p_2,p_3)
  \right]\nonumber\\
  &+&&\int\limits_{p_1}^{\infty} dp_2
  \left.\left[\int\limits_0^{p_1}dp_3\,FJ_1(p_1,p_2,p_3)
  +\int\limits_{p_1}^{p_2}dp_3\,FJ_2(p_2,p_1)
  +\int\limits_{p_2}^{p_1+p_2}dp_3\,FJ_3(p_1,p_2,p_3)
  \right]\right\}.\label{trans-app:collterm7}
\end{alignat}
\end{widetext}
In our nomenclature, the $\int dp_2$ in Eq.\ \eqref{trans-app:collterm7} is the
outer integral and the $\int dp_3$ is the inner integral.  Notice that for
$\int_{p_1}^{\infty} dp_2$, the arguments of $J_2$ and the limits of
integration for each $\int dp_3$ are permuted in $p_1$ and $p_2$.


\section{Other collision terms}\label{trans-app:other}

This appendix gives the reduction of the collision integral for the other
processes in Table \ref{trans-tab:ssamps}.  Notice that the indexing of the
particle species may be different than that presented in
Table \ref{trans-tab:ssamps}, yielding different \ssamp.  We adopted the changes
to simplify the mathematics involved in computing the collision integral.

\subsection{$\nu_i+\nu_j\leftrightarrow\nu_i+\nu_j$}

The summed--squared amplitude for this process is identical to the process in
Appendix \ref{trans-app:nunu} except for a factor of $1/4$.  Because the
symmetrization factor is $S=1$, there is an overall factor of $1/2$ on the
collision integral.  Therefore, the collision integral for this process has the
same form as the collision integral in Appendix \ref{trans-app:nunu}.

\subsection{$\nu_i+\overline{\nu}_i\leftrightarrow\overline{\nu}_i+\nu_i$}\label{trans-app:nunubar}

All of the neutrinos and anti-neutrinos have the same flavor for this process.
We write the reaction with the following indices:
\beq
  \nu(1)+\overline{\nu}(2)\leftrightarrow\overline{\nu}(3)+\nu(4),
\eeq
and simplify \ssamp as:
\beq
  \ssamp=2^5G_F^2(P_1\cdot P_3)^2.
\eeq
The collision integral is:
\begin{widetext}
\begin{alignat}{3}
  I=\frac{G_F^2}{2^2(2\pi)^3p_1^2}\left\{\vphantom{\int\limits_0^{p_1}}\right.
  &{}&&\int\limits_0^{p_1} dp_2\left[
  \int\limits_0^{p_2}dp_3\,FK_1(p_1,p_3)
  +\int\limits_{p_2}^{p_1}dp_3\,FK_2(p_1,p_2,p_3)
  +\int\limits_{p_1}^{p_1+p_2}dp_3\,FK_3(p_1,p_2,p_3)
  \right]\nonumber\\
  &+&&\left.\int\limits_{p_1}^{\infty} dp_2\left[
  \int\limits_0^{p_1}dp_3\,FK_1(p_1,p_3)
  +\int\limits_{p_1}^{p_2}dp_3\,FK_1(p_3,p_1)
  +\int\limits_{p_2}^{p_1+p_2}dp_3\,FK_3(p_1,p_2,p_3)
  \right]\right\}
\end{alignat}
In the above expression, $F=F(p_1,p_2,p_3,p_1+p_2-p_3)$.  The $K$ functions are the following:
\begin{align}
  K_1(p_1,p_3) &\equiv
  \int\limits_{p_1-p_3}^{p_1+p_3} dy\,[(p_1-p_3)^2-y^2]^2
  =\frac{16}{15}p_3^3[10p_1^2-5p_1p_3+p_3^2],\\
  \label{trans-app:j2}K_2(p_1,p_2,p_3) &\equiv
  \int\limits_{p_1-p_3}^{p_1+2p_2-p_3} dy\,[(p_1-p_3)^2-y^2]^2
  =\frac{16}{15}p_2^3[10(p_1-p_3)^2+15(p_1-p_3)p_2+6p_2^2],\\
  K_3(p_1,p_2,p_3) &\equiv
  \int\limits_{p_3-p_1}^{p_1+2p_2-p_3} dy\,[(p_1-p_3)^2-y^2]^2
  =\frac{16}{15}[(p_1-p_3)^5+10(p_1-p_3)^2p_2^3+15(p_1-p_3)p_2^4+6p_2^5].
\end{align}
\end{widetext}

\subsection{$\nu_i+\overline{\nu}_j\leftrightarrow\overline{\nu}_j+\nu_i$}\label{trans-app:nunubarij}

In this process, the neutrino and anti-neutrino have different flavors.  \ssamp
is identical to the process in Sec.\  \ref{trans-app:nunubar} except for a
factor of $1/4$.  Therefore, the collision integral for this process has the
same form as the collision integral in Sec.\  \ref{trans-app:nunubar}.

\subsection{$\nu_i+\overline{\nu}_i\leftrightarrow\overline{\nu}_j+\nu_j$}

In this process, a neutrino/anti-neutrino pair annihilate into another
neutrino/anti-neutrino pair of different flavor.  \ssamp is identical to the
process in Sec.\  \ref{trans-app:nunubarij} and so the collision integral is
the same.

\subsection{$\nue+e^-\leftrightarrow e^-+\nue$}\label{trans-app:nueem}

We write the reaction with the following indices:
\beq
  \nue(1)+e^-(2)\leftrightarrow e^-(3)+\nue(4),
\eeq
and simplify \ssamp as:
\begin{widetext}
\begin{alignat}{3}
  \ssamp=&{}&&2^5G_F^2(2\sin^2\theta_W+1)^2
  \left[(P_1\cdot Q_2)^2
  -\frac{2\sin^2\theta_W}{2\sin^2\theta_W+1}m_e^2(P_1\cdot Q_2)\right]\nonumber\\
  &+&&2^7G_F^2\sin^4\theta_W
  \left[(P_1\cdot Q_3)^2+\frac{2\sin^2\theta_W+1}{2\sin^2\theta_W}m_e^2(P_1\cdot Q_3)\right]\\
  \equiv&{}&& M_1^{'}(P_1\cdot Q_2) + M_2^{'}(P_1\cdot Q_3).
\end{alignat}
\end{widetext}
We will consider the collision integrals for $M_1^{'}$ and $M_2^{'}$ separately.

\subsubsection{$R_1$ collision integral}

We consider two cases for the collision integral for $M_1^{'}$: $p_1<m_e/2$ and $p_1>m_e/2$.

\noindent{\it Case 1}: $p_1<m_e/2$.  The collision integral is:
\begin{widetext}
\begin{alignat}{3}
  \nonumber R_1^{(1)} =\frac{1}{2^4(2\pi)^3p_1^2}\left[\vphantom{\int\limits_{m_e}^{\ecutiii}}\right.
  &{}&&\int\limits_{m_e}^{\ecutiii} dE_2\left(
  \int\limits_{m_e}^{E_2} dE_3\,FM_1^{(1)}
  +\int\limits_{E_2}^{\etransii} dE_3\,FM_1^{(2)}
  +\int\limits_{\etransii}^{\elimi} dE_3\,FM_1^{(3)}\right)\nonumber\\
  &+&&\int\limits_{\ecutiii}^{\ecuti} dE_2\left(
  \int\limits_{m_e}^{\etransii} dE_3\,FM_1^{(1)}
  +\int\limits_{\etransii}^{E_2} dE_3\,FM_1^{(4)}
  +\int\limits_{E_2}^{\elimi} dE_3\,FM_1^{(3)}\right)\nonumber\\
  &+&&\left.\int\limits_{\ecuti}^{\infty} dE_2\left(
  \int\limits_{\elimii}^{E_2} dE_3\,FM_1^{(4)}
  +\int\limits_{E_2}^{\elimi} dE_3\,FM_1^{(3)}\right)\right].
\end{alignat}

\noindent{\it Case 2}: $p_1>m_e/2$.  The collision integral is:
\begin{alignat}{3}
  \nonumber R_1^{(2)} =\frac{1}{2^4(2\pi)^3p_1^2}\left[\vphantom{\int\limits_{m_e}^{\ecutiii}}\right.
  &{}&&\int\limits_{m_e}^{\ecutiii} dE_2\left(
  \int\limits_{m_e}^{E_2} dE_3\,FM_1^{(1)}
  +\int\limits_{E_2}^{\etransii} dE_3\,FM_1^{(2)}
  +\int\limits_{\etransii}^{\elimi} dE_3\,FM_1^{(3)}\right)\nonumber\\
  &+&&\left.\int\limits_{\ecutiii}^{\infty} dE_2\left(
  \int\limits_{m_e}^{\etransii} dE_3\,FM_1^{(1)}
  +\int\limits_{\etransii}^{E_2} dE_3\,FM_1^{(4)}
  +\int\limits_{E_2}^{\elimi} dE_3\,FM_1^{(3)}\right)\right].
\end{alignat}

The following definitions apply to both cases:
\begin{align}
  F&\equiv F(p_1,E_2,E_3,p_1+E_2-E_3),\\
  \ecuti&\equiv m_e+\frac{2p_1^2}{m_e-2p_1},\\
  \ecutiii&\equiv\sqrt{p_1^2+m_e^2},\\
  \etransii&\equiv\frac{1}{2}\left(2p_1+E_2 - q_2+\frac{m_e^2}{2p_1+E_2 - q_2}\right),\\
  \elimi&\equiv\frac{1}{2}\left(2p_1+E_2+q_2+\frac{m_e^2}{2p_1+E_2+q_2}\right),\\
  \elimii&\equiv\etransii,\\
  M_1^{(1)}&\equiv\int\limits_{p_1+E_2-E_3-q_3}^{p_1+E_2-E_3+q_3}dyM_1^{'}\left\{
  \frac{1}{2}[(p_1+E_2)^2-m_e^2-y^2]\right\},\\
  M_1^{(2)}&\equiv\int\limits_{p_1-q_2}^{p_1+q_2}dyM_1^{'},\\
  M_1^{(3)}&\equiv\int\limits_{E_3+q_3-p_1-E_2}^{p_1+q_2}dyM_1^{'},\\
  M_1^{(4)}&\equiv\int\limits_{q_2-p_1}^{p_1+E_2-E_3+q_3}dyM_1^{'}.
\end{align}
\end{widetext}
The argument for $M_1^{'}$ is the same for each $M_1^{(i)}$.  The integral
expressions for $M_1^{(i)}$ are all analytic, but we do not write them out here
for the sake of brevity.

\subsubsection{$R_2$ collision integral}

We consider four cases for the collision integral for $M_2^{'}$:
\begin{alignat}{5}
  && \frac{p_1}{m_e} &<&&\frac{\sqrt{5}-1}{4},\\
  \frac{\sqrt{5}-1}{4} &<&\frac{p_1}{m_e} &<&&\frac{1}{2\sqrt{2}},\\
  \frac{1}{2\sqrt{2}} &<&\frac{p_1}{m_e} &<&&\frac{1}{2},\\
  \frac{1}{2} &<&\frac{p_1}{m_e}. &&&
\end{alignat}

\noindent{\it Case 1}: $p_1/m_e<(\sqrt{5}-1)/4$.  The collision integral is:
\begin{widetext}
\begin{alignat}{3}
  R_2^{(1)}=\frac{1}{2^4(2\pi)^3p_1^2}\left[\vphantom{\int\limits_{m_e}^{E_3}}\right.
  &{}&&\int\limits_{m_e}^{\ecutiii} dE_3\left(
  \int\limits_{m_e}^{E_3} dE_2\,FM_2^{(1)}
  +\int\limits_{E_3}^{\etransii} dE_2\,FM_2^{(2)}
  +\int\limits_{\etransii}^{\elimi} dE_2\,FM_2^{(3)}
  \right)\nonumber\\
  &+&&\int\limits_{\ecutiii}^{\ecutii} dE_3
  \left(\int\limits_{m_e}^{\etransii} dE_2\,FM_2^{(1)}
  +\int\limits_{\etransii}^{E_3} dE_2\,FM_2^{(4)}
  +\int\limits_{E_3}^{\elimi} dE_2\,FM_2^{(3)}\right)\nonumber\\
  &+&&\int\limits_{\ecutii}^{\ecuti} dE_3
  \left(\int\limits_{\elimii}^{E_3} dE_2\,FM_2^{(4)}
  +\int\limits_{E_3}^{\elimi} dE_2\,FM_2^{(3)}\right)\nonumber\\
  &+&&\left.\int\limits_{\ecuti}^{\infty} dE_3\left(
  \int\limits_{\elimii}^{E_3} dE_2\,FM_2^{(4)}
  +\int\limits_{E_3}^{\infty} dE_2\,FM_2^{(3)}
  \right)\right].
\end{alignat}

\noindent{\it Case 2}: $(\sqrt{5}-1)/4<p_1/m_e<1/(2\sqrt{2})$.  The collision integral is:
\begin{alignat}{3}
  R_2^{(2)}=\frac{1}{2^4(2\pi)^3p_1^2}\left[\vphantom{\int\limits_{m_e}^{E_3}}\right.
  &{}&&\int\limits_{m_e}^{\ecutiii} dE_3\left(
  \int\limits_{m_e}^{E_3} dE_2\,FM_2^{(1)}
  +\int\limits_{E_3}^{\etransii} dE_2\,FM_2^{(2)}
  +\int\limits_{\etransii}^{\elimi} dE_2\,FM_2^{(3)}
  \right)\nonumber\\
  &+&&\int\limits_{\ecutiii}^{\ecuti} dE_3
  \left(\int\limits_{m_e}^{\etransii} dE_2\,FM_2^{(1)}
  +\int\limits_{\etransii}^{E_3} dE_2\,FM_2^{(4)}
  +\int\limits_{E_3}^{\elimi} dE_2\,FM_2^{(3)}\right)\nonumber\\
  &+&&\int\limits_{\ecuti}^{\ecutii} dE_3
  \left(\int\limits_{m_e}^{\etransii} dE_2\,FM_2^{(1)}
  +\int\limits_{\etransii}^{E_3} dE_2\,FM_2^{(4)}
  +\int\limits_{E_3}^{\infty}dE_2\,FM_2^{(3)}\right)\nonumber\\
  &+&&\left.\int\limits_{\ecutii}^{\infty} dE_3\left(
  \int\limits_{\elimii}^{E_3} dE_2\,FM_2^{(4)}
  +\int\limits_{E_3}^{\infty} dE_2\,FM_2^{(3)}
  \right)\right].
\end{alignat}

\noindent{\it Case 3}: $1/(2\sqrt{2})<p_1/m_e<1/2$.  The collision integral is:
\begin{alignat}{3}
  R_2^{(3)}=\frac{1}{2^4(2\pi)^3p_1^2}\left[\vphantom{\int\limits_{m_e}^{E_3}}\right.
  &{}&&\int\limits_{m_e}^{\ecuti} dE_3\left(
  \int\limits_{m_e}^{E_3} dE_2\,FM_2^{(1)}
  +\int\limits_{E_3}^{\etransii} dE_2\,FM_2^{(2)}
  +\int\limits_{\etransii}^{\elimi} dE_2\,FM_2^{(3)}
  \right)\nonumber\\
  &+&&\int\limits_{\ecuti}^{\ecutiii} dE_3
  \left(\int\limits_{m_e}^{E_3} dE_2\,FM_2^{(1)}
  +\int\limits_{E_3}^{\etransii} dE_2\,FM_2^{(2)}
  +\int\limits_{\etransii}^{\infty} dE_2\,FM_2^{(3)}\right)\nonumber\\
  &+&&\int\limits_{\ecutiii}^{\ecutii} dE_3
  \left(\int\limits_{m_e}^{\etransii} dE_2\,FM_2^{(1)}
  +\int\limits_{\etransii}^{E_3} dE_2\,FM_2^{(4)}
  +\int\limits_{E_3}^{\infty}dE_2\,FM_2^{(3)}\right)\nonumber\\
  &+&&\left.\int\limits_{\ecutii}^{\infty} dE_3\left(
  \int\limits_{\elimii}^{E_3} dE_2\,FM_2^{(4)}
  +\int\limits_{E_3}^{\infty} dE_2\,FM_2^{(3)}
  \right)\right].
\end{alignat}

\noindent{\it Case 4}: $p_1/m_e>1/2$.  The collision integral is:
\begin{alignat}{3}
  R_2^{(4)}=\frac{1}{2^4(2\pi)^3p_1^2}\left[\vphantom{\int\limits_{m_e}^{E_3}}\right.
  &{}&&\int\limits_{m_e}^{\ecuti} dE_3\left(
  \int\limits_{m_e}^{E_3} dE_2\,FM_2^{(1)}
  +\int\limits_{E_3}^{\infty} dE_2\,FM_2^{(2)}
  \right)\nonumber\\
  &+&&\int\limits_{\ecuti}^{\ecutiii} dE_3
  \left(\int\limits_{m_e}^{E_3} dE_2\,FM_2^{(1)}
  +\int\limits_{E_3}^{\etransii} dE_2\,FM_2^{(2)}
  +\int\limits_{\etransii}^{\infty} dE_2\,FM_2^{(3)}\right)\nonumber\\
  &+&&\int\limits_{\ecutiii}^{\ecutii} dE_3
  \left(\int\limits_{m_e}^{\etransii} dE_2\,FM_2^{(1)}
  +\int\limits_{\etransii}^{E_3} dE_2\,FM_2^{(4)}
  +\int\limits_{E_3}^{\infty}dE_2\,FM_2^{(3)}\right)\nonumber\\
  &+&&\left.\int\limits_{\ecutii}^{\infty} dE_3\left(
  \int\limits_{\elimii}^{E_3} dE_2\,FM_2^{(4)}
  +\int\limits_{E_3}^{\infty} dE_2\,FM_2^{(3)}
  \right)\right].
\end{alignat}

The following definitions apply to all cases.  Note that some of the below definitions are incongruous with the definitions for $R_1$:
\begin{align}
  F&\equiv F(p_1,E_2,E_3,p_1+E_2-E_3),\\
  \ecuti&\equiv p_1+\frac{m_e^2}{4p_1},\\
  \ecutii&\equiv p_1+m_e\frac{p_1+m_e}{2p_1+m_e},\\
  \ecutiii&\equiv\sqrt{p_1^2 + m_e^2},\\
  \etransii&\equiv\frac{1}{2}\left(
  E_3+q_3-2p_1+\frac{m_e^2}{E_3+q_3-2p_1}\right),\\
  \elimi&\equiv\frac{1}{2}\left(
  E_3-q_3-2p_1+\frac{m_e^2}{E_3-q_3-2p_1}\right),\\
  \elimii&\equiv\etransii,\\
  M_2^{(1)}&\equiv\int\limits_{p_1-E_3+E_2-q_2}^{p_1-E_3+E_2+q_2}dy\,M_2^{'}\left\{
  \frac{1}{2}[y^2+m_e^2-(p_1-E_3)^2]\right\},\\
  M_2^{(2)}&\equiv\int\limits_{p_1-q_3}^{p_1+q_3}dy\,M_2^{'},\\
  M_2^{(3)}&\equiv\int\limits_{E_3-p_1-E_2+q_2}^{p_1+q_3}dy\,M_2^{'},\\
  M_2^{(4)}&\equiv\int\limits_{q_3-p_1}^{p_1-E_3+E_2+q_2}dy\,M_2^{'}.
\end{align}
\end{widetext}

\subsection{$\nu_{\mu(\tau)}+e^-\leftrightarrow e^-+\nu_{\mu(\tau)}$}

The \ssamp in this case is identical to the \ssamp in Sec.\
\ref{trans-app:nueem}, except for the transformation
$2\sin^2\theta_W+1\rightarrow2\sin^2\theta_W-1$.  Therefore, the structure of
the collision integral for this process is the same as Sec.\
\ref{trans-app:nueem}.

\subsection{$\nue+e^+\leftrightarrow e^++\nue$}\label{trans-app:nueep}

We write the reaction with the following indices:
\beq
  \nue(1)+e^+(2)\leftrightarrow e^+(3)+\nue(4),
\eeq
and simplify \ssamp as:
\begin{widetext}
\begin{alignat}{3}
  \ssamp=&{}&&2^5G_F^2(2\sin^2\theta_W+1)^2
  \left[(P_1\cdot Q_3)^2
  -\frac{2\sin^2\theta_W}{2\sin^2\theta_W+1}m_e^2(P_1\cdot Q_3)\right]\nonumber\\
  &+&&2^7G_F^2\sin^4\theta_W
  \left[(P_1\cdot Q_2)^2+\frac{2\sin^2\theta_W+1}{2\sin^2\theta_W}m_e^2(P_1\cdot Q_2)\right]\\
  =&{}&& M_1^{'}(P_1\cdot Q_3) + M_2^{'}(P_1\cdot Q_2),
\end{alignat}
\end{widetext}
where $M_1^{'}$ and $M_2^{'}$ are the same functions as in Sec.\
\ref{trans-app:nueem}.  Therefore, we can use the same collision integrals as
Sec.\  \ref{trans-app:nueem} but use $M_2^{'}$ in the integrands of $R_1$,
and $M_1^{'}$ in the integrands of $R_2$.

\subsection{$\nu_{\mu(\tau)}+e^+\leftrightarrow e^++\nu_{\mu(\tau)}$}

The \ssamp is the same as in Sec.\  \ref{trans-app:nueep} except for the
transformation $2\sin^2\theta_W+1\rightarrow2\sin^2\theta_W-1$.  Therefore, the
structure of the collision integral for this process is the same as Sec.\
\ref{trans-app:nueep}.

\subsection{$\nue+\bnue\leftrightarrow e^-+e^+$}\label{trans-app:nueepma}

We write the reaction with the following indices:
\beq
  \nue(1)+\bnue(4)\leftrightarrow e^+(2)+e^-(3),
\eeq
and simplify \ssamp as:
\begin{widetext}
\begin{alignat}{3}
  \langle|\mathcal{M}|^2\rangle=&{}&&2^5G_F^2(1+2\sin^2\theta_W)^2\left[
  (P_1\cdot Q_2)^2+
  \frac{2\sin^2\theta_W}{1+2\sin^2\theta_W}m_e^2(P_1\cdot Q_2)\right]\nonumber\\
  &+&&2^7G_F^2\sin^4\theta_W\left[
  (P_1\cdot Q_3)^2
  +\frac{1+2\sin^2\theta_W}{2\sin^2\theta_W}m_e^2(P_1\cdot Q_3)\right]\\
  \equiv&{}&& L_1^{'}(P_1\cdot Q_2) + L_2^{'}(P_1\cdot Q_3).
\end{alignat}
\end{widetext}
We consider four cases for the collision integral:
\begin{alignat}{5}
  && \frac{p_1}{m_e} &<&&\frac{1}{2},\\
  \frac{1}{2} &<&\frac{p_1}{m_e} &<&&\frac{1+\sqrt{5}}{4},\\
  \frac{1+\sqrt{5}}{4} &<&\frac{p_1}{m_e} &<&&1,\\
  1 &<&\frac{p_1}{m_e} &.&&
\end{alignat}

\noindent{\it Case 1}: $p_1/m_e<1/2$.  The collision integral is:
\begin{alignat}{3}
  I^{(1)} = \frac{1}{2^4(2\pi)^3p_1^2}&{}&&\int\limits_{\ecuti}^{\infty}d\eout
  \int\limits_{\elimi}^{\infty}d\ein
  (F_{oi}L_1^{(1)}+ F_{io}L_2^{(1)}).
\end{alignat}

\noindent{\it Case 2}: $1/2<p_1/m_e<\frac{1+\sqrt{5}}{4}$.  The collision integral is:
\begin{widetext}
\begin{alignat}{3}
  I^{(2)}= \frac{1}{2^4(2\pi)^3p_1^2}\left\{\vphantom{\int\limits_{m_e}^{\ecuti}}\right.
  &{}&&\int\limits_{m_e}^{\ecuti}d\eout\left[
  \int\limits_{\elimi}^{\etransi}d\ein(F_{oi}L_1^{(1)}+F_{io}L_2^{(1)})
  +\int\limits_{\etransi}^{\infty}d\ein(F_{oi}L_1^{(2)}+F_{io}L_2^{(2)})
  \right]\nonumber\\
  &+&&\int\limits_{\ecuti}^{\ecutii}d\eout\left[
  \int\limits_{\elimi}^{\infty}d\ein(F_{oi}L_1^{(1)}+F_{io}L_2^{(1)})
  \right]\nonumber\\
  &+&&\left.\int\limits_{\ecutii}^{\infty}d\eout\left[
  \int\limits_{m_e}^{\etransii}d\ein(F_{oi}L_1^{(3)}+F_{io}L_2^{(3)})
  +\int\limits_{\etransii}^{\infty}d\ein(F_{oi}L_1^{(1)}+F_{io}L_2^{(1)})
  \right]\right\}.
\end{alignat}

\noindent{\it Case 3}: $\frac{1+\sqrt{5}}{4}<p_1/m_e<1$.  The collision integral is:
\begin{alignat}{6}
  I^{(3)}= \frac{1}{2^4(2\pi)^3p_1^2}\left\{\vphantom{\int\limits_{m_e}^{\ecuti}}\right.
  &{}&&\int\limits_{m_e}^{\ecutii}d\eout\left[\vphantom{\int\limits_{\elimi}^{\etransi}}\right.&{}&&&\left.
  \int\limits_{\elimi}^{\etransi}d\ein(F_{oi}L_1^{(1)}+F_{io}L_2^{(1)})
  +\int\limits_{\etransi}^{\infty}d\ein(F_{oi}L_1^{(2)}+F_{io}L_2^{(2)})
  \right]\nonumber\\
  &+&&\int\limits_{\ecutii}^{\ecuti}d\eout\left[\vphantom{\int\limits_{\elimi}^{\etransi}}\right.&{}&&&
  \int\limits_{m_e}^{\etransii}d\ein(F_{oi}L_1^{(3)}+F_{io}L_2^{(3)})
  +\int\limits_{\etransii}^{\etransi}d\ein(F_{oi}L_1^{(1)}+F_{io}L_2^{(1)})\nonumber\\
  &{}&&{}&+&&&\left.\int\limits_{\etransi}^{\infty}d\ein(F_{oi}L_1^{(2)}+F_{io}L_2^{(2)})
  \right]\nonumber\\
  &+&&\int\limits_{\ecuti}^{\infty}d\eout\left[\vphantom{\int\limits_{\elimi}^{\etransi}}\right.&{}&&&\left.\left.
  \int\limits_{m_e}^{\etransii}d\ein(F_{oi}L_1^{(3)}+F_{io}L_2^{(3)})
  +\int\limits_{\etransii}^{\infty}d\ein(F_{oi}L_1^{(1)}+F_{io}L_2^{(1)})
  \right]\right\}.
\end{alignat}

\noindent{\it Case 4}: $p_1/m_e>1$.  The collision integral is:
\begin{alignat}{6}
  I^{(4)}= \frac{1}{2^4(2\pi)^3p_1^2}\left\{\vphantom{\int\limits_{m_e}^{\ecuti}}\right.
  &{}&&\int\limits_{m_e}^{\ecutii}d\eout\left[\vphantom{\int\limits_{\elimi}^{\etransi}}\right.&{}&&&\left.
  \int\limits_{\elimii}^{\etransii}d\ein(F_{oi}L_1^{(4)}+F_{io}L_2^{(4)})
  +\int\limits_{\etransii}^{\infty}d\ein(F_{oi}L_1^{(2)}+F_{io}L_2^{(2)})
  \right]\nonumber\\
  &+&&\int\limits_{\ecutii}^{p_1}d\eout\left[\vphantom{\int\limits_{\elimi}^{\etransi}}\right.&{}&&&
  \int\limits_{m_e}^{\etransi}d\ein(F_{oi}L_1^{(3)}+F_{io}L_2^{(3)})
  +\int\limits_{\etransi}^{\etransii}d\ein(F_{oi}L_1^{(4)}+F_{io}L_2^{(4)})\nonumber\\
  &{}&&{}&+&&&\left.\int\limits_{\etransii}^{\infty}d\ein(F_{oi}L_1^{(2)}+F_{io}L_2^{(2)})
  \right]\nonumber\\
  &+&&\int\limits_{p_1}^{\ecuti}d\eout\left[\vphantom{\int\limits_{\elimi}^{\etransi}}\right.&{}&&&
  \int\limits_{m_e}^{\etransii}d\ein(F_{oi}L_1^{(3)}+F_{io}L_2^{(3)})
  +\int\limits_{\etransii}^{\etransi}d\ein(F_{oi}L_1^{(1)}+F_{io}L_2^{(1)})\nonumber\\
  &{}&&{}&+&&&\left.\int\limits_{\etransi}^{\infty}d\ein(F_{oi}L_1^{(2)}+F_{io}L_2^{(2)})
  \right]\nonumber\\
  &+&&\int\limits_{\ecuti}^{\infty}d\eout\left[\vphantom{\int\limits_{\elimi}^{\etransi}}\right.&{}&&&\left.\left.
  \int\limits_{m_e}^{\etransii}d\ein(F_{oi}L_1^{(3)}+F_{io}L_2^{(3)})
  +\int\limits_{\etransii}^{\infty}d\ein(F_{oi}L_1^{(1)}+F_{io}L_2^{(1)})
  \right]\right\}.
\end{alignat}

The above expressions use the following definitions:
\begin{align}
  \qout&\equiv\sqrt{\eout^2-m_e^2},\\
  \qin&\equiv\sqrt{\ein^2-m_e^2},\\
  F_{oi}&\equiv F(p_1,\eout+\ein-p_1,\eout,\ein),\\
  F_{io}&\equiv F(p_1,\eout+\ein-p_1,\ein,\eout),\\
  \ecuti&\equiv p_1+\frac{m_e^2}{4p_1},\\
  \ecutii&\equiv\frac{1}{2}\left(2p_1-m_e+\frac{m_e^2}{2p_1-m_e}\right),\\
  \etransi&\equiv\frac{1}{2}\left(2p_1-\eout-\qout+\frac{m_e^2}{2p_1-\eout-\qout}\right),\\
  \etransii &\equiv \frac{1}{2}\left(2p_1-\eout+\qout+\frac{m_e^2}{2p_1-\eout+\qout}\right),\\
  \elimi&\equiv\etransii,\\
  \elimii&\equiv\etransi,
\end{align}
and
\begin{align}
  L_1^{(1)}&\equiv\int\limits_{\eout-p_1+\ein-\qin}^{p_1+\qout}dy
  \,L_1^{'}\left\{\frac{1}{2}[y^2+m_e^2-(p_1-\eout)^2]\right\},\\
  L_1^{(2)}&\equiv\int\limits_{p_1-\qout}^{p_1+\qout}dy\,L_1^{'},\\
  L_1^{(3)}&\equiv\int\limits_{\eout-p_1+\ein-\qin}^{\eout-p_1+\ein+\qin}dy\,L_1^{'},\\
  L_1^{(4)}&\equiv\int\limits_{p_1-\qout}^{\eout-p_1+\ein+\qin}dy\,L_1^{'},\\
  L_2^{(1)}&\equiv\int\limits_{\eout-p_1+\ein-\qin}^{p_1+\qout}dy
  \,L_2^{'}\left\{\frac{1}{2}[y^2+m_e^2-(p_1-\eout)^2]\right\},\\
  L_2^{(2)}&\equiv\int\limits_{p_1-\qout}^{p_1+\qout}dy\,L_2^{'},\\
  L_2^{(3)}&\equiv\int\limits_{\eout-p_1+\ein-\qin}^{\eout-p_1+\ein+\qin}dy\,L_2^{'},\\
  L_2^{(4)}&\equiv\int\limits_{p_1-\qout}^{\eout-p_1+\ein+\qin}dy\,L_2^{'}.
\end{align}

\end{widetext}

\subsection{$\nu_{\mu(\tau)}+\overline{\nu}_{\mu(\tau)}\leftrightarrow e^-+e^+$}

The \ssamp is the same as in Sec.\  \ref{trans-app:nueepma} except for the
transformation $2\sin^2\theta_W+1\rightarrow2\sin^2\theta_W-1$.  Therefore, the
structure of the collision integral for this process is the same as Sec.\
\ref{trans-app:nueepma}.

\bibliography{master}

\end{document}